\documentclass{tcibook}
\usepackage{fancyhea}
\usepackage{work}
\usepackage{bm}       
\usepackage{graphicx}
\usepackage{multirow}
\usepackage{lineno}
\usepackage{threeparttable}
\usepackage[normalem]{ulem}
\usepackage{color}
\usepackage[colorlinks = true,
            linkcolor = blue,
            urlcolor  = blue,
            citecolor = blue,
            anchorcolor = blue,
           hypertexnames=false]{hyperref}

\def\beq{\begin{equation}}
\def\eeq#1{\label{#1}\end{equation}}
\def\eeqn{\end{equation}}

\newenvironment{Eqnarray}%
   {\arraycolsep 0.14em\begin{eqnarray}}{\end{eqnarray}}
\def\beqa{\begin{Eqnarray}}
\def\eeqa#1{\label{#1}\end{Eqnarray}}
\def\eeqan{\end{Eqnarray}}

\let\bar=\overbar

\def\lsim{\mathrel{\raise.3ex\hbox{$<$\kern-.75em\lower1ex\hbox{$\sim$}}}}
\def\gsim{\mathrel{\raise.3ex\hbox{$>$\kern-.75em\lower1ex\hbox{$\sim$}}}}

\def\del{\partial}
\def\Dslash{\not{\hbox{\kern-4pt $D$}}}
\def\dslash{\not{\hbox{\kern-2pt $\del$}}}
\def\pslash{\not{\hbox{\kern-2pt $p$}}}
\def\ETmiss{\not{\hbox{\kern-4pt $E$}}_T}

\def\Dlr{\mathrel{\raise1.5ex\hbox{$\leftrightarrow$\kern-1em\lower1.5ex\hbox{$D$}}}}

\def\MSB{{\bar{M \kern -2pt S}}}
\def\msb{{\bar{\scriptsize M \kern -1pt S}}}

\def\drb{{\bar{\scriptsize D \kern -1pt R}}}

\makeatletter 
\def\myfootnote{\xdef\@thefnmark{}\@footnotetext} 
\makeatother 

\begin{document}


\pagenumbering{roman}

\parindent=0pt
\parskip=8pt
\setlength{\evensidemargin}{0pt}
\setlength{\oddsidemargin}{0pt}
\setlength{\marginparsep}{0.0in}
\setlength{\marginparwidth}{0.0in}
\marginparpush=0pt



\renewcommand{\chapname}{chap:intro_}
\renewcommand{\chapterdir}{.}
\renewcommand{\arraystretch}{1.25}
\addtolength{\arraycolsep}{-3pt}

\thispagestyle{empty}
\begin{center}

 \mbox{\null}
\vskip -1.0in
 \rightline{\begin{tabular}{l}
 FERMILAB-CONF-23-008\\
SLAC-PUB-17717\\
January 2023\\
 \end{tabular}}

\begin{Large}
 {\bf Report of the 
2021  U.S.  Community Study   \\
     on the Future of Particle Physics \\
(Snowmass 2021)} 

\medskip

{\bf  Summary Chapter  }

\end{Large}


\begin{large} {\bf 2021 -- 2022 Snowmass Steering Group:}  

\smallskip

{Joel N. Butler$^1$, 
R. Sekhar Chivukula$^2$, \\
Andr{\'e} de Gouv{\^e}a$^3$, 
Tao Han$^4$, 
Young-Kee Kim$^5$, \\ Priscilla~Cushman$^6$  (APS Division of Particles
and Fields),\\
Glennys~R.~Farrar$^7$ (APS Division of Astrophysics),\\   Yury~G.~Kolomensky$^8$ (APS  Division of Nuclear Physics),
\\ Sergei~Nagaitsev$^1$ (APS  Division of Physics of Beams),\\
Nicol{\'a}s~Yunes$^{9}$ (APS Division of Gravitational Physics)} \end{large}

\medskip

\begin{large} {\bf Snowmass 2021 Frontier Conveners:}

\smallskip

  Stephen~Gourlay$^{10}$, Tor~Raubenheimer$^{11}$, Vladimir~Shiltsev$^1$ (Accelerator),
K{\'e}t{\'e}vi~A.~Assamagan$^{12}$, Breese~Quinn$^{13}$  (Community
Engagement), 
V.~Daniel~Elvira$^{1}$, Steven~Gottlieb$^{14}$, Benjamin~Nachman$^{10}$ (Computational),
Aaron~S.~Chou$^1$, Marcelle~Soares-Santos$^{15}$, Tim M.~P.~Tait$^{16}$ (Cosmic),
Meenakshi~Narain$^{17 \dagger}$\myfootnote{$^\dagger$deceased, Jan. 1, 2023.}, Laura~Reina$^{18}$, Alessandro Tricoli$^{12}$
(Energy), 
Phillip S. Barbeau$^{19}$, Petra Merkel$^1$, Jinlong Zhang$^{20}$
(Instrumentation),  
Patrick~Huber$^{21}$, Kate~Scholberg$^{19}$, Elizabeth~Worcester$^{12}$  (Neutrino),
Marina~Artuso$^{22}$, Robert H. Bernstein$^1$, Alexey A. Petrov$^{23}$
(Rare Processes), 
Nathaniel~Craig$^{24}$, Csaba~Cs{\'a}ki$^{25}$, Aida~X.~El-Khadra$^{9}$ (Theory), 
Laura~Baudis$^{26}$, Jeter~Hall$^{27}$, Kevin~T.~Lesko$^{10}$, John~L.~Orrell$^{28}$  (Underground Facilities), 
Julia~Gonski$^{29}$, Fernanda~Psihas$^1$, Sara~M.~Simon$^{1}$ (Early
Career) \end{large}

\medskip

\begin{large} {\bf Editor: }   Michael E. Peskin$^{11}$  \end{large}
\smallskip


\begin{large}{\em  Abstract } 
\end{large}
\end{center}

 \smallskip

The 2021-22 High-Energy Physics Community Planning Exercise (a.k.a. “Snowmass 2021”) was organized by the Division of Particles and Fields of the American Physical Society. Snowmass 2021 was a scientific study that provided an opportunity for the entire U.S. particle physics community, along with its international partners, to identify the most important scientific questions in High Energy Physics for the following decade, with an eye to the decade after that,  and the experiments, facilities,  infrastructure, and R\&D needed to pursue them. This Snowmass summary report synthesizes the lessons learned and the main conclusions of the Community Planning Exercise as a whole and presents a community-informed synopsis of U.S. particle physics at the beginning of 2023. This document, along with the Snowmass reports from the various subfields, will provide input to the 2023 Particle Physics Project Prioritization Panel (P5) subpanel of the U.S. High-Energy Physics Advisory Panel (HEPAP), and will help to guide and inform the activity of the U.S. particle physics community during the next decade and beyond.

\vfill

\thispagestyle{empty}

\mbox{\null}


$^1$  Fermi National Accelerator Laboratory, Batavia IL 60510  USA \\
$^2$  University of California, San Diego, La Jolla CA 92093 USA \\
$^3$  Northwestern University, Evanston IL 60208  USA \\
$^4$  University of Pittsburgh, Pittsburgh PA 15260  USA \\
$^5$  University of Chicago, Chicago IL60637  USA \\
$^6$  University of Minnesota, Minneapolis MN 55455  USA  \\
$^7$  New York University, New York NY 10027  USA \\
$^8$  University of California, Berkeley CA 94720  USA \\
$^9$  University of Illinois at Urbana-Champaign, Urbana IL 61801  USA \\
$^{10}$  Lawrence Berkeley National Laboratory, Berkeley CA 94720  USA \\
$^{11}$  SLAC National Accelerator Laboratory, Menlo Park CA 94025 USA \\
$^{12}$  Brookhaven National Laboratory, Upton NY 11973  USA \\
$^{13}$  University of Mississippi, University MS 38677 USA \\
$^{14}$  University of Indiana, Bloomington IN 47405 USA \\
$^{15}$  University of Michigan, Ann Arbor MI 48109  USA \\
$^{16}$  University of California, Irvine CA 92697 USA \\
$^{17}$  Brown University, Providence RI 02912  USA \\
$^{18}$  Florida State University, Tallahassee FL 32306  USA \\
$^{19}$  Duke University, Durham, NC 27708  USA \\
$^{20}$  Argonne National Laboratory, Lemont, IL 60439  USA \\
$^{21}$  Virginia Tech, Blacksburg, VA 24061  USA \\
$^{22}$  Syracuse University, Syracuse NY 13244  USA \\
$^{23}$  University of South Carolina, Columbia, SC 29208 USA \\
$^{24}$  University of Calilfornia, Santa Barbara CA 93106 USA \\
$^{25}$  Cornell University, Ithaca NY 14853 USA \\
$^{26}$  University of Z\"urich, 8006 Z\"urich SWITZERLAND \\
$^{27}$  SNOLAB, Lively, ON P3Y 1N2  CANADA \\
$^{28}$  Pacific Northwest National Laboratory, Richland, WA 99354  USA \\
$^{29}$  Columbia University, New York NY 10027  USA \\

\vfill

\newpage

\tableofcontents

\newpage

 \pagenumbering{arabic}



\chapter{The Future of U.S. Particle Physics: Summary of Snowmass 2021} 

\bigskip

\section{Introduction}

Particle physics seeks to identify the elementary constituents of matter and the forces and principles by which they interact.  By looking at matter at the smallest distance scales, physicists have identified the interactions that operate at the subatomic level: the strong, weak, and electromagnetic forces -- the last two of which are unified into a single electroweak theory.  The force carriers of these interactions,
the photon, the $W$, and $Z$ bosons of the electroweak theory, and the gluons of the strong interactions, along with the quarks and leptons, the fermionic components of matter, form the basis of the Standard Model (SM) of particle physics. 

The last missing element in the SM was the agent of electroweak symmetry breaking, which gives mass to the $W$ and $Z$ bosons as well as the quarks and leptons. The new scalar particle discovered at the Large Hadron Collider (LHC) by the ATLAS and CMS experiments in 2012 is produced and decays in a manner that suggested that it is the ``Higgs boson'', the excitation of an elementary scalar Higgs field which accommodates the simplest mechanism for electroweak symmetry breaking. In the decade since then, ATLAS and CMS have shown that the couplings of this new boson to the heaviest SM particles, the top quark, the tau lepton, and the $b$-quark, all of which are third-generation fermions, as well as to the $W$ and $Z$, are all consistent with its interpretation as the Higgs boson. While there is plenty of room for new physics 
and there is much work to be done to fully study this remarkable, unique scalar particle, its discovery demonstrates that the SM is at least approximately correct and accurately describes elementary particles and their interactions at energies up to a few hundred GeV. 

While the SM explains many aspects of the fundamental physics world with impressive accuracy, it is nonetheless silent on several issues of fundamental importance, including the ``small'' mass, 125 GeV, of the  Higgs boson, which is unstable to radiative corrections from interactions at higher energy scales;   the  excess of matter over antimatter (the baryon asymmetry of the universe); why the strong interaction does not exhibit CP violation;
the mechanism behind the tiny yet nonzero neutrino masses;
the issue of flavor, namely why are there three generations of quarks and leptons with disparate masses and mixings and why is the mixing pattern observed in the lepton sector so different from that in the quark sector; the origin and nature of dark matter and dark energy; 
the mechanism for cosmic inflation in the early Universe;  
how gravity fits into the other fundamental interactions and how our general relativistic understanding of space and time can be reconciled with quantum mechanics. 
These are profound questions whose answers will fundamentally change our understanding of the physical world. 
 Since the answers to these questions cannot be found within the SM, 
there must be new Beyond the  Standard Model (BSM) particles and interactions that explain these aspects of nature. 

We have many theoretical ideas that address some or all of these problems but, for now, no unambiguous experimental guidance  on which ones are correct.  Hints do exist, including the tension between some theoretical calculations of the anomalous magnetic moment of the muon and the measured value, anomalies in neutrino physics exhibited in short-baseline experiments, a possible disagreement between the most recently published, most precise measurement of the $W$ boson mass and its theoretical value, and several anomalies in quark flavor physics.
With more and more precise data, some of these may disappear or have explanations within the SM, while others may evolve into puzzles whose solutions provide vital hints about BSM physics. Along the way, new anomalies might emerge, potentially pointing in unforeseen directions. The next big clue may be hiding within data already collected at the LHC or about to be collected  during the planned High Luminosity LHC (HL-LHC) era or in the rare decays and properties of known particles. 
Ongoing or future oscillation experiments may reveal more surprises in the neutrino sector. The new physics may be at masses higher than we can currently reach or weakly coupled enough that it has yet to manifest itself in a way we are prepared to notice. 

At present we have come to understand that we are confronted by a vast range of possibilities for BSM physics. For example, the particle that might constitute dark matter can have a mass that spans 90 orders of magnitude. 
This is an exciting challenge. There is room for innovative ideas and 
a vast unexplored territory to investigate. There are extraordinary new accelerators and detectors that we can build, new and different observations we can make, along with enormous computing power and new techniques that permit us to collect and analyze unprecedented amounts of data using innovative, more powerful software tools. These new techniques enable us to probe the boundaries of the SM to unprecedented accuracy and extend the search for BSM physics to higher energy scales and weaker couplings. For these reasons, we believe that we will find the trail of the new physics that lies beyond the SM. 
This will take the efforts, imagination, and creativity of the community that participated in Snowmass 2021, joined by a large and even more diverse new generation of bold thinkers and intrepid doers. We, therefore, present the following vision that emerged from Snowmass:
\begin{quote} 
{\em Lead the exploration of the fundamental nature of matter, energy, space and time, by using ground-breaking theoretical, observational, and experimental methods; developing state-of-the-art technology for fundamental science and for the benefit of society; training and employing a diverse and world-class workforce of physicists, engineers, technicians, and computer scientists from universities and laboratories across the nation; collaborating closely with our global partners and with colleagues in adjacent areas of science; and probing the boundaries of the Standard Model of particle physics to illuminate the exciting terrain beyond, and to address the deepest mysteries in the Universe.}
\end{quote}
We conclude from our long community study that, in the spirit of Snowmass, we stand ready with great enthusiasm to “delve deep, search wide, and aim high” to advance our understanding of the Universe!

\subsection{The Snowmass Process, Introduction and History} 
This report  presents the main conclusions of  Snowmass 2021, the most recent of the U.S. High Energy Physics (HEP) Community Planning Exercises, sponsored by the Division of Particles and Fields (DPF) of the American Physical Society (APS), with strong consultation from the aligned APS Divisions of Nuclear Physics, Astrophysics, Gravitational Physics, and Physics of Beams. These exercises have occurred roughly  every seven to twelve years since 1982. The goal of these community studies has been to identify the most important scientific questions in HEP for the following decade, with an eye to the decade after that,  and the facilities, infrastructure, and R\&D needed to pursue them. Although focused on the U.S. HEP program, HEP is a global enterprise  and international participation was encouraged. Many scientists from outside the U.S. have contributed to Snowmass 2021. HEP has important connections to many other fields of physics and  physicists from all related fields have participated. For many years these efforts concluded with a workshop in the summer in Snowmass, Colorado and so these exercises became known as the Snowmass Summer Study and Workshops. The last such exercise was held in  2013 \cite{Rosner:2014pja}. Its concluding workshop was for the first time held outside of Snowmass at the University of Minnesota. However, we retained the name ``Snowmass” to connect it to and commemorate the previous studies.

In 2013  and all previous editions, the planning exercise focused on scientific issues without considering detailed budget constraints or establishing project or funding priorities. All scientifically credible ideas and proposals were welcome. These science exercises were each followed by a subpanel of High Energy Physics Advisory Panel (HEPAP) that advised the funding agencies on which of the many proposed experiments and projects should be carried out, given realistic budget constraints. In 2013, the output of the Snowmass study provided input to the HEPAP subpanel called the Particle Physics Project Prioritization Panel, “P5”, which was charged by the U.S. Department of Energy (DOE) and the National Science Foundation (NSF) with the task of recommending plans for U.S. investments in particle physics under specific budget scenarios for the coming decade. P5 began its work in the fall of 2013 and produced a report \cite{P5-HEPAP}  with its recommendations in the spring of 2014.

\begin{figure}
\begin{center}
\includegraphics[width=0.80\hsize]{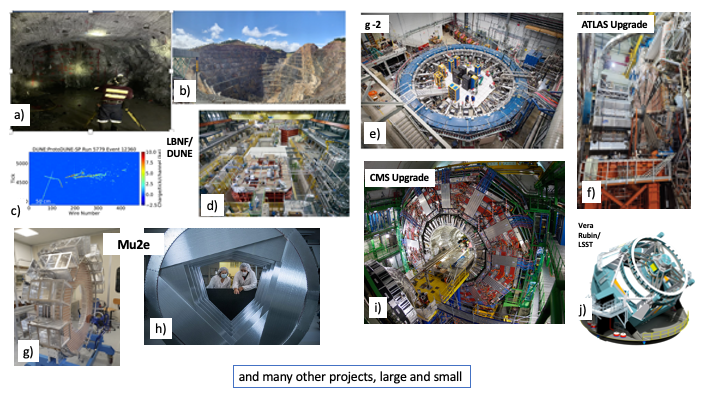}
\end{center}
\caption{Examples of some of the projects prioritized in the Snowmass 2013/P5 2014 process. {\bf LBNF/DUNE:} a) cavern excavation; b) ``open cut" to receive excavated rock; c) interactions recorded by Proto-DUNE-SP; d) Neutrino Platform at CERN;  {\bf Muon g-2:}  e) muon storage ring; {\bf ATLAS Upgrade:} f) New Small Wheel; {\bf Mu2e:} g) calorimeter support frame; h) Tracker planes; {\bf CMS Upgrade:} i) Phase 1 upgrade pixel installation; {\bf Vera Rubin Observatory:} j) schematic.}
\label{fig:projects}
\end{figure}

At the start of its deliberations, P5 examined the findings of the Snowmass study and identified five main science Drivers for HEP:
\begin{enumerate}
\item Use the Higgs Boson as a Tool for Discovery, 
\item Pursue the Physics Associated with Neutrino Mass, 
\item Identify the New Physics of Dark Matter, 
\item Understand Cosmic Acceleration: Dark Energy and Inflation, 
\item Explore the Unknown: New Particles, Interactions, and Physical Principles.
\end{enumerate}
 It also made 29 recommendations concerning the projects and we have been constructing the favored ones, some of which are shown in Fig.~\ref{fig:projects}, since 2015. The two-step approach to planning, the community Snowmass study, centered on the science,  and the subsequent P5 prioritization, also taking into account budget constraints,  led to a clear understanding and broad acceptance of the 2014 P5 report by the HEP community and beyond. 

Among the projects assigned the highest priority by P5 was participation in the High Luminosity Upgrade of the Large Hadron Collider and the ATLAS and CMS experiments at CERN; the LSST/Rubin Observatory; and the construction of PIP-II SRF proton linac at Fermilab and the LBNF/DUNE (Long Baseline Neutrino Facility/Deep Underground Neutrino Experiment) long baseline neutrino experiment.
The full suite of experiments enabled by the projects recommended by the 2013/14 Snowmass/P5 process was chosen to enable U.S. high energy physicists and their partners to  examine the properties of matter at the tiniest distance scales ever probed and set the stage for the exciting experimental investigations of the decade just past and into the decade to come.

\subsection{Introduction to Snowmass 2021}

\begin{figure}
\begin{center}
\includegraphics[width=0.90\hsize]{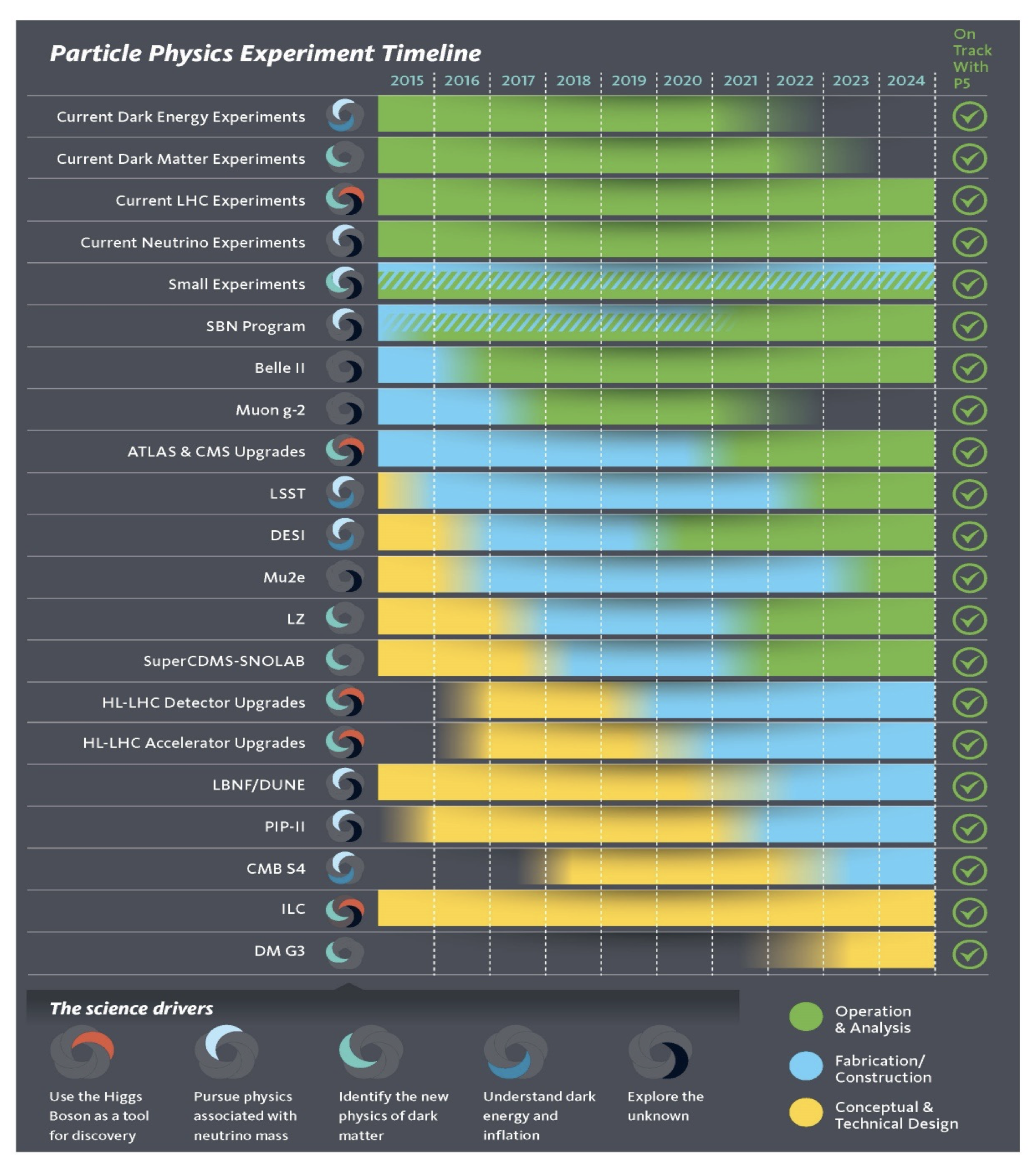}
\end{center}
\caption{Status of projects supported all or in part by DOE in operation (green), under construction or in production (blue), or still under design and project definition (yellow), shown at the Snowmass Community Summer Study, July 2022.}
\label{figures:Current_project_status}
\end{figure}

In 2019, with the construction of many of the projects supported by the 2013/14 planning process well underway or in an advanced stage of planning, as shown in Fig.~\ref{figures:Current_project_status}, 
the U.S. High Energy Physics Advisory Panel (HEPAP) recognized the need for a new community study to develop the initiatives in the decade starting in 2025. DPF began preparation for Snowmass 2021 community study in 2019. 
This ``Snowmass 2021" HEP Community Planning Exercise formally began with a ``kick-off" meeting at the 2020 APS April Meeting and was followed by a  Community-wide Planning Meeting (CPM), in October 2020, in which many activities, meetings, and workshops were mapped out. The exercise was to have concluded in July of 2021 with a final workshop in Seattle hosted by the University of Washington. The COVID-19 pandemic severely disrupted our plans and forced us to pause work between January and September of 2021, especially to lighten the burden on our younger scientists, many of whom had to take care of and school their young children.  The final workshop in Seattle was delayed until the summer of 2022 and the  start of the P5 panel was also delayed by one year by the DOE and NSF.  We resumed work  for Snowmass in September of 2021 and, despite the continuing, but now diminished, challenges of COVID-19, our community accomplished an amazing amount of work and was well prepared for the Seattle meeting in July of 2022.

Snowmass 2021 was organized into ten working groups, or “Frontiers”:  Accelerator (AF),  Community Engagement (CEF), Computational (CompF), Cosmic (CF), Energy (EF), Instrumentation (IF), Neutrinos (NF), Rare Processes and Precision Measurements (RPF), Theory (TF), and Underground Facilities and Infrastructure (UF). These Frontiers comprise a broad array of ground-breaking scientific research topics and the underlying technology and infrastructure needed to execute them, as well as, through the CEF,  a forum to examine how the U.S. HEP community can become more representative of and responsive to all members of our community and can engage with society as a whole.  Each Frontier divided its work into several Topical Groups (TGs).  The number of TGs varied by Frontier from 6 to 11 and there were approximately 80 overall. A call for two-page ``Letters of Interest (LOIs)'', due by August 31, 2020, resulted in over 1500 submissions. These were assigned to the TGs (recognizing overlaps). Each Topical Group encouraged its members to produce contributed papers more completely  describing their ideas and plans. Often, several  submitters of LOIs joined together to produce a single contributed paper. Over five hundred contributed papers, also known as white papers, were produced! The work  incorporated input from the ``European Strategy for Particle Physics" \cite{Eurostrat} and  other national and international planning exercises. 

Early Career physicists have been formally represented at Snowmass since 2001 and gained even more formal recognition in 2013. 
In 2021/22, a Snowmass Early Career (SEC) organization was formed to assist young physicists in contributing to the Snowmass process and to bring their issues into the community study.  The SEC was treated comparably to a full Frontier. There were also important connections among all the Frontiers, and we implemented a strong team of about 80 cross-frontier liaisons to make sure nothing fell between the cracks. International and interdisciplinary participation was encouraged. Our goal was for everyone to be able to participate and for all voices to be heard. 

Leading up to the Seattle meeting, both before and after the pandemic pause, there were many TG workshops/meetings, cross-frontier/cross-TG workshops/meetings, and many studies carried out by groups and individuals. By the spring of 2022, nearly 500 white papers had already been submitted. By January 2022, it was possible for people to travel to and attend meetings with a relatively low risk of contracting COVID, and about half the Frontiers held ``all-frontier" meetings with both in-person and remote participation, with the other Frontiers relying completely on videoconferencing. The purpose of these meetings was to begin the synthesis of the white papers and the formulation of the topical group reports.

\begin{figure}
\begin{center}
\includegraphics[width=0.75\hsize]{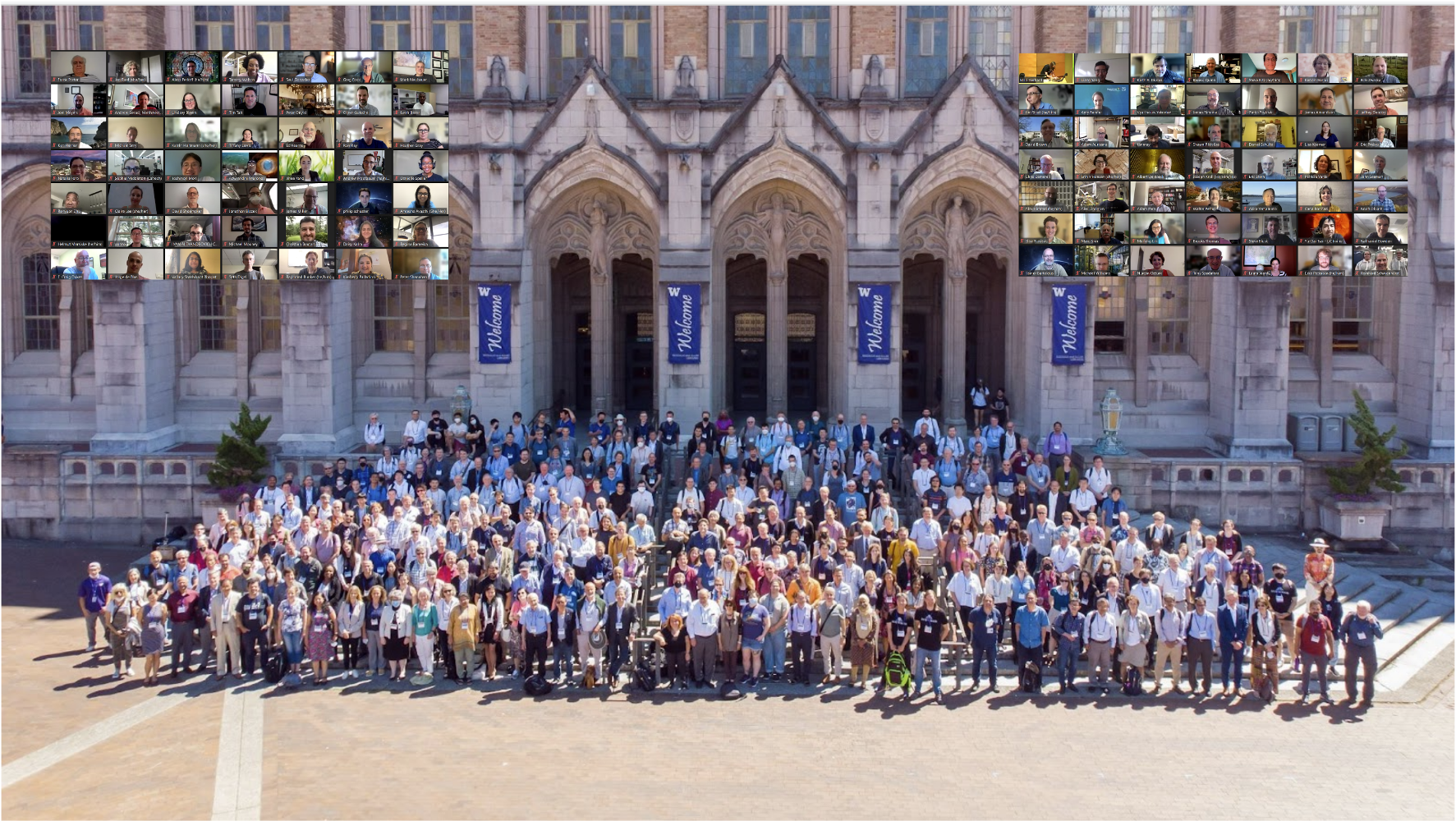}
\end{center}
\caption{Snowmass Community Summer Study, University of Washington, Seattle, a photograph showing in-person attendees  in the center and panels in the upper right- and left-hand corners showing some of the remote participants.}
\label{figures:css_group_photo}
\end{figure}

The final discussions and synthesis of all the white papers and other materials developed for Snowmass took place at the Community Summer Study and Workshop (CSS) at the University of Washington in Seattle, from July 17-26, 2022. Vaccines and other measures enabled us to conduct a safe, hybrid meeting, with an in-person attendance of about 700 and about 650 participants connecting remotely, as pictured in Fig.~\ref{figures:css_group_photo}.

The daily CSS program typically started at 8 am and ran until 7 pm. Days 2-8 were packed with parallel sessions dedicated to Frontier and cross-frontier meetings in the mornings and three 90-minute-long plenary sessions in the afternoons in which each Frontier presented talks designed to inform the entire community of its main plans and conclusions. There were also panel discussions of issues of common interest to the whole community.  Day 1 and days 9 and 10 consisted of all-plenary sessions that included special presentations of plans and planning processes by many leaders of U.S. and international institutes, laboratories, and funding agencies. 
Also, on days 8 and 9, each Frontier summarized its vision for the future and its main conclusions and recommendations, including the projects and facilities that would be needed to realize the vision.

On the final day of the CSS, Priscilla Cushman of the University of Minnesota put it all in perspective with her inspiring talk,  “Snowmass Highlights and Message Distillation".  As the audience held its breath, JoAnne Hewett of SLAC National Accelerator Laboratory and chairperson of the U.S. High Energy Physics Advisory Panel, announced that the new P5 chairperson is Hitoshi Murayama from UC Berkeley.

A full schedule for the Seattle meeting, along with talks and eventually recordings, can be found on the Snowmass 
Indico site \cite{CSS-Indico}.
Short versions of the Frontier Summaries \cite{Gourlay:2022odf} -- \cite{Baudis:2022qjb}, the  report of the Snowmass Early Career Scientists \cite{Agarwal:2022ldf}, and discussions of selected cross-cutting and interdisciplinary aspects of HEP \cite{Llatas:2022goe,Black:2022cth,Roser:2022sht,Boveia:2022adi}
are included below.

\subsection{Overview of the Snowmass Report}

This Snowmass Report represents the summary of the Snowmass 2021 Community Planning Exercise. It is organized around the work of the ten Frontiers and the SEC. Each Frontier and the SEC conducted its work in meetings of various kinds, mostly using Zoom because of COVID-19, but with some hybrid meetings with an in-person component  as the pandemic began to abate in the spring of 2022. By the late spring, the Topical Groups in each Frontier and the SEC began preparing their reports. All participants in the work were encouraged to read and comment on the various drafts. Revisions were made and eventually, a final consensus report emerged. By the time of the CSS, Topical Group reports of each Frontier were being synthesized into Frontier Reports. These also went through a process of drafting, followed by meetings, commentary, and iteration to achieve a consensus. 

The remainder of this Chapter consists of some very high-level conclusions, a discussion and assessment of the five science Drivers defined by P5/2014 and their applicability to HEP in the next decade and beyond, a brief summary of each of the ten Frontier reports and the SEC report, some examples of cross-cutting HEP activities, and some concluding observations.   In the full Snowmass Report, this summary chapter is followed by eleven chapters presenting  the full reports of the ten Frontiers and the SEC. These are followed  by chapters that discuss options for future particle colliders; the search for dark matter as an example of how  the Frontiers are working together, and with adjacent disciplines,  to solve one of  our most important puzzles; and the close connections between particle physics, astrophysics, and nuclear physics. Finally, there is a brief summary and conclusion for the whole Snowmass process. Key contributors are acknowledged at the end of the report. 

All Snowmass Frontier summaries, Topical Group summaries, and white papers are accessible from  the Snowmass Proceedings website \cite{Snowmass-Proceedings-Website}. The  Snowmass Report \cite{Snowmass-full-report} is also available from that site.

\section{High-Level Conclusions from Snowmass 2021}

There was broad agreement at Snowmass 2021 on the general principles needed to have a successful U.S. HEP program in the future.

\begin{itemize}

\item In 2014, P5 formulated five science Drivers that were meant to summarize and organize the scientific questions that HEP seeks to answer. 
These five Drivers have guided U.S. HEP for nearly a decade with great success. We have made tangible progress toward addressing them, and have used that progress to formulate the steps for the next decade that are outlined in this report. There was consensus in Snowmass 2021 that these Drivers were still appropriate for the next decade.
\begin{itemize}
\item  A proposal was made by the Rare Processes and Precision Measurements Frontier that the physics of flavor, currently included without explicit acknowledgment in the science Drivers,
be specifically named, given its unique characteristics and opportunities. 
One way of accomplishing this would be
for flavor physics to become a sixth Driver.
\end{itemize}

\item The completion of existing experiments and the construction and operation of approved projects, including those prioritized by the 2014 P5 such as the HL-LHC, LBNF/DUNE, PIP-II,
LSST/Rubin Observatory, and Mu2e programs, along with our many midsize and small experiments, are critical for addressing the science Drivers for the near term and for much of the next two decades.

\item As existing approved construction projects come to a completion, a broad and complementary set of  new projects should be considered a high priority to maintain the continuity of our investigations and opportunities for new discoveries.
\begin{itemize}
\item 
A list of the large-scale (estimated total project costs of approximately \$500M or larger) new projects or programs discussed at Snowmass is shown in Table~\ref{table:major-projects} and is discussed in section \ref{section:major-projects}  and in the Frontier summaries that follow. The proposals are examined in detail in the Snowmass 2021 Frontier and Topical Group reports.

\item The portfolio of HEP projects should continue to include a healthy breadth and balance of physics topics, experiment sizes, and timescales, supported by a dedicated, robust, ongoing funding process.  Medium- and smaller-scale projects are discussed more fully in section \ref{section:mid-scale-projects}, the reports in this volume, and the corresponding Topical Group reports.

\end{itemize}

\item Robust support for physics research programs at universities and national laboratories is essential to operate existing and planned experiments, analyze the data they collect, plan and construct upgrades and new experiments and projects, and educate the next generation of researchers and technical experts.

\item Greater support for the infrastructure and enabling technologies, namely accelerators, computation, detectors, and instrumentation, is essential. Both R\&D directed to specific future projects and generic research should be supported in these critical enabling technologies, as well as in new areas such as quantum science and machine learning.

\begin{itemize}
\item Some guidance to HEPAP  and the funding agencies in how the work of the enabling Frontiers should be coordinated and advanced would be very valuable. 
\item When R\&D projects produce successful tools, where appropriate, plans should be developed to convert them to products and support them throughout their useful lives.
\end{itemize}

\item HEP benefits from collaboration with adjacent scientific disciplines, such as 
nuclear physics, accelerator technology and beam physics, astronomy and astrophysics, gravitational physics, and atomic and molecular physics, and from interactions with industry, and contributes to them in return. Opportunities to strengthen and expand such collaborations are mutually beneficial and should be pursued, and new opportunities for collaboration are arising in other areas such as artificial intelligence and machine learning, and quantum information science and sensing.

\item The HEP community should strengthen connections with its early career researchers, foster their professional success both within and outside of academia, and ensure their voices are well-represented in physics and community planning.

\item HEP should take a cohesive and strategic approach to promote diversity, equity, and inclusion in high-energy physics, in collaboration with funding agencies, universities, adjacent scientific disciplines, APS/DPF, and others.
\begin{itemize}
\item The community should institute a broad array of practices and programs to reach and retain the diverse talent pool needed for success in achieving our scientific vision and to address the persistent under-representation of women scientists, LGBT+ scientists, scientists who are Black, Indigenous, and people of color (BIPOC), and scientists with disabilities.
\end{itemize}
\item The HEP community must engage in a coordinated way with five other interrelated communities: academia, the education communities providing instruction from kindergarten through postdoctoral
training, private industry, government policymakers, and the broader society. A structure for formulating a coordinated approach to achieve these goals should be created and provided with the resources needed for success.
\end{itemize}

\subsection{Large-Scale Future Proposals Discussed at Snowmass 2021}
\label{section:major-projects}

Although costs and budgets of future projects or programs were not considered in detail during the Snowmass process, the new proposals discussed were divided into categories based on the approximate total projected cost 
– “large”, greater than $\sim$\$500M, “medium”, between of order \$75M$-$\$500M, and “small”, less than or of order \$75M.\footnote{These monetary thresholds were used in the Snowmass study for discussion purposes only, and are not aligned with any budgeting process. The costs are intended to represent the U.S. share of the projects.} The identification of large proposals, and their related timescales, is particularly important for the combined Snowmass/P5 process since the impact that they have on planning is so significant. In contrast,  “medium” and “small” projects and proposals, though extremely important, of great scientific value, and essential to maintain the diversity and vibrancy of the U.S. HEP program (see section \ref{section:mid-scale-projects}), have a much smaller effect on overall resources. In Table~\ref{table:major-projects} we list the future large proposals discussed during the Snowmass process for the coming two decades to address the essential scientific goals that were identified. 

As illustrated in  Table~\ref{table:major-projects}, the future large-scale projects/programs that were discussed during Snowmass 2021 fall into four scientific themes – broadly corresponding to the scientific questions addressed by the energy, neutrino, cosmic, and rare process Frontiers – and can be classified as occurring in the coming decade or the following one. The specific scientific goals and capabilities of the projects included in this table are discussed further in this document and the Frontier Reports that follow – and even more thoroughly in the related Topical Group Reports. This table is not a timeline, rather large projects are listed by the decade in which the preponderance of their activity is projected to occur.  Projects may start sooner than indicated or may take longer to complete, as described in the frontier reports. The projects listed were considered during Snowmass 2021 without prioritization or consideration of any future budget scenarios.

\begin{centering}
\begin{table}[t]
\begin{tabular}{|l|l|l|} \hline
\multicolumn{3}{|c|}{Decadal Overview of Future Large-Scale Projects}  \\  \hline
Frontier/Decade & \multicolumn{1}{c|}{2025 - 2035} & \multicolumn{1}{c|}{2035 -2045} \\ \hline \hline
\multirow{2}{*}{Energy Frontier} & \multicolumn{2}{|c|} {U.S. Initiative for the Targeted Development of Future Colliders  and their Detectors} \\ \cline{2-3} 
 &     &  Higgs Factory \\ \hline
Neutrino Frontier & LBNF/DUNE Phase I \& PIP- II & DUNE Phase II (incl. proton injector) \\ \hline
\multirow{3}{*}{Cosmic Frontier} & Cosmic Microwave Background - S4&
Next Gen. Grav. Wave Observatory$^{*}$ \\
 &  Spectroscopic Survey - S5$^{*}$ & Line Intensity Mapping$^{*}$ \\ \cline{2-3}
 &\multicolumn{2}{|c|}{Multi-Scale Dark Matter Program (incl. Gen-3 WIMP searches)}  \\ \hline
 Rare Process Frontier &     &  Advanced Muon Facility \\ \hline
\end{tabular}
\caption{
An overview, binned by decade, of future large-scale projects or programs (total projected costs of \$500M or larger) endorsed by one or more of the Snowmass Frontiers to address the essential scientific goals of the next two decades. This table is not a timeline, rather large projects are listed by the decade in which the preponderance of their activity is projected to occur.  Projects may start sooner than indicated or may take longer to complete, as described in the frontier reports. Projects were not prioritized, nor examined in the context of budgetary scenarios. In the observational Cosmic program, project funding may come from sources other than HEP, as denoted by an asterisk.
}
\label{table:major-projects}
\end{table}
\end{centering}

Some notes on each of the scientific categories in the table:

\begin{itemize}

\item In the Cosmic Frontier, a coordinated multi-scale dark matter program would combine direct, indirect, and cosmic probe experiments to explore the large dark-matter landscape (and, in total, rise to the “large” project category). Note that an expansion of underground facilities at SURF may be required as a component of this program. In the observational Cosmic program, projects may leverage funding from sources outside of HEP itself, as denoted by the asterisks in the table. Both CMB-S4 and Gen-3 WIMP searches (previously DM-G3) were endorsed as promising future directions by the previous Snowmass/P5 process.

\item In the case of the Energy Frontier, and as emphasized by the Accelerator and Theory Frontiers as well, the goal should be to position the U.S. HEP program to 
 support construction of
an Higgs Factory as early as 2030, and to subsequently be prepared to host or participate in the construction of a multi-TeV (muon or hadron) collider. In total, these investments (referred to as a ``U.S. Initiative for the Targeted Development of Future Colliders and their Detectors" in Table \ref{table:major-projects})
rise to the level of a large-scale project. In the early phase accelerator work should prioritize an $e^+e^-$  Higgs Factory (such as 
ILC, CLIC, FCC-ee, CEPC, C$^3$, or HELEN), a parallel effort 
should focus on multi-TeV colliders for the longer term, and some work on advanced accelerator R\&D should continue. Targeted detector R\&D for the Higgs Factory is required in the early phase, with a smaller detector R\&D component related to multi-TeV colliders. In the later phase, as an $e^+e^-$ Higgs Factory construction is taking place, accelerator and detector R\&D effort on multi-TeV colliders will need to increase.

\item For the Neutrino Frontier the highest priorities are the completion of LBNF/DUNE Phase I in the coming decade (with the corresponding PIP-II upgrade), and the construction of DUNE Phase II (with the corresponding proton source upgrade) in the decade following. DUNE Phase I and Phase II are described briefly in section 4.7 and more completely in the Neutrino Frontier report. The completion of the DUNE science program was identified as a high priority of the previous Snowmass/P5 process.

\item For the Rare Process and Precision Measurements Frontier, the “Advanced Muon Facility” for studies of muon decays, muon conversion, and muonium transitions, may require coordinated improvements or modifications to the FNAL proton complex.

\end{itemize}

\subsection{Medium- and Small-Scale Future Experiments and Projects}

\label{section:mid-scale-projects}

Medium- and small-size experiments and projects are an important component of the current and proposed program. 
In the past, experiments with these scales have made significant measurements and important discoveries, opening
up new areas of scientific exploration. Furthermore, because of their timescale and size, these experiments 
offer unique leadership and training opportunities for younger physicists and allow for greater
diversity in the experimental particle physics ecosystem.

In the Snowmass process, many medium- and small-scale experiments have been presented  
-- too many  and too diverse  a group to easily categorize and summarize. 
Here we mention a few broad classes corresponding to the four major experimental directions,
though this list is not comprehensive.

\begin{itemize}

\item{\bf Cosmic Frontier:} Many small- and mid-scale dark matter searches are needed to cover the 
huge parameter space, including models of ultralight dark matter such as the QCD axion, in which dark matter could reside. Various expansions of astroparticle observatories such as IceCube Gen2, AugerPrime, the Cerenkov Telescope Array, and the Southern Wide-field Gamma Ray Observatory are also underway or planned, to probe fundamental physics beyond accelerator energies. While mostly supported outside of DOE, these can benefit HEP by probing fundamental physics in regimes not accessible to accelerators.  

\item{\bf Energy Frontier:} A series of medium-scale “auxiliary experiments” are being pursued
using the products of collisions that will already be available in the ATLAS, CMS, and LHCb 
interaction regions to study parts of phase space that are not covered by those experiments. 
Examples of experiments being considered include: FASER$\nu$, Mathusla, CODEX-b, FACET (a forward extension of CMS), and those associated
with the Forward Physics Facility (FPF). 

\item{\bf Neutrino Frontier:} 
A case was made at Snowmass for a comprehensive short-baseline neutrino program;  a program to measure many basic neutrino scattering processes that we have never before been able 
to measure precisely; a campaign to measure the proton-nucleus cross sections needed to understand fluxes in neutrino 
beams; a search for new neutrino interactions; the use of astrophysical neutrinos as messengers of physics and astrophysics; and several other topics. 

\item{\bf Rare Processes and Precision Measurements Frontier:} Upgrades are proposed to two projects to study $b$-quarks in which the  U.S. currently participates, LHCb at CERN and Belle II at KEK in Japan.
The U.S. community working in heavy-flavor physics has a long tradition of leadership and innovation. LHCb is currently commissioning its Upgrade I, which features an innovative purely software trigger, and is planning a second upgrade that should lead to an even higher sensitivity and a broader physics reach in the HL-LHC era. Belle II and SuperKEKB are considering staged upgrades to reach higher luminosity. Individually and even more in combination, the synergistic efforts of LHCb and Belle II will lead to an unprecedented breadth of exploration of heavy quark physics as a tool of discovery.   The pair add unique studies of rare tau decays, exotic QCD states, and other fundamental quantities with an overall program that spans our Frontier.  

New projects are proposed
to study the  decays of particles containing light or strange quarks, charged lepton flavor experiments to
study flavor violation and lepton flavor non-universality, and dark sector beam dump experiments
to search for new feebly-interacting particles. Additional directions discussed include
an upgrade to the Mu2e experiment, and the development of storage rings 
to search for a proton electric dipole moment.

\end{itemize}

A balanced portfolio of these mid- and small-scale experiments from all four Frontiers is a vital component  of  a healthy and  vibrant U.S. HEP program. An important recommendation of Snowmass is that resources be allocated to allow for a sufficiently large and diverse selection of important small- and medium-scale experiments to go forward in the coming decade.

\section{A Science Driver-Oriented View}

These are exciting and challenging times for fundamental particle physics. During the last several decades of intense experimental and theoretical research, we have come to rely on the Standard Model of particle physics --- a very predictive, relativistic, spontaneously-broken quantum gauge field theory ---  in order to describe microscopic phenomena at the smallest accessible distance scales and highest accessible energy scales. Its most remarkable prediction -- the existence of a Higgs particle associated with a fundamental scalar field responsible for electroweak symmetry breaking -- was established in the last decade, while innumerable  other nontrivial predictions and relationships have been confirmed, sometimes with extraordinary precision. 

There are, however, big questions in fundamental particle physics to which we currently do not know the answer. These cannot be addressed with the current version of the SM  and call for new particles, new interactions, and, perhaps, a qualitatively different description of nature at very small distance scales. There are several potential answers, but at this point, we know very little about these potential new particles and interactions. There are different research directions and inquiries we expect, in tandem, to guide the way. These were organized by the last P5 into science Drivers \cite{P5-HEPAP}. Here, we make use of the science Drivers in order to discuss the progress made on the big questions in particle physics over the last decade and to guide the research directions particle physics should pursue in this decade. In the last subsection, we present a proposal from the Rare Process and Precision Measurements Frontier to highlight the physics of flavor, possibly as another Driver.

\subsection{Use the Higgs Boson as a Tool for Discovery}


A fundamental aspect of the SM of particle physics is that the electroweak gauge interactions are spontaneously broken to electromagnetism, resulting in masses for the $W$ and $Z$ bosons, while the photon remains massless. The model requires that, in addition to the gauge bosons themselves, along with the quarks and leptons, new particles and interactions must also be present to account for electroweak symmetry breaking. The simplest possibility for a symmetry-breaking sector, hypothesized in the 1960s when the standard electroweak theory was developed, is to introduce a fundamental scalar field (a Higgs doublet) whose potential energy is adjusted so that it acquires a vacuum expectation value. The simplest one-doublet Higgs model makes very specific, testable, predictions. In particular, the model predicts that there will be a single neutral scalar particle (the Higgs boson) in the spectrum which couples to all massive particles -- the quarks, leptons, and $W$ and $Z$ bosons -- with definite strengths related to the masses of those particles.  

A decades-long campaign in pursuit of the physics behind electroweak symmetry breaking resulted in a breakthrough in 2012, with the discovery of a new scalar particle at the LHC. The properties of this new particle are consistent with the Higgs boson of the simplest one-doublet Higgs model. In the last decade, the LHC experiments have measured many of the properties of this putative Higgs particle, including its charge, spin, CP properties, and mass, while putting stringent limits on its decay width. Existing measurements of the particle's couplings with the quarks, leptons, and electroweak gauge bosons are all consistent, within current experimental errors (typically of order $10\%-20\%$), with the predictions of the standard Higgs models.

These experimental results indicate that the physics of electroweak symmetry breaking is, at least approximately, consistent with the simplest Higgs model hypothesized in the 1960s. However, is this minimal model the whole story? Several questions remain unanswered and we have only just begun to reveal how nature chooses to spontaneously break the electroweak symmetry. For example, we currently lack experimental knowledge of the form of the Higgs potential. This information is encoded in the couplings of the Higgs particle to gauge bosons, fermions, and itself. Implications from these investigations directly inform our understanding of the universe. For example, whether our vacuum is fundamentally stable or only metastable is tied to the Higgs potential. On the other hand, the nature of the electroweak phase transition in the early universe, which may provide the key to understanding why the universe has so much matter in it, also depends on Higgs properties and interactions. 

The existence of what appears to be a fundamental scalar invites other questions. For example, how are the Higgs particle mass and electroweak symmetry breaking scale so different -- by sixteen orders of magnitude -- from the scales associated with gravitational phenomena? In the absence of new particles and interactions, it was expected that these two energy scales would naturally be the same. On the other hand, in the SM, the masses of the quarks and leptons arise from their couplings to the Higgs boson. Hence, the physics of the Higgs boson is associated with establishing the different ``flavors" of matter which, as discussed below, is a mystery in and of itself. Information on what might lie beyond the SM is also potentially encoded in the couplings of the Higgs boson if they are measured with enough precision. For example, increased accuracy in the measurement of the SM Higgs couplings to other SM particles can be naively translated into sensitivity to heavy new physics; 10\% precision is sensitive to new scales up to 1--2~TeV while a precision below the percent level would be sensitive to new physics with masses of order 10~TeV of higher. Alternatively, the Higgs boson may serve as a portal to a so-far hidden sector of new physics and be uncovered through rare ``invisible" Higgs decays.

The HL-LHC will reach percent level precision on most couplings and aims at constraining the Higgs potential through a measurement of the Higgs-boson self-coupling with a projected precision of about 50\%. Future colliders, both Higgs factories (mainly $e^+ e^-$ colliders with energies at the $Z$-pole, near the $ZH$ threshold, and up to 1~TeV) and multi-TeV colliders (either muon or hadron colliders providing partonic center of mass energies of 10~TeV or above) are the two main avenues that will allow precision measurements of Higgs properties and interactions and, at the highest energies, potentially provide direct access to such physics. Studying the Higgs at Higgs factories, coupled with commensurate improvements in precision collider phenomenology, may uncover the scale of new BSM physics and the minimum target energy for multi-TeV machines.

\subsection{Pursue the Physics Associated with Neutrino Mass}

By the end of the twentieth century, decades-old puzzles associated with the measurements of the fluxes of neutrinos from the Sun and the atmosphere were decisively solved by a collection of gargantuan detectors of neutrinos from different sources. These revealed, rather surprisingly, that neutrino masses are tiny but definitively not zero, contradicting the assertions of the minimal version of the SM. Our understanding of nature is such that all fundamental masses are dynamical -- they arise as a consequence of interactions among the massive particles and the mechanism of electroweak symmetry breaking --  and we currently know nothing about the origin of nonzero neutrino masses other than the fact that new particles and interactions must exist to explain it. These new particles associated with neutrino masses can be very light or very heavy -- from sub-eV to YeV ($10^{24}$ eV) -- and either charged or neutral, opening an avenue to discovery in a broad range of experiments from low-energy neutrino oscillation experiments to astrophysical neutrino detectors to high-energy colliders.

Nonzero neutrino masses were established through the discovery of neutrino oscillations, the phenomenon that describes how lepton-flavor quantum numbers are not conserved but evolve -- oscillate -- in space and time.  The early twenty-first century has witnessed rapid experimental and theoretical progress in our understanding of neutrino oscillations. We are quickly moving from the exploratory phase to the precision era, in which oscillation experiments will be able to answer the questions of how the neutrino masses are ordered and whether neutrinos and antineutrinos oscillate differently. The mass ordering directly informs models of neutrino mass generation, and CP violation in neutrinos may play a role in the matter-antimatter asymmetry of the universe.
 Precise, over-constraining measurements of neutrino oscillations, which will require improved theoretical predictions of neutrino-nucleus cross sections, will test our current understanding of neutrino masses and mixings and allow us to look for more new physics in the neutrino sector. Moreover, neutrino experiments have a unique sensitivity to a wide range of weakly- or feebly-interacting new particles due to the high beam luminosity and sensitive, large detectors. 

Beyond measurements of neutrino oscillations,  another fundamental question we need to address is the nature of the neutrinos themselves -- are they Dirac or Majorana fermions? If neutrinos have Majorana masses, that would imply that lepton number is not an exact symmetry of nature.  Searches for neutrinoless double-beta decay are the most powerful probes of lepton-number violation, and an ambitious program aimed at improving existing sensitivity by at least one order of magnitude is  underway. 
On another front, precision measurements of the kinematics of beta decay continue to probe the absolute neutrino mass scale and there are vigorous R\&D efforts aiming at an order-of-magnitude improvement in sensitivity by the end of the decade. Finally, neutrinos impact the evolution of the universe and their masses leave an imprint on the large-scale structure of the universe. Cosmology-informing large-scale surveys already constrain the magnitude of the sum of neutrino masses. Future enterprises (including CMB-S4) aim at ``discovering'' non-zero neutrino mass by being sensitive to the minimum possible sum of neutrino masses consistent with oscillation measurements  -- modulo more surprises -- by the end of the decade.

The degrees of freedom responsible for nonzero neutrino masses can manifest themselves in a variety of observables, many of which are not directly related to neutrinos. For example, next-generation precision measurements of charged-lepton properties (e.g., the muon $g-2$ and the electron electric-dipole moment) and the measurement of exceedingly rare charged-lepton processes (e.g., $\mu\to eee$ and $\mu \to e \gamma$ decays or $\mu\to e$-conversion in nuclei) will inform the neutrino mass puzzle; the same might also be true of precision measurements of leptonic neutral-current decays of heavy mesons. The current and next-generation of high-energy colliders are also capable of directly producing the new particles responsible for neutrino masses as long as they are neither too heavy nor too weakly coupled. Finally, cosmic surveys are not only sensitive to the values of the neutrino masses but are also impacted by new neutrino properties and interactions. 

\subsection{Identify the New Physics of Dark Matter}

Precision measurements of the properties of the universe and its constituents at different distance scales -- from galactic scales to clusters of galaxies, to the cosmic microwave background -- combined with our understanding of gravitational interactions, indicate that most of the matter in the universe is not made SM out of  particles (protons, neutrons, electrons, and neutrinos). We know very little about this so-called dark matter but there are many hypotheses regarding its nature. It can be stable or unstable and it can be a fundamental particle or a composite object. It can be very light (less than a zeV) or macroscopically heavy (dozens of solar masses). It can interact with ordinary matter a lot, a little, or only gravitationally.

Given the enormous parameter space of possibilities, a diverse experimental and theoretical effort is required in order to learn the identity and nature of dark matter and to measure its properties and interactions. If the dark matter mass is, roughly, between 100~MeV and a TeV and it interacts with  SM particles via some new ``weaker-than-weak'' interactions, it can be detected in very large, very quiet underground experiments. So-called WIMP searches have been pursued for a few decades  and the technologies have evolved to allow for multi-ton experiments that have examined many orders of magnitude of parameter space in detail with more improvements to come. The next generation of these direct-detection searches will aim at ${\cal O}(10^2)$ tons of target material. These are multi-purpose experiments that are also sensitive to neutrinos from the Sun, the atmosphere, and supernova explosions. Dark matter is also expected to accumulate in the centers of galaxies and stellar objects, where it can self-annihilate or decay. These processes may lead to cosmic rays, gamma rays, neutrinos, or other cosmic messengers that can be detected on Earth (or in orbit around the Earth or the Sun) in a diverse variety of observatories. For lighter dark matter particles, new detector technologies and search strategies have opened the door to a wealth of  distinct experiments. The last decade witnessed an explosion of qualitatively different searches for the QCD axion and ultra-light axion-like dark matter particles, often taking advantage of the development of new quantum devices. Meanwhile, new cosmic observational probes have been developed to study the properties of dark matter via its gravitational effects on cosmology and astrophysics.

Indirect searches for dark-matter-mediated processes in the cosmos have also evolved and diversified since the last P5, making use of the current generation of gamma-ray telescopes, X-ray telescopes, cosmic-ray detectors, and neutrino telescopes. Some of these facilities are underground, some are on the Earth's surface, and some are in orbit. Many of these detectors have upgrade programs underway, in different stages of R\&D or implementation.

Alternatively, rather than searching for the dark matter present in the universe, the new particle(s) associated with the dark matter can also be produced in particle physics experiments. Experiments at high energy colliders, including the LHC, the upcoming High Luminosity LHC, and next-generation lepton and hadron colliders are well-positioned to inform the dark matter puzzle through their ability to produce and detect new particles. High-intensity experiments, including the facilities where intense neutrino beams are produced (for example, J-PARC, in Japan, or the upcoming LBNF/DUNE target and near-detector complex at FNAL), allow one to search for light new, long-lived particles that are produced in different (via weaker than SM interactions) ways and which may be observed in the near-detectors that serve the neutrino oscillation experiments. Our understanding of different ways to search for light, very weakly coupled dark matter particles has evolved dramatically in the last decade and multiple  experiments sensitive to ``dark sectors'' have emerged and are under consideration.

\subsection{Understand Cosmic Acceleration: Dark Energy and Inflation}

Cosmic surveys allow us to measure the expansion rate of the universe as a function of time. In the 1990s, these unexpectedly revealed that today, the expansion rate of the universe is accelerating. We know very little about the fundamental physics -- referred to as dark energy -- responsible for this phenomenon. The simplest solution is the existence of a nonzero cosmological constant. Our current understanding of fundamental physics is essentially silent about the cosmological constant; simple estimates yield a value that is 120 orders of magnitude too large. There are also dynamical hypotheses for the origin of dark energy, associated with the existence of new fundamental fields. Learning more about the nature of dark energy is a challenge and a priority for fundamental physics in this and the upcoming decade. 

Precise cosmic surveys of different cosmological eras using different tracers are needed in order to measure the expansion rate of the universe as a function of cosmological time. These include large-scale-structure measurements that will be facilitated by the Vera Rubin Observatory and next-generation measurements of the cosmic microwave background -- including its polarization. These observations are positioned to study the possible dynamical evolution of dark energy over a wide range of redshifts. On the theory side, there are several attempts to explain and interpret the physics of dark energy, which may be related to our efforts to understand quantum gravity.

There is also very compelling indirect evidence for inflation, a period of exponential expansion of the universe almost immediately after the Big Bang. The dynamics behind inflation is also virtually unknown. Measurements of the polarization patterns in the cosmic microwave background may determine the absolute energy scale of inflation, etched in the form of primordial gravitational waves, while optical spectroscopic and CMB measurements of the matter power spectrum will reveal clues about the scattering dynamics of this ultra high energy epoch as imprinted in the large scale structure of the universe.
The direct observation of gravitational waves, a revolutionary discovery in 2015, has motivated studies to broaden the sensitivity and frequency reach of gravitational wave experiments. Applications to fundamental physics, including uncovering the physics of inflation along with the exploration of other early-universe phenomena (phase transitions, the imprint of ultra-heavy fields, etc.), are under intense investigation both theoretically and experimentally.  

\subsection{Explore the Unknown: New Particles, Interactions, and Physical Principles}

Particle Physics research operates at the boundary between the known and the unknown. New facilities allow us to explore new distance and energy scales or unveil unexpected very rare phenomena. Broadly speaking, all particle physics experimental enterprises are sensitive to unanticipated new phenomena and have the potential to uncover new particles, new interactions, or even new principles that guide our description of fundamental physics. Exploring the unknown is at the heart of experimental and theoretical particle physics research. 

The direct exploration of nature at the smallest distance scales requires high-energy collisions of fundamental particles. Very high energy collisions allow for the production of new, heavy particles, as long as these interact with the known fundamental particles. The highest laboratory center-of-mass energies are currently found at the LHC, and experiments there at the 1 TeV partonic energy scale are a central piece of particle physics research for the next decade and beyond. Access to higher center-of-mass energies in the future requires extensive R\&D efforts in accelerator technologies, superconducting magnets, detector technologies, computing and data analysis, etc. Two options to reach the 10 TeV partonic energy scale have been identified: muon colliders with multi-TeV center-of-mass energies and proton colliders with one order of magnitude higher energy (around 100~TeV). 

New, light particles can also be discovered in lower-energy experiments. For example, new neutral particles that mix with the electroweak neutrino partners of the SM charged leptons can be discovered via neutrino oscillations -- they allow for new oscillation frequencies --  and are the subject of intense experimental exploration in this decade (building on measurements made during the last few decades). Precision measurements of neutrino oscillations are also sensitive to new neutrino--matter interactions. New, light particles can also be discovered through their influence on the evolution of the Universe, as revealed by precision cosmology.

Precision measurements and the searches for rare processes are sensitive to the quantum effects of hypothetical new particles and interactions. One special class of rare-process searches involves those that are forbidden in the SM. In the  SM (when neutrino masses are zero) baryon number and lepton number conservation are the consequence of accidental global symmetries. The conservation of baryon number forbids proton decay and $n\leftrightarrow\bar{n}$ oscillations while lepton-number conservation forbids neutrinoless double-beta decays. While neutrino masses may ultimately be evidence that the lepton number is not conserved -- experimental evidence is needed to prove this, as discussed above -- there are no indirect or direct hints that baryon number conservation is violated. Searches for baryon-number-violating processes (which are predicted by most Grand Unified theories) serve as, sometimes, the deepest probes we have of new phenomena at very small distance scales.  Very large neutrino detectors also allow the most sensitive searches for proton decay and other baryon-number-violating nuclear processes. With bigger and better-instrumented neutrino detectors our sensitivity to proton decay is expected to increase by at least one order of magnitude in the coming decade.

\subsection{Flavor as a Tool for Discovery $-$ Possible New Driver}

At Snowmass 2021, a new Driver was proposed by the Rare Processes and Precision Measurements Frontier: {\it Flavor as a tool for discovery}. In the last P5, this topic was understood as included in a subset of the other science Drivers. However, new experimental capabilities and the ability for flavor probes to be sensitive to ultra-high-energy physics suggest that new physics in the flavor sector might be uncovered in the coming decade, as described below. 

All matter particles come in sets of three, called generations. There are three types of charged leptons, three types of neutral leptons (neutrinos), and  three different pairs of quarks whose members have charges $+2e/3$ and $-e/3$. These are postulated to have all the same properties relative to one another except for their masses, which are known to span several orders of magnitude. We don't understand this triplication of the matter particles nor is there a compelling mechanism behind the disparate masses of the different generations. In the SM, only the weak interactions allow for interactions among different generations. In the quark sector, inter-generational mixing is known to be small or very small. In the lepton sector, instead, inter-generational mixing is large or very large (albeit difficult to access in practice because neutrino masses are tiny). We don't understand how quark mixing and lepton mixing are related, if at all. Since the discovery of the muon many decades ago, the solutions to these flavor puzzles continue to elude the particle physics community. 

A known consequence of the three generations, plus nontrivial mixing, is that SM interactions can account for the violation of CP invariance, a phenomenon that is well established in the quark sector. However, while the search for new CP-violating phases is still an active field of investigation in quark flavor,
determining whether (and how much) CP invariance is violated in the lepton sector is an open experimental challenge for the current and next generation of neutrino oscillation experiments and among the top priorities for the LBNF/DUNE and Tokai-to-Hyper-Kamiokande projects. Understanding the physics of CP violation is a necessary ingredient for addressing another big question in fundamental physics: why is there so much more matter in the universe than antimatter? Cosmological observations and our theoretical interpretations indicate that, in the absence of new particles and interactions, the universe should be mostly empty, except for photons and neutrinos. The presence of an overabundance of matter implies a primordial matter-antimatter asymmetry. Furthermore, since the electroweak interactions are known to violate baryon number at temperatures above the electroweak phase transition, this asymmetry must be generated dynamically in the early universe, through a process known as baryogenesis. CP violation is a necessary ingredient for baryogenesis. The CP violation present in the quark sector is not sufficient to explain the matter-antimatter asymmetry in the universe. Whether the potential violation of CP invariance in the lepton sector is sufficient to account for baryogenesis is a  fundamental particle physics problem that can be definitively addressed in the coming two decades.   

Precision measurements of rare processes involving different generations, when complemented by precise theoretical calculations, are powerful probes of new phenomena. Searches for the virtual effects of new particles in flavor or CP-violating observables can, in many cases, provide access to the physics at scales higher than those directly accessible at high-energy collider experiments. These include precision measurements of CP-violation in Kaon, $B$-meson, and $D$-meson mixing, and flavor-changing neutral current processes involving Kaons (e.g., $K^0\to\pi^0\nu\bar{\nu}$), $B$-mesons (e.g., $B^+\to K^+\mu^+\mu^-$), muons (e.g., $\mu\to e \gamma$), and taus (e.g., $\tau^-\to \mu^+\mu^- e^-$). These observables are the subject of active research at accelerator facilities around the world and are important targets for next-generation experiments. 
Some can be probed with current experiments like LHCb and the Muon $g-2$ experiment at Fermilab. Others will be investigated in the near future at Belle II. Depending on what the near future brings, dedicated next-generation experiments will be required. 

At lower energies still, searches for permanent electric dipole moments of fundamental particles -- electrons and muons -- or baryons -- protons and neutrons are also very clean probes of new particles and interactions and would reveal the existence of new sources of CP violation.

\section{Brief Summaries of Frontier Reports}

The main body of this Snowmass Report consists of the  full reports of each of the ten Frontiers and the SEC report. These range from about 10 to 70 pages in length. In this section, we present a brief summary of each Frontier report, giving an overview of the material and the main high-level conclusions. For the main physics Frontiers, they discuss the large- and mid-range projects that are proposed for consideration by P5 for construction or development in the next two decades. 
While the brief summaries of the full Frontier reports are presented in alphabetical order, the Frontiers may be thought of as divided into two groups,  ``enabling frontiers" and ``physics frontiers". 

The enabling frontiers, namely Accelerator, Community Engagement, Computational, Instrumentation, and Underground,  provide the basic foundation, infrastructure, R\&D, and support on which facilities and experiments are constructed and operated and that ultimately are necessary to produce the physics. These Frontiers include experts in accelerator physics, instrumentation, and computation. These experts may be engineers, skilled technicians, computer scientists, or physicists. The line between these Frontiers and the physics frontiers is not sharp and individuals and projects may  cross the boundaries. Since working scientists are the bedrock of our field, we have included the Community Engagement Frontier as an enabling technology. The Theory Frontier is both an enabling technology for HEP experiments and a highly significant physics research endeavor in its own right, producing fundamental new knowledge. For that reason, we include it among the physics frontiers.  

The physics frontiers are the ones that carry out the high energy physics program, from proposal to final physics analysis and publication,  and whose projects and experiments have been the main subject of P5's prioritization.  These include the Cosmic, Energy, Neutrino, and Rare Processes and Precision Measurements  Frontiers, and for the reasons presented above, Theory. These correspond closely to the Three Frontiers of Snowmass 2013, Energy, Cosmic, and Intensity, except that in 2021 the Intensity Frontier was split into  the Neutrino Frontier and the Rare Processes and Precision Measurements Frontier to reflect the growth in the Neutrino program worldwide and in the U.S. in the last decade. These Frontiers set the physics goals and work with the enabling frontiers to ensure that  the necessary infrastructure and facilities  will be available to achieve them. An excellent example of this is the interaction between the Energy Frontier and the Accelerator Frontier, apparent in the two corresponding summaries in this section and further described in Section 5. 

With ten Frontiers, we were concerned that communications would be difficult and 
stove-piping could be a problem. We implemented a very complete system of 
inter-frontier liaisons  to ensure good interactions among the Frontiers. At the Seattle meeting, many morning parallel sessions were dedicated to bringing together subsets of the Frontiers or their Topical Groups that benefited from  interacting with each other. The work presented here is informed by the studies done by the Frontiers both individually and in collaboration.

\subsection{Accelerator Frontier~\cite{Gourlay:2022odf}}

High-energy accelerators have been a major enabling technology for HEP and allied fields since the 1950s. The experiments envisioned by the Rare Process and Precision Measurement  and Neutrino Frontiers rely on high-intensity proton beams to provide controlled sources of pions, kaons, neutrinos, and other particles for study. These same facilities also provide the opportunity to produce new feebly interacting particles related to the dark sector (a focus also of the Cosmic Frontier) in beam dump experiments. In contrast, the Energy Frontier requires accelerator complexes to fill storage rings of high-energy colliding beams of protons, as well as electrons and possibly muons (and their antiparticles), to allow for a thorough exploration of the properties of matter at the shortest possible distances.  Despite past successes, the increasing size, cost, and timescale required for modern and future accelerator-based HEP projects arguably distinguish them as the most challenging scientific research endeavors being considered. 
    
The goal  for the Accelerator Frontier is to answer questions such as:  What is needed to advance the physics? What is currently available (state of the art) around the world? What new accelerator facilities could be available in the next decade (or next-to-next decade)? What R\&D would enable these future opportunities? What are the time and cost scales of the R\&D and associated test facilities, as well as the time and cost scale  of the final facilities?

Since the previous Snowmass/P5 exercise, the LHC Phase-1 accelerator upgrades at CERN have been completed and the planning for Phase-2 upgrades (HL-LHC) is well underway. Construction of the PIP-II proton linac for the Phase I LBNF/DUNE neutrino program at FNAL is underway, and planning for a multi-MW beam power upgrade for Phase II is beginning. In addition, numerous new high-energy colliders in various stages of maturity have been proposed. There are several $e^+e^-$ Higgs Factory concepts,  including linear colliders such as the ILC (which has a complete TDR) and others such as 
CLIC, C$^3$, and HELEN, and storage-ring proposals such as FCC-ee at CERN and CEPC in China. Finally, there is increasing interest in energy-frontier ``discovery machines" with partonic center-of-mass energies of order 10 TeV, including a muon collider and a proton-proton collider, FCC-hh.

 To properly support HEP scientific goals the {\bf planning of accelerator development and research should be aligned with the strategic planning for particle physics and should be part of the P5 prioritization process.} The U.S.~HEP accelerator R\&D portfolio presently is presently managed by the general accelerator R\&D program (GARD) and 
 contains no collider-specific scope. This creates a gap in our knowledge base and accelerator/technology capabilities. It also limits our national aspiration for a leadership role in particle physics in that the U.S.~cannot lead or even contribute to proposals for HEP collider facilities. To address the gap, the accelerator community proposes that the U.S.~establish a {\bf National Integrated R\&D Program on future colliders} in the DOE Office of High Energy Physics (OHEP) to carry out technology R\&D and accelerator design for future collider concepts. This program would aim to enable synergistic engagement in projects proposed abroad (e.g. FCC-ee/hh, ILC, and IMCC). It would also support the development of design reports on collider options by the time of the next Snowmass and P5, perhaps 7 -- 10 years from now, particularly for options that can be hosted in the U.S., and the creation of R\&D plans for the decades past 2030.
       
In addition to the collider-specific program envisioned above, general accelerator R\&D is also needed. Specific needs identified include developing high-power targets for multi-MW beams for the Phase II LBNF/DUNE upgrade and for muon collider sources; continuing development of high-gradient super-conducting and normal-conducting  RF cavities for particle acceleration at linear colliders; high-field magnets for high-energy proton and muon colliders; wake-field acceleration methods to enable ultra-high energy colliders; and basic experimental, computational, and theoretical studies of accelerators and beam physics to enable the optimized design and control of future facilities. These topics require a substantial increase in accelerator R\&D support, and the Accelerator Frontier recommends that the U.S. strategic goals and roadmaps of GARD program thrusts be updated.
       
Finally, workforce development through the strengthening of recruitment, education, and training, especially of women and other underrepresented groups, and the expansion of U.S. accelerator beam test facilities to maintain their competitiveness with respect to worldwide capabilities, are essential to realizing the vision of the Accelerator Frontier
 
The Accelerator Frontier of Snowmass 2021 has undertaken a thorough investigation of the current state and possible future plans that we summarize below, for which details may be found in the Frontier and Topical Group Reports (and references therein). The questions facing the frontier were addressed  in seven Topical Groups led by internationally recognized researchers.

\smallskip\noindent {\bf Beam Physics and Accelerator Education} Funding for fundamental beam physics needs to be reinvigorated, including enhancing investment in both general and directed accelerator research. These investments should be paired with strengthening and expanding U.S. accelerator test beam capabilities. Furthermore, the U.S. accelerator physics community must comprehensively and strategically build a strong and diverse workforce of accelerator physicists to realize HEP's long-term plans.

\smallskip\noindent {\bf Accelerators for Neutrinos} The neutrino physics program is calling for 
next-generation, higher-power, megawatt and multi-MW-class super-beam facilities, such as $>$2MW beam power upgrade of the Fermilab proton accelerator complex. Challenges needing to be addressed include developing extremely robust targets to withstand the enormous thermal and mechanical stresses from  
multi-megawatt beams; major improvements to the performance and power efficiency of the linacs, synchrotrons, and accumulator rings needed for a super-beam accelerator complex; and new approaches to control instabilities and minimize beam losses.

\smallskip\noindent {\bf Accelerators for Electroweak/Higgs Physics}
      The world HEP community has called for a Higgs/Electroweak  Factory as the next large accelerator project. At present, a dozen Higgs/ElectroWeak Factory proposals are under consideration, including circular $e^+e^-$ colliders such as CEPC in China and FCC-ee at CERN 
       and linear colliders such as 
       CLIC (CERN) and the ILC.
       Two recently proposed Higgs factories can potentially be shorter than 7 km and fit the Fermilab site $-$ C$^3$ 
       and HELEN. 
       Major conclusions of this topical group (AF3) are: of all the Higgs/EW Factory proposals considered, only a small number of them are ready or close to a construction phase $-$ namely, the ILC, CLIC, FCC-ee, and CEPC. Most of the other proposals are in the conceptual design stage. These proposals should focus on the main R\&D tasks to move forward to a  CDR and, later, to a TDR. A detailed estimation of the Technical Readiness Level, risk factors, technology validation, cost reduction impact, performance achievability, and timescale 
       are presented in detail in the Collider Implementation Task Force report \cite{Roser:2022sht}.

\smallskip\noindent {\bf Multi-TeV Colliders} The Snowmass process has emphasized the emerging need to explore the Energy Frontier with particle collisions at significantly higher energies than LHC. At present only a few design concepts have achieved a level of design maturity to have reliable performance evaluations based on prior R\&D and design efforts. These options reach only slightly above the 1 TeV: the linear $e^+e^-$ collider, the ILC (1 to 3 TeV), FCC-eh ($\sim$1.2 TeV), and  CLIC (3 TeV). Critical project risks have been identified and subsystem-focused R\&D is underway where necessary. Other emerging accelerator complexes offer unique opportunities to reach the very much sought 10 TeV parton energy scale, such as FCC-hh (100 TeV p-p) and a high-energy muon collider (6 -- 14 TeV), which is more compact and has the potential to be more power efficient. Both concepts still require significant basic R\&D and design effort to reach a maturity level to be properly evaluated (see the Collider Implementation Task Force report \cite{Roser:2022sht}).

 \smallskip\noindent {\bf Accelerators for Physics Beyond Colliders and Rare Processes} The highest priority here is the optimum exploitation of the PIP-II 800-MeV superconducting RF CW-compatible linac under construction at Fermilab.  The LBNF/DUNE experiment  will need only a small fraction of the total potentially available PIP-II proton beam power (bunched). The proposed RPF experiments, such as rare muon decays and dark matter searches with beam dumps, could make use of the PIP-II CW beam capability, after facility upgrades, but the bunches would have to be reconfigured to meet the varying needs of each experiment. Additionally, a new 
 0.8 -- 1 GeV proton bunch compressor ring (PAR) has been proposed to accumulate the beam from the PIP-II linac and extract it in intense bunches at rates of 100 -- 1000 Hz.

\smallskip\noindent {\bf Advanced Accelerator Concepts} This Topical Group focused on understanding the limits, the state of the design concepts, and the status of R\&D for plasma and  structure-based wakefield acceleration. At this time, there are no parameter sets for a plasma or structure-based linear collider that self-consistently address the known accelerator physics challenges. To move forward, a vigorous R\&D program will be required. That should be motivated by a better understanding of the desired parameter ranges for a collider. The U.S. is in a good position in this respect with several state-of-the-art beam test facilities mainly dedicated to research in beam physics and the advanced accelerator field, including FACET-II, BELLA, ATF, AWA, and FAST-IOTA as well as at numerous universities.

\smallskip\noindent {\bf Accelerator Technology - RF, Magnets, and Targets/Sources}
The GARD roadmap continues to serve as community-developed guidance for RF technology R\&D, but based on recent progress it would benefit from some mid-course corrections detailed in the topical group report. The U.S. Magnet Development Program (U.S.-MDP) focuses on fundamental accelerator magnet R\&D 
providing key magnet technologies that benefit all future accelerators. Discussions and initiatives emerging from Snowmass 2021  motivate enhanced scope and a significant increase in resources to prepare accelerator magnet technology for applications in future hadron and muon colliders and for fast-ramping accelerator facilities. The next generation of multi-MW high-power targets for future accelerators will use more complex geometries, novel materials, and new concepts. There are challenges in high-intensity high-brightness beam sources for future accelerators. They are particularly formidable for sources of beams with special characteristics, such as polarized electrons and ions or ultra-small emittances, and for tertiary and secondary particles, such as muons and positrons.
Normal-conducting and superconducting RF technology R\&D will develop toward higher gradient and higher quality factor cavities and novel, more power-efficient RF power sources.  

\medskip

U.S. High-Energy Physics depends crucially on the success of the Accelerator Physics Frontier, and the analyses and plans presented here provide a framework for a bright future.

\subsection{Community Engagement Frontier~\cite{Assamagan:2022yii}}

     In Snowmass 2013, the U.S. high-energy physics community incorporated a working group on Communication, Education, and Outreach (CE\&O). The 2013 experience made clear that it is crucial to address community engagement for the health of U.S. HEP. Therefore, in Snowmass 2021 community engagement was expanded into a full-fledged Community Engagement Frontier (CEF), co-equal with the physics Frontiers and with a much broader scope than CE\&O. 
     
     A major conclusion of the Community Engagement Frontier is that the climate of U.S. High-Energy Physics needs to be improved: {\bf The HEP community should institute a broad array of practices and programs to reach and retain the diverse talent pool needed for success in achieving our scientific vision.}

Of particular concern is broadening the representation of the U.S. high-energy physics community. Those from marginalized communities who do join the U.S. HEP community are, all too often, subject to alienation, microaggressions, and other forms of subtle and overt discrimination. {\bf HEP needs to address these persistent issues of underrepresentation by employing the use of robust strategic planning procedures including a full re-envisioning of our workplace norms and culture to prioritize eliminating the barriers and negative experiences faced by our marginalized colleagues.} In addressing the needs of marginalized physicists, HEP communities must engage in partnership with scholars, professionals, and other experts in several disciplines, including but not limited to anti-racism, critical race theory, and social science. Progress on these issues will require the implementation of new modes of community organization and decision-making procedures that promote agency and leadership from all stakeholders within the scientific community.  In addition, all HEP communities should create structures to fully open career path opportunities to everyone and {\bf conduct event planning to ensure events are accessible to all community members, including those with disabilities.} Finally, as demonstrated vividly during the past two years of  COVID-19 lockdowns, to retain our junior colleagues, {\bf research institutes and universities should do more to maintain the highest standard in work-life balance and mental health of staff.}

{\bf A diverse pool of candidates cannot be expected in the HEP community if educational efforts are not made as far back as the K -- 12 and university undergraduate levels, to nurture the pipeline.} In this regard, the HEP community should encourage stronger participation in HEP collaborations by faculty and students from non-R1 academic institutions; continue and expand work with K-12 teachers and students to create supportive local communities to nurture student interest in math and science; foster expanded pre-university and university programs for international student training, especially in partnership with colleagues in developing countries.

Engagement with developing countries needs improvement. Universities, laboratories, and HEP societies should improve and sustain international outreach, partnerships, schools, workshops, conferences, training, short visits for research, and development of research consortia; mechanisms should be
developed to facilitate the participation of colleagues from developing countries. Large international research collaborations should improve efforts to facilitate the integration and participation of research groups from developing countries and support efforts to foster HEP in these countries.

Our field cannot absorb all the early career members that it produces, so {\bf funding agencies,  national laboratories and universities should work together to provide more education and career opportunities for engineering and industry-focused research within and outside HEP}, and update degree programs to match better the skills needed and career opportunities required in today's HEP and related fields. These steps should be complemented by developing effective alumni networking tools and programs, to deepen connections with industry. 

{\bf Technology transfers between HEP and industry are necessary for society to receive the benefits of HEP research.} Funding agencies and laboratories should improve policies and programs to foster technology transfers and facilitate collaborative programs with industry on targeted technology development beneficial for both HEP and society. 

While HEP has many excellent Communication, Outreach, and Engagement programs, our community needs to enact {\bf structural changes to foster broader, deeper, and more effective participation in community engagement}, through strengthened policies such as considering community engagement work in hiring, promotion, and grant decisions. Individual scientists should encourage others, including peers, mentees, and students, by participating in public engagement and discussing its importance. Our community also needs to {\bf transition from an ethos of conducting outreach and communication with the public, to a culture of engagement in relationships with the public.} This should be done by building lasting relationships with the full breadth of all our supporting communities (especially those that have been historically excluded) that are not based on transactional interactions, but rather real two-way partnerships that consider the needs and interests of the audience and include its members in program design.

{\bf The HEP community should provide resources to sustain and grow the annual HEP Congressional advocacy efforts.} HEP groups should coordinate efforts by laboratories and universities to {\bf extend advocacy to the federal executive branch, and state and local governments.} These efforts are particularly important in disseminating the results of the current Snowmass/P5 exercise. HEP should strengthen partnerships with other science and physics societies on advocacy for non-HEP funding issues.

Finally, {\bf HEP must take greater responsibility for its impacts on climate change} by addressing
and mitigating these impacts through DOE project policies and individual community
member actions.

     The expanded Community Engagement Frontier included seven Topical Groups defined primarily by the broad and diverse issues to be addressed, and a brief summary of their work is given below.
     
     \begin{itemize}
     
     \item{\bf Applications and Industry:} The charge for the topical group was to develop strategies to strengthen HEP/Industry relationships. Members of this topical group produced five contributed papers examining three different modes of collaborative partnership, and cooperation on three different specific areas of technology: ``Programs enabling deep technology transfer from national labs"; ``Application-driven engagement with universities, synergies with other funding agencies"; ``Big industry engagement to benefit HEP: microelectronics support from large CAD companies"; ``Transformative technology for FLASH radiation therapy"; and ``Nurturing the industrial accelerator technology base in the U.S.".

     \item{\bf Career Pipeline and Development:} This working group was not simply about making early career scientists aware of different opportunities, but also about changing the culture of HEP career paths.  Two topics were studied, namely: ``Facilitating Non-HEP Career Transition" and ``Enhancing HEP research in predominantly undergraduate institutions and community colleges".
     
     \item{\bf Diversity, Equity, and Inclusion:} The DEI topical group focused on issues and projects related to Diversity, Inclusion, Equity, and Accessibility; all of which are essential not only to professional success in our field but to develope a better society at large.  Ultimately, twelve contributed papers were developed,  which may be categorized as follows: ``Accessibility in High Energy Physics: Lessons from the Snowmass"; ``Lifestyle and personal wellness in particle physics research"; ``Climate of the Field: Snowmass 2021"; ``Why should the United States care about high energy physics in Africa and Latin America"; ``Experiences of Marginalized Communities in HEP"; ``In Search of Excellence and Equity in Physics"; and ``Strategies in Education, Outreach, and Inclusion to Enhance the U.S. Workforce in Accelerator Science and Engineering".
     
     \item{\bf Physics Education:} The Physics Education topical group, examined the role that physics education at all levels plays in advancing the field of HEP. Two goals were identified as critical for the long-term health of the field: attracting students across all demographics to the study of physics, and providing them the education, training, and skills they will need to pursue any career in STEM or related fields. The working group contributed papers are: ``Opportunities for Particle Physics Engagement in K-12 Schools and Undergraduate Education"; ``Transforming U.S. Particle Physics Education: A Snowmass 2021 Study"; ``Broadening the Scope of Education, Career and Open Science in HEP"; ``The Necessity of International Particle Physics Opportunities for American Education"; ``Data Science and Machine Learning in Education"; and ``U.S. CMS - PURSUE (Program for Undergraduate Research Summer Experience)".
     
     \item{\bf Public Education and Outreach:} The Public Education and Outreach working group focused on enabling members of the physics community to effectively communicate about scientific research through public engagement. The contributions to this working group resulted in  two contributed paper topics, namely, ``The need for structural changes to create impactful public engagement in U.S. particle physics” and “Particle Physics Outreach at Non-traditional Venues”.

     \item{\bf Public Policy and Government Engagement:} The topical group on Public Policy and Government Engagement was tasked with conducting a review of all current interactions between the HEP community and government offices and individuals. Three contributed papers produced by the topical group document this work: ``Congressional Advocacy for HEP Funding"; ``Congressional Advocacy for Areas Beyond HEP Funding"; and ``Non-congressional Government Engagement".
     
     \item{\bf Environmental and Societal Impacts:} This topical group focused on ideas and projects related to how particle physics research impacts society and the environment.  Five contributed papers were developed and submitted to this group, focusing on three areas: environmental impacts of particle physics; interactions of different laboratories with their local communities;  and nuclear non-proliferation.

\end{itemize}

The recommendations made by CEF are to improve and sustain strategic engagements with our communities in order to draw support for and strengthen the field of particle physics (an inward-focused goal carried forward from Snowmass 2013) while playing key roles in serving those communities (an outward-focused goal added for Snowmass 2021). {\bf It is critical that we as individuals agree on the importance of all working together to address CEF issues in HEP. DPF should work in collaboration with all HEP stakeholders to develop a structure for  implementing the recommendations contained in the CEF report.}

\subsection{Computational Frontier~\cite{Elvira:2022wyn}}
    
Software and computing (S\&C) is essential to all high-energy physics experiments, accelerator and detector design, and many theoretical studies. It is a key enabler of all the other Frontiers and all science Drivers, requiring physics research along with expertise in computer science to address the complex and unique challenges of our field.  The escalating demand for computing resources is a result of the need to do more sensitive and more precise experiments, using higher intensity beams, higher luminosity colliders, and stronger sources,  to collect more astrophysical data over wider  fields and farther into the past (higher red-shifts), and to do more precise theoretical calculations.   While experiments may last for many decades and their experimental hardware  may be upgraded a few times during this period, the computing hardware they use, driven by the commercial sector,  may change every half-decade or even more frequently. The software for the detectors  and facilities evolves continuously to respond to operational issues, but undergoes larger changes to respond to major detector upgrades, and must also adapt  on shorter time scales to utilize and exploit the latest computing hardware and software and computing infrastructure changes. In the last decade, limited computing resources have gone much farther than expected due to methodological innovation, but it is highly  likely that analyzing all the data to be acquired in the next decades will stress the community's financial and human  resources. While we cannot predict with any accuracy what the specific S\&C environment will look like in 2035 or 2045, we can try to create an R\&D environment for computing hardware and software that will allow HEP to adapt to whatever changes occur so it can utilize the most cost-effective methods.

S\&C had a prominent  role in the 2013 Snowmass report. However, in the ten years since then, many new features and trends appeared that were absent or hardly noticeable then. These include 
\begin{itemize}
    \item The use of heterogeneous accelerators such as GPUs and FPGAs, off-loading parts of calculations. These devices come from the commercial sector, with significant implications for the way we develop and maintain software. Computing has entered a new “post-Moore’s law” phase. Speeds-ups in processing now come from  the use of parallelism within computations rather than higher clock speeds. This paradigm shift has brought up the enormous challenge of adapting or re-engineering almost every piece of HEP software, a process that is heavy in labor and requires rare and expensive expertise.
     \item The  reliance on community hardware resources such as High-Performance Computing (HPC) centers and the Cloud rather than dedicated experiment resources to do production and very compute-intensive application. HPC facilities are not specifically designed for HEP experiments,  thus using them effectively requires significant investment in R\&D to address technical challenges related to task factorization, I/O, storage, and utilization of external services.
    \item  The pervasive use of artificial intelligence and machine learning, AI/ML, in nearly every aspect of our software. Hardly mentioned in the 2013 report, these revolutionary machine-learning approaches are transforming the way we work. 
    \item The appearance of potentially disruptive technologies, such as quantum computing. These are R\&D projects, and we still need to understand which, if any, HEP computing needs they satisfy.
    
\end{itemize}

It continues to be true today, as it was in 2013, that S\&C efforts are neither funded nor managed like projects in the manner that major facilities and experiments are. They are funded by various streams including dedicated resources for large experiments, shared resource pools provided for general use by small and large groups, and national computing centers provided by national funding agencies for large, multidisciplinary projects and communities. Increasingly, HEP researchers are being encouraged to use these computing centers, as well as commercial resources available in the Cloud. 

This new computing environment requires R\&D and novel approaches to address the long-term development, maintenance, and user support of essential software packages and cross-cutting R\&D efforts supported from proof of concept to prototyping and production. Additionally, strong investment in career development for software and computing researchers is needed to ensure future success. All of our computing tasks in HEP are likely to be affected by these changes, so common solutions  across HEP, other sciences, government, and industry are going to be necessary.

\medskip

The high-level conclusions of the Computational Frontier  are:

\begin{itemize}
\item {\bf The  creation of a standing ``Coordinating Panel for Software and Computing"} under the auspices of the APS Division of Particles and Fields, mirroring the Coordinating Panel for Advanced Detectors (CPAD) established in 2012. As with CPAD, this would, among other things, require a modification of the DPF bylaws. A suggested mandate is:

\parbox{15cm} { \em{Purpose: Promote, coordinate, and assist the HEP community on Software and Computing, working with scientific collaborations, grassroots organizations, institutes and centers, community leaders, and funding agencies on the evolving HEP Software and Computing needs of experimental, observational, and theoretical aspects of the HEP programs. The scope should include research, development, maintenance, and user support.}}

\item {\bf Support of software packages} The U.S. HEP Community should take a leading role in the long-term development, maintenance, and user support of essential software packages with targeted investment. Support of software for HEP has been uncertain in the U.S. An excellent example is GEANT4, which does not receive dedicated support in the U.S. at present and yet is critical to most HEP programs, where it is used in facilities prospect studies, detector design and optimization, computing infrastructure stress tests, software development, and physics analysis. Event generators are another example.

\item {\bf Support for R\&D} Through existing, reshaped, and expanded programs, R\&D efforts cutting across project or discipline boundaries should be supported from proof of concept to prototype to production.

\item  {\bf Support for necessary personnel to enable the use of heterogeneous resources} Support for computing professionals/researchers and physicists to conduct code 
re-engineering and adaptation will enable the most effective use of heterogeneous resources.

\item {\bf Training, career development, and EDI} Strong investment in career development for HEP S\&C researchers will ensure future success.
\end{itemize}

These  recommendations of the Snowmass Computational Frontier resonate with the European Strategy, spelled out in their recent report\cite{Eurostrat}.

In 2013,  Computing was organized along frontier lines. In Snowmass 2021, Computing instead organized itself in Topical Groups that are all cross-cutting in nature to reflect the need for common solutions to computing challenges. The seven Topical Groups are Experimental Algorithm Parallelization, Theoretical Calculations and Simulation, Machine Learning, Storage and Processing Resource Access, End User Analysis, Quantum Computing,  and Reinterpretation and Long-term Preservation of Data and Code. 
The following is a summary of important observations and conclusions from each of the seven Topical Groups of the Computational Frontier.

\smallskip\noindent {\bf Experimental Algorithm Parallelization} This topical Group covers  event and physics object reconstruction for all experiments and comparable work for surveys. Workable and portable solutions must be found and supported  to exploit heterogeneous computing platforms.  In parallel, targeted optimizations of key algorithms for experiments, which often provide the largest speedups,  must also be developed and supported. Software frameworks and common tools must evolve and adapt to the new computing landscape to enable the usage of parallel algorithms in production environments of big and small experiments. Interdisciplinary collaborations and programs will play a critical role. Training opportunities for early-career researchers are essential, and career opportunities for researchers who focus on algorithm development need to be created. To enable the long-term sustainability of HEP software and
provide job security to researchers in this area, software development should be supported according to a model similar to that for detector projects, or as part of detector projects. Funding should continue beyond the R\&D phase of algorithm exploration and extend into the development, production, and maintenance phases. Funding agencies and research teams have to commit to sustainable long-term effort for mission-critical software products:  big experiments/surveys need multi-year, rather than annual, computing resource allocation planning for the use of national
facilities.

\smallskip\noindent {\bf Theoretical Calculations and Simulation} Simulations and theory calculations are used in nearly all aspects of HEP.  For this study, the topic was divided into six domains: cosmic calculations, particle accelerator modeling, detector simulation, event generators, perturbative calculations, and lattice quantum chromodynamics (QCD).
Each area has specific needs, but there is also significant commonality, leading to the following general conclusions. Hardware accelerator-friendly portable common tools are essential. Universal programming interfaces should be developed to facilitate portability. CPU clusters will still be necessary to provide 
general-purpose computational cycles for tasks that do not map well to accelerators but these must be carefully sized to the actual needs. Automation of memory hierarchy will remove a burden from the scientific programmer and enhance productivity. Best practices for the development of common software should be encouraged. Working with funding agencies, new arrangements should be established for long-term support and maintenance of common software tools. Collaboration of computer scientists, applied mathematicians, nuclear physicists, and others relevant to HEP computational and physics model development should be encouraged and supported. Programs should be developed, including
joint lab-university tenure-track appointments in S\&C, to generate academic opportunities for young researchers in computation in academia. Training for graduate students and postdocs in AI/ML should be provided.

\smallskip\noindent {\bf Machine Learning}
Machine Learning ML/AI techniques are firmly embedded in HEP
and their use is still growing. Areas where dedicated resources are required for ML to continue to expand and flourish include:
\begin{itemize}
\item {\bf Dedicated HEP-ML Research}  While industry is driving research in ML, there are specific areas where its use in HEP requires dedicated solutions, such as in  uncertainty quantification, validation, and interpretability. Ideas such as anomaly detection and simulation-based inference may enable analyses that are not currently possible. ML may also be of use in
the design, quality assurance, and control and calibration of instrumentation and in control at accelerator facilities and experiments, especially where ultra-low latency is required for correction. Much of this work is inherently interdisciplinary.
\item {\bf Software and hardware needs} HEP now  relies on industry standards such as TensorFlow, SciPy, and PyTorch. There is also a
diverse set of hardware that supports ML, most involving GPUs, which are mainly available at computing centers. Small hardware setups available through research grants would help stimulate research. At the same time, industry continues to develop more advanced hardware accelerators whose development the HEP community should track and participate in so that its needs can be met. The HEP community should be trained in the use of demand services as a way to access the latest custom hardware.
\item {\bf Training and Personnel}  ML/AI should be part of the standard training program. Courses should be available in graduate school and in summer schools. Training materials and paths should be readily available.  Many of the issues connected to the use of ML are  statistical in nature  so researchers need formal training in that as well. It is also
important that career paths be available to researchers with these broad and cross-cutting capabilities.
\end{itemize}

\smallskip\noindent {\bf Storage and Processing Resource Access}
This group  concentrated  on the R\&D challenges that must be
addressed in the next decade so that HEP can take advantage of the newest developments, which will most likely come from industry.
It did not focus on the required capacity for computing or storage.
The complex HEP workflow is divided into six areas – storage,
processing, edge services, AI/ML hardware, analysis facilities, and networking. Each of these is
examined in detail in the report and four overarching themes have emerged: efficiently exploit specialized compute architectures and systems; invest in portable and reproducible software and
computing solutions to allow exploitation of diverse facilities; embrace disaggregation of systems and facilities; and extend common interfaces to diverse facilities. HPC facilities, often several different ones, will be needed to do major processing workflows for the analysis of experiments. There are four recommendations for work that will facilitate the use of these diverse capabilities for processing HEP data: HPC facilities should revisit their resource access policies to allow more flexible allocations and job executions so that more HEP projects can benefit from the large computing facilities;
investment in the development of portable software is key to maximizing the efficient utilization of diverse
processing resources; research is needed to determine the trade-off between dedicated HEP computing facilities and
general access computing facilities such as the HPC center, Grid and Cloud resources; and
infrastructure development will be needed to support better data management frameworks across different types of facilities.

\smallskip\noindent {\bf End User Analysis} Two analysis ecosystems are used in HEP, one based on ROOT, hosted by CERN,  with U.S. support, especially in the area of I/O, and the other based on Python, which is a set of analysis tools developed primarily to host analysis software, including ML/AI, developed outside of HEP. One challenge is that increasingly running on massive amounts of data will be done at HPCs. Particle physics analysis problems usually require relatively light-weight computations to be done
repeatedly on massive numbers of events, often referred to as high-throughput computing (HTC). High-Performance Computing center execution environments are usually not needed for HEP analysis and may introduce constraints and overheads that make it difficult to use them efficiently.
Actions that need to be taken to serve the needs of the HEP community for long-term, sustainable end-user analysis are identified:
\begin{itemize}
\item Develop both ROOT-based and Python-based analysis ecosystems and   maintain interoperability between them. 
\item Critical elements of the ROOT evolution plans include the development of improved columnar data formats and  multiple  I/O implementations.
\item Analysis facilities for low latency columnar analysis in the context of the Python ecosystem will be essential for future experiments.
\end{itemize}

User-friendly data provenance and metadata storage systems that can be easily integrated into typical analysis tasks should be developed. Scaleable analysis models should be developed so users can perform small interactive tests and also can run over large datasets using a single interface. Pipelines compatible with long-term preservation should be built into the structure of analysis systems. Collaborative software is an important element of the analysis software stack of an experiment and includes documentation, messaging between users, discussion forums, software version
control, bug tracking, and document workflow management. Host laboratories should provide a full stack of these services to their experiments, large and small. Documentation and training must be produced for analysis software understood as an ecosystem, not as a disconnected set of packages.

\smallskip\noindent {\bf Quantum Computing}
Quantum computing is an exciting new technology that can be shown to outperform classical computers in certain applications. However, current quantum computers (QCs) are relatively small and suffer from vulnerability to noise. This Topical Group focused solely on the use of re-programmable quantum hardware. It is not clear where it will be advantageous to employ QC in HEP, but there are a number of benchmark problems where QC is expected to be useful and perhaps even transformational, including lattice QCD, event generation, and data analysis. Currently, there is no consensus on the best hardware for implementing quantum computers and there are several software packages available. Access to QC hardware is also a problem. Nevertheless, we should support
research by HEP physicists, in collaboration with industry
and academia, to explore this rapidly developing field which may eventually offer substantial benefits to HEP. The National QIS Research Centers are a significant step in this direction.

\smallskip\noindent {\bf Reinterpretation and Long-term Preservation of Data and Code} This group considered how data, simulations, analyses, and codes needed to do an analysis can be preserved so that the analysis could, for example, be repeated well after the experiment that took the data ended. The purpose might be to verify the original analysis, try to improve it using new methods or concepts, use the data for some new analysis, or combine it with some other analysis.
Recommendations in this area include:
\begin{itemize}
\item Ensuring that all running and in-preparation experiments have a strategy and resources for the long-term preservation of data and analysis capabilities.
 \item Investing in shared cyber-infrastructure to preserve these data and support a comprehensive analysis from various experiments and surveys – both active and completed – in order to
realize their full scientific impact. The infrastructure should support the requisite theoretical
inputs and computational requirements for analysis as well as metadata and APIs to track provenance and provide incentives for participation.
\item  As a specific example, there should be a data archive center to preserve
Cosmic Frontier datasets and simulations, and facilitate their joint analysis across different computing centers. Preservation needs are not restricted to experimental data and analyses. As an example, for many years, the lattice QCD community has had a tradition of sharing the expensive-to-produce gauge configurations that can be used for many physics analyses.
\end{itemize}

Software and computing technologies traverse all areas of HEP and are evolving rapidly. It is hard to imagine what the state of the art will be at the time of the next Snowmass in about a decade. It is our hope that the establishment of a Coordinating Panel for Software and Computing will serve the community on timescales commensurate with the rapid evolution of S\&C technologies.

 \subsection{ Cosmic Frontier~\cite{Chou:2022luk}}

The Cosmic Frontier is focused on understanding how elementary particle physics shapes the behavior and evolution of the universe, and how observations of the universe inform our understanding of physics beyond the SM. In the moments just after the Big Bang, fundamental physics governs the  production of the particles and energy fluctuations that give rise to the current universe. The properties of elementary particles also govern the behavior of the most energetic processes observed in the universe today, including supernovas. The physics questions which guide the cosmic frontier include:

\begin{itemize}
    \item {\bf What is the nature of dark matter?}
    
    Does dark matter interact with ordinary matter, perhaps via the Higgs boson or another mediator?
    Is dark matter bosonic and does it exhibit wave-like properties? Does the dark sector itself have non-trivial dynamics, and does dark matter interact with itself? How was dark matter produced in the early universe? Does the nature of dark matter have general implications for physics beyond the SM?
    
    \item {\bf What is the nature of dark energy?}
    
    Is the observed acceleration in the expansion rate of the universe due to a new energy component or does it require modifications to general relativity? If dark energy is a new energy density component, is it a cosmological constant or is it a dynamical quantity changing in time? When did dark energy become important in the history of the universe?

    \item {\bf How is the inflationary paradigm realized in nature?}
    
     What is the energy scale of primordial inflation? Does inflation have dynamics that manifest themselves as an observable imprint on the primordial distribution of matter fluctuations? Detect or place limits on ``B-modes" to discriminate among theories of the Big Bang. Why were there two eras of acceleration in the history of the universe, one early and one late?
    
    \item {\bf How can we use cosmic observations to learn about BSM physics?}
    
    Did BSM degrees of freedom influence the thermal history of the universe? What is the scale of neutrino masses? Can cosmological observations distinguish between the ordinary and inverted neutrino mass hierarchies? What do gravitational waves reveal about early Universe dynamics and phase transitions? What do Nature's highest energy particles produced in cosmic sources tell us about BSM particles and interactions?
    
\end{itemize}

A major thrust of the future Cosmic Frontier program is building the next generation of cosmological probes. Cosmic surveys “aim high” at observables spanning almost the entire 13.8 billion-year history of our Universe. The next big project in this arena is CMB-S4, a system of telescopes to study the cosmic microwave background and address the mystery of cosmic inflation, which is expected to operate through to at least 2036. Additional projects that would start after 2029 are Spec-S5 (the follow-on spectroscopic device to DESI), a project to carry out line intensity mapping (LIM), and planning efforts to increase the sensitivity of gravity wave detection by at least a factor of 10 ($10^3$ in sensitive volume) beyond what will be achieved by LIGO/Virgo.

In the Cosmic Frontier survey ecosystem, each type of program brings unique strengths to the portfolio, including shape measurements that can only come from imaging, precise and robust redshift measurements that require spectroscopy, and multi-messenger data that exploits new gravitational wave detection capabilities; by also spanning a range in wavelength (e.g., Optical/IR vs. mm-wave vs. 21-cm), information from different redshift ranges can be accessed. While each survey will deliver groundbreaking results on its own, no single experiment can meet all of the discovery thresholds without complementary information from other programs.

A second thrust of the Cosmic Frontier is a suite of experiments to explore the physics of dark matter (DM). The space of DM models encompasses a dizzying array of possibilities representing many orders of magnitude in mass and couplings, making the DM program one of the most ``interdisciplinary" investigations in high-energy and particle physics. The Cosmic Frontier's DM program will ``delve deep, search wide" by employing a broad portfolio of small/medium-scale direct and indirect detection experiments, as is required to search optimally for each decade in dark matter mass. Furthermore, an expanded Cosmic Probes program will strive to identify the properties of dark matter via cosmic surveys, which is highly synergistic with the other science targeted by those surveys.

The DM research program enhances adjacent areas of science. New DM detector technologies will require cross-disciplinary collaborations to access the expertise, technology, and facilities of neighboring fields of study such as AMO, condensed matter physics, quantum information science, and gravitational physics. Similar synergy between fields is present when searching for indirect effects of dark matter scattering, annihilation, or decay in galactic halos, where HEP provides a critical role in developing instrumentation for astroparticle observatories. Equally important, collaboration with the broader non-HEP astronomy and astrophysics community is essential for understanding the astrophysical backgrounds in DM searches, while at the same time alerting those communities to possible non-gravitational DM impacts.

The conclusions of this Frontier arise from the work of the seven Topical Groups of this Frontier, which were organized into three general categories as described below.
\\

\noindent{\bf Dark Matter}

Our understanding of the landscape of dark matter theories has evolved significantly in the past several years, as theoretical exploration has better defined the boundaries of what models are consistent with observations. As of Snowmass 2013, the classification of dark matter candidates was largely based on the particle physics features of the underlying models. Since then, the focus has shifted toward exploring a wide range of possible phenomena in an effort to understand how well existing experimental searches cover the space of possibilities, and how new experimental opportunities provide sensitivity to regions of theory space that are not captured by the current program. 

\begin{itemize}

\item{\bf Particle-like Dark Matter:}
In conventional models of GeV-TeV mass dark matter associated with solving the electroweak gauge hierarchy problem, WIMPs with electroweak couplings to SM particles would be naturally populated at approximately the correct density by the freeze-out mechanism. Large second-generation WIMP detectors based on a variety of scattering targets, combined with indirect searches for high-energy annihilation products, have excluded $Z$-mediated couplings to SM particles for WIMP masses up to $\sim$TeV. Experiments are now probing weaker couplings such as those mediated by Higgs boson exchange and exploring new techniques to be sensitive to light particle dark matter (1 eV to 1 GeV).

\item{\bf Wave-like Dark Matter:}
Provided that gravitational clumping of a single dark matter species is responsible for the formation of all galaxies and galactic substructures, the sizes of the dark matter halos provide a quantum-mechanical lower bound on the mass of dark matter of order $10^{-22}$ eV. For dark matter mass less than 100 eV, the mode occupation number must exceed unity in order for there to be enough dark matter to account for the local gravitational well. Hence dark matter in the mass range from $10^{-22}$ eV to $10^2$ eV must be wave-like bosonic fields. A high-priority target in this mass range is the QCD axion, and perhaps the most significant technological development in the field of terrestrial dark matter searches since the last Snowmass is the demonstration by ADMX and HAYSTAC of experimental sensitivity to the invisible QCD axion made possible through the use of quantum sensing technologies.  A comprehensive search for the QCD axion is now within reach, but this program will require significant investment in high-field magnets which serve as the axion scattering target.

\item{\bf Cosmic Probes of Dark Matter:}
Cosmic probes of dark matter, which seek to determine the fundamental properties of dark matter through observations of the cosmos, have emerged as a promising means to reveal the nature of dark matter, and are valid even when the coupling between dark matter and normal matter is extremely weak (e.g., as weak as gravity).  Rubin LSST has enormous potential to discover new physics beyond the prevailing cold dark matter paradigm. Microlensing measurements will directly probe primordial black holes as a component of dark matter. The planned CMB-S4 and future Spec-S5 and CMB-S5 projects will also extend access to rich dark matter particle physics. These probes also constrain neutrino physics. The current cosmological limit on the sum of the neutrino masses is $\sum m_\nu < 0.12$ eV as obtained using cosmic microwave background, baryon acoustic oscillations, supernova Ia, and large-scale structure measurements. CMB-S4 will reach sensitivity $\sum m_\nu < 0.02$ eV,  and if the data indicate a mass sum less than 0.11 eV, this would rule out the inverted neutrino mass ordering which predicts a higher mass sum.

\end{itemize}

\noindent{\bf Dark Energy and Cosmic Inflation}

Developing an understanding of the cause of cosmic acceleration in the modern universe would revolutionize our understanding of fundamental physics. Cosmology research in the twenty-first century is dominated by large cosmic survey experiments carried out by worldwide collaborations comprising hundreds of members from dozens of institutions. The goal is to study both early and late universe phase transitions and the new particles and interaction dynamics governing new fundamental scales of nature.  New measurements will establish the values of key parameters, including the equation of state parameter of dark energy, the rate of expansion of the universe today, and the amplitudes of fluctuations in the density of matter in the universe.

\begin{itemize}

\item{\bf The Modern Universe:}
The community’s input to this Topical Group jointly describes a multi-probe experimental program that provides a rich, deep, and flexible portfolio that provides powerful constraints on cosmic acceleration. The program includes a powerful new Stage V spectroscopic facility (also referred to here as Spec-S5) that would pursue larger and deeper surveys enabling transformational advances in our understanding of both eras of accelerated expansion in the history of the universe – the early inflationary epoch and the late dark energy-driven one.

\item{\bf Cosmic Dawn and Before:}
The scientific goals of this Topical Group are to study inflation and to search for new physics through precision measurements of the Cosmic Microwave Background (CMB) from the early universe. 
Polarization-sensitive experiments like CMB-S4 offer unique windows on the B-mode patterns, which reveal the absolute energy scale of inflation and will continue to provide precision measurements of the other key observables. Upcoming cosmic surveys including DESI and the envisioned Spec-S5 project will provide much larger datasets, using precise redshift determinations to map the matter distribution in all 3 dimensions instead of being constrained to a single 2-dimensional slice at the surface of the last scattering. A new technique of line intensity mapping (LIM) of the 3-d distributions of neutral hydrogen or other gas will possibly access the largest range of redshifts to provide the highest statistics measurements of the large-scale structure.

\item{\bf Complementarity of Probes and New Facilities:}
This topical group presented a multi-faceted vision for the cosmic survey program which provided a path to understanding the causes of early- and late-time acceleration by leveraging ongoing surveys, developing and demonstrating new technologies, and constructing and operating new instruments The previously mentioned large CMB-S4 project and the next large Stage V Spectroscopic Facility (Spec-S5) should play a key role in advances over the next decade and beyond. The vision also includes potential future surveys that will be enabled by near-term R\&D and small pathfinder initiatives; these projects would employ new technologies and techniques such as 21-cm and mm-wave Line Intensity Mapping (LIM), Gravitational Wave Observatories (GWO), and future broadband CMB imaging (CMB-S5).

\end{itemize}

\noindent{\bf Cosmic Probes of Fundamental Physics}

Cosmic Probes of Fundamental Physics encompass a multitude of approaches:
\begin{itemize} 
\item {\bf Very high energy particles (ultra-high energy cosmic rays, gamma rays, and neutrinos):}  The discovery of 
TeV$-$PeV astrophysical neutrinos opens up unique opportunities to probe the neutrino sector at energy scales not accessible with laboratory neutrino beams; high-energy neutrinos from IceCube were used to discover the Glashow resonance and measure the high-energy neutrino-nucleon cross section and inelasticity distribution. 
UHECR air showers probe hadronic interactions at center-of-mass energies an order of magnitude beyond the LHC;  they have been used to determine the p-p cross section and have established significant failings in state-of-the-art modeling of hadronic interactions.  Upper limits on UHE gammas and neutrinos constrain GUT-scale physics.  
\item  {\bf Gravitational waves and neutron star observations: } The LIGO-Virgo collaboration has observed many examples of binary black holes, BH-neutron stars, and binary NS mergers, revealing an unexpected population of black holes with masses up to several hundred $M_\odot$.  LV and NICER have measured the radii of neutron stars, forcing a re-evaluation of the QCD equation of state at very high density.  Future measurements with a next-generation gravitational wave observatory, will nail down crucial aspects of QCD and probe some scenarios of inflation and phase transitions in the early universe.  

\item  {\bf Multi-messenger astrophysics:} The dawn of a new astrophysical multi-messenger era has been heralded by the recent co-detection of gamma rays and gravitational waves in a binary neutron star merger, the co-detection of gamma rays and neutrinos in a blazar flare, and several examples of neutrinos consistent with production in tidal disruption events.  Over the next decade, simultaneous observations with different techniques promise to reveal where these extreme-energy cosmic messengers come from, and how they came to be.
\end{itemize}

\smallskip

To summarize, the Cosmic Frontier strategy detailed in this report encompasses a science-rich program with small, medium, and large experiments which are planned on both near-term and long-term timescales. The community consensus is that this coming decade must feature strong support for this program in full to ensure that HEP will enjoy the new era of discovery and groundbreaking precision measurements that have been promised for the next two decades, as well as building even more powerful experiments to reach key science thresholds.

\subsection {Energy Frontier~\cite{Narain:2022qud}}

After decades of pioneering explorations and milestone discoveries, the SM of particle physics has been confirmed as the theory that describes electroweak and strong interactions up to energies of a few hundred GeV with great accuracy. However, the SM also leaves several fundamental questions unexplained such as the details of the evolution of the early universe, the origin of the matter-antimatter asymmetry of the universe, the nature of dark matter, the origin of neutrino masses, the origin of the electroweak scale, and the origin of flavor dynamics. The answers to these questions must lie in BSM physics.

In this context, colliders at the energy frontier offer the unique opportunity to study a huge number of phenomena and explore the connection between many of the fundamental questions we want to answer. The Energy Frontier (EF) aims at advancing the investigation of such questions with a broad and strongly motivated physics program that will push the exploration of particle physics to the TeV energy scale and beyond. The sharply focused EF agenda includes in-depth studies of the SM as well as the exploration of physics beyond the SM to discover new particles and interactions. The EF vision keeps its focus on big questions and provides opportunities to examine them from as many angles as possible while also continuing to pursue the exploration of the unknown, a leading driver of the EF physics program.

The EF currently has a top-notch program with the LHC and the HL-LHC at
CERN, which sets the basis for the EF vision. The fundamental lessons learned from the LHC thus far are that a Higgs-like particle exists at 125 GeV and there is no obvious and unambiguous signal of BSM physics. This implies that either there is a gap in the scale of new physics, or BSM physics must be weakly coupled to the SM or hidden in backgrounds at the LHC. The HL-LHC will either strengthen these conclusions further or  point us in a particular direction for discovery. At the same time, the discovery of the Higgs boson at the LHC has provided us with the unique possibility of exploring the most mysterious aspects of electroweak  interactions  because of a thorough program of Higgs-boson precision physics that has been very successfully started by the LHC. The studies of the Higgs-boson properties will reach a new level of precision at the HL-LHC and have the potential to indirectly probe new physics in the 10-TeV range and beyond, due to the higher precision and energy reaches of future lepton and hadron colliders. Auxiliary forward-physics facilities will further extend the physics potential of the HL-LHC both for SM measurements and  BSM discoveries. In view of all these considerations, {\bf the EF supports continued strong U.S. participation in the success of the LHC, and the HL-LHC construction, operations, and physics programs, including auxiliary experiments.}
 
Colliders are the ultimate tool to carry out the EF program thanks to the broad and complementary set of measurements and searches they enable. With a combined strategy of precision measurements and high-energy exploration, future lepton colliders starting at energies as low as a few hundred GeV up to a few TeV can shed substantial light on some of these key questions. Ultimately, it will be crucial to find a way to carry out experiments at higher energy scales, directly probing new physics at the 10 TeV energy scale and beyond. The EF community has proposed several opportunities for pursuing its scientific goals, among them the most prominent ones are {\bf Higgs-boson factories} and {\bf multi-TeV colliders} at the energy frontier. These projects have the potential to be truly transformative as they will push the boundaries of our knowledge by testing the limits of the SM and directly discovering new physics beyond the SM. Thus, {\bf the EF affirms that it is essential to complete the HL-LHC program, to support the construction of a Higgs-boson Factory, and to ensure the long-term viability of the field by developing a multi-TeV energy-frontier facility such as a muon collider or a hadron collider.}

The discussions on projects that extend the reach of the HL-LHC underlined that preparations for the next collider experiments have to start now to maintain and strengthen the vitality and motivation of the community. Several projects have been proposed such as ILC, CLIC, FCC-ee, CEPC, C$^3$, or HELEN for $e^+e^-$ Higgs factories, and CLIC at 3 TeV center-of-mass energy, FCC-hh, SPPC and Muon Collider for multi-TeV colliders. {\bf The EF supports a fast start for construction of an $e^+e^-$ Higgs Factory (linear or circular), and a significant R\&D program for multi-TeV colliders (hadron and muon). The realization of a Higgs Factory will require an immediate, vigorous, and targeted Detector R\&D program, while the study towards multi-TeV colliders will need significant and long-term investments in a broad spectrum of R\&D programs for accelerators and detectors.}

The EF aims to facilitate U.S. leadership in an innovative, comprehensive, and international program of collider physics. The timescales to fully realize the EF vision extend to the end of this century, and the ultimate goals can only be realized if our actions foster a vibrant, diverse, and intellectually rich U.S. EF community. Maintaining and strengthening such a community is only possible if our plans reflect the aspirations of and provide a rich set of opportunities for our Early Career physicists. {\bf The U.S. EF community has also expressed renewed interest and ambition to bring back EF collider physics to U.S. soil, maintaining its international collaborative partnerships and obligations.}

The proposed plans in five-year periods starting in 2025 are given below.
\medskip

{\bf For the five-year period starting in 2025:}

\begin{enumerate}
    \item Prioritize the HL-LHC physics program, including auxiliary experiments,
    \item Establish a targeted $e^+e^-$ Higgs Factory Detector R\&D program,
    \item Develop an initial design for a first-stage TeV-scale Muon Collider in the U.S.,
    \item Support critical Detector R\&D towards EF multi-TeV colliders.
\end{enumerate}

{\bf  For the five-year period starting in 2030:}

\begin{enumerate}
    \item Continue strong support for the HL-LHC physics program,
    \item Support the construction of an $e^+e^-$ Higgs Factory,
    \item Demonstrate principal risk mitigation for a first-stage TeV-scale Muon Collider.
\end{enumerate}

{\bf Plan after 2035:}
\begin{enumerate}
    \item Continuing support of the HL-LHC physics program to the conclusion of archival measurements,

      \item Support completing construction and establishing the physics program of the Higgs factory,
    \item Demonstrate readiness to construct a first-stage TeV-scale Muon Collider,
    \item Ramp up funding support for Detector R\&D for energy frontier multi-TeV colliders.
\end{enumerate}

These conclusions were derived from the analyses of the ten Topical Groups in the Energy Frontier, which were divided into three major areas broadly defined as Electroweak Physics (Higgs-boson physics, top-quark and heavy-flavor physics, electroweak gauge bosons physics), Strong Interactions (precision QCD, hadronic structure and forward QCD, heavy ions), and BSM (model-specific explorations, general explorations, dark matter at colliders), which have focused on three main key questions:

\begin{itemize}

\item {\bf What can we learn about the origin of the electroweak scale and the electroweak phase transition from an in-depth study of the SM?}
\item {\bf What can we learn about the nature of strong interactions in different regimes?}
\item {\bf How is a complete program of BSM searches built that includes both model-specific and model-independent explorations?}

\end{itemize}

\noindent A summary of the outcomes from the ten Topical Groups is included below.

\medskip

\noindent{\bf Electroweak Physics}

\begin{itemize}

\item{\bf Higgs Boson Properties and Couplings:} The Higgs boson is central to the electroweak sector of the SM and, in the ten years since the discovery of the Higgs boson, measurements of its mass are now at the per-mille level and several couplings are now measured at the $5\%-10\%$ level. The precision measurement of the Higgs boson properties is the central component of the physics of Higgs factories, with a goal of reaching coupling accuracies at or below the percent level. Of particular importance in the future is the measurement of the Higgs triple-coupling, a direct measure of the dynamics associated with electroweak symmetry-breaking, and which (for SM the  prediction, in which this self-coupling is directly related to the Higgs mass) should be measured at the 50\% level at the HL-LHC, and the $10\%-20\%$ level at high-energy $e^+e^-$ colliders. Precision measurements (at the few percent levels) of the Higgs self-coupling, however, require colliders with partonic center-of-mass energies of order 10 TeV, such as the FCC-hh or a muon collider.

\item{\bf The Higgs Boson as a Portal to New Physics:} The scalar sector of the  SM could be extended in many ways, such as through the addition of a scalar singlet (real or complex) or through one or more electroweak doublets as in Two Higgs Doublet Models (2HDMs). The phenomenology of these and related models are rich and complex, and new colliders can extend limits on these models beyond what is possible at the HL-LHC through either precision measurements of Higgs and electroweak couplings or via direct production of new heavy scalar states. For example, new techniques for strange-quark tagging at the ILC could allow measurements to directly probe the Higgs coupling to the second generation, providing a window into flavor physics.

\item{\bf Heavy Flavor and Top Quark Physics:} The top quark is the heaviest SM particle and is therefore the quark that couples most strongly to the electroweak symmetry-breaking sector. Accurate measurements of the top-quark mass, production cross sections, and couplings at future colliders provide incisive tests of the SM and probes for new physics. For example, the ultimate precision in the top-quark mass can be reached in a scan of the top-quark production threshold at a future lepton collider, where uncertainties on the top-quark mass are estimated to fall to 40 MeV$-75$ MeV $-$ approximately an order of magnitude lower than achievable today.

\item{\bf Electroweak Precision Physics and Constraining New Physics:} Precision measurements of electroweak observables have and will continue to provide stringent tests of the SM and windows into physics beyond the SM. While the HL-LHC will provide accurate measurements of the $W$-boson mass, the top-quark mass, and the effective weak mixing angle through Drell-Yan production, future high-intensity $e^+e^-$ running at the $Z$-pole, as well as at the $WW$ and $t\bar{t}$ thresholds, have the potential to reduce precision electroweak measurements to the per mille level or beyond, reducing current uncertainties by an order of magnitude. Future high-energy lepton colliders can also greatly constrain the effects of new physics as summarized through the Standard Model Effective Field Theory (SMEFT), reaching energy scales in four-fermion interactions of $50-100$ TeV or higher depending on the particular process and assumed center-of-mass energy.

\end{itemize}

\noindent{\bf QCD and Strong Interactions}

\begin{itemize}

\item{\bf Precision QCD:} Tremendous progress has been made since the last Snowmass exercise on precision calculations in QCD, determinations of the 
strong coupling constant, and use of jet substructure techniques to understand and predict experimental results $-$ with implications for the HL-LHC and beyond. The non-perturbative input needed for track-based jet observables can now be computed at high precision. Energy correlation functions have emerged as a powerful tool to explore the scaling behavior of weakly coupled quarks and gluons, and the transition to the regime of free hadrons. The strong coupling constant at the $Z$-boson mass can be measured in the next decade with 0.4\% precision, currently 0.8\%, and with dedicated measurements at the $Z$-pole at future $e^+e^-$ colliders, the precision can be reduced to 0.1\%, enabling searches for small deviations from SM predictions that could signal the presence of new physics.

\item{\bf Hadronic Structure and Forward QCD:} Parton distribution functions (PDFs) are crucial for the interpretation of all hadron-collider experiments and often are the dominant source of uncertainty. New developments in lattice QCD hold the promise of first-principles computation of PDFs, and also related distributions, that can augment and enhance current phenomenological extractions of PDFs and provide input to new experiments. LHC experiments have opened access to a wide range of forward and diffractive processes, driving advances in relevant QCD theory, such as charting the gluon at very low $x$, revealing dynamics at high partonic densities, and testing Monte Carlo models for forward hadron production. The proposed Forward Physics Facility would extend the coverage at small $x$ by almost two orders of magnitude at low Q, reaching down to $x \simeq 10^{-7}$. 

\item{\bf Heavy Ions:} The relativistic heavy-ion program prioritizes the studies of the quark-gluon plasma (QGP), the partonic structure of nuclei, collectivity in small collision systems, and nuclear electromagnetic interactions. These measurements will greatly benefit from planned detector upgrades for ALICE, ATLAS, and CMS increasing, for example, capabilities for the detection of heavy-flavor mesons and quarkonia. Observations of heavy-flavor mesons elucidate mechanisms for their thermalization, Debye color screening, and recombination inside the quark-gluon plasma. 

\end{itemize}

\noindent{\bf Physics Beyond the Standard Model}

\begin{itemize}

\item{\bf Model Specific Explorations:} Motivated by the question of the origin of the electroweak scale, composite Higgs and SUSY models remain very compelling and viable (though increasingly constrained) possibilities for physics beyond the SM. New high-energy colliders provide access to the new states predicted in these models. The highest mass reach is given by the FCC-hh or a 10 TeV scale muon collider, which provides opportunities to produce new physics particles with masses of order 10 TeV. These results complement and extend the precision measurements at lower energies which can be made at $e^+e^-$ Higgs factories. 

\item{\bf More General Explorations:} New states beyond the SM, including new gauge bosons, new fermions, or other resonances, are a feature of many extensions of the SM. High-energy colliders with a 10 TeV partonic center-of-mass energy scale allow for the extension of expected HL-LHC bounds by an order of magnitude. For example, a 10 TeV muon collider will have the highest mass reach for a universal $Z'$ with large couplings, uniquely probing masses $M_{Z'} > 100$ TeV. Simultaneously, high-luminosity Higgs factories and high-energy colliders can also probe light, exotic, and long-lived particle states with small coupling to  SM particles. The direct new physics searches, together with the Higgs and electroweak physics will deepen our understanding of Nature to the next level.

\item{\bf Dark Matter at Colliders:} Collider searches for dark matter are complementary to astrophysical observations and direct-detection experiments, and potentially allow for a more precise determination of dark matter properties than available through those detection modes. Consistent with the results discussed above, the high luminosities and 
high energies of proposed new colliders provide a significant extension of results possible at the LHC to higher masses and/or lower couplings. For example, depending on the  search strategies employed, WIMPs can be probed at high-energy lepton colliders up to close the kinematic limit for production, and at a 100 TeV hadron collider into the multi-TeV range, which can cover some of the most compelling benchmarks.

\end{itemize}

In summary, to pursue its scientific goals, the EF community proposes a combined strategy of precision measurements and high-energy exploration. Future lepton colliders starting at energies as low as the $Z$-pole up to a few TeV can shed substantial light on some of the key questions discussed in the ten topical group discussions. Ultimately, it will be crucial to find a way to carry out experiments at higher energies, directly probing and discovering  new physics at the 10 TeV energy scale and beyond.

\subsection{Instrumentation Frontier~\cite{Barbeau:2022muf}}

HEP experiments are always trying to collect larger data samples to improve the precision of measurements and the sensitivity of searches for previously unobserved phenomena. For colliding beam experiments, this is enabled by the ability to produce more intense beams and a consequently larger number of collisions per unit of time, which must be matched by improved and new instrumentation to deal with the larger, more complex data samples and the challenges of operating in harsh environments. In astrophysics, the development of enhanced instrumentation with new or improved capabilities is often the key to uncovering new regions of phase space.

Many solutions developed for experiments in this decade will not scale to meet the requirements of future experiments. Improved instrumentation is the key to progress in neutrino physics, collider physics, and the physics of the Cosmic and Rare Processes and Precision Measurement Frontiers. Many aspects now at the cutting edge of detector development were hardly present 10 years ago, including quantum sensors, machine learning, and picosecond-level timing. In many areas of HEP experiments, we are now at a point where completely new concepts are needed in order to enable scientific advancements. Evolutionary developments are no longer sufficient. This leads to a paradigm shift in which much more blue-sky R\&D is needed. This is R\&D with a high risk of failure, but also the potential to open completely new realms of sensitivity. Blue-sky R\&D has been neglected over the last decade and funding and recognition for it, as well as the entire U.S. detector program, needs to increase in order to enable the science of the future. Funding for HEP instrumentation in the U.S., however, is actually declining.	

In order to rejuvenate the instrumentation effort and to meet the challenging requirements for future scientific discoveries, the following actions are proposed:	

\begin{itemize}
\item  Advance performance limits of existing technologies and push new techniques and materials, nurture enabling technologies for new physics, and scale new sensors and readout electronics to large, integrated systems using 
co-design methods.
\item Develop and maintain the critical and diverse technical workforce, and enable careers for technicians, engineers, and scientists across disciplines working in HEP instrumentation, at laboratories and universities.
\item Double the U.S. Detector R\&D budget over the next five years and modify existing funding models to enable R\&D Consortia along critical key technologies for the planned 
long-term science projects, sustaining the support for such collaborations for the needed duration and scale.
\item Expand and sustain support for blue-sky, table-top R\&D, and seed funding. Establish a separate review process for such pathfinder R\&D.
\item Develop and maintain critical facilities, centers, and capabilities for the sharing of common knowledge and tools, as well as develop and maintain close connections with international technology road maps, other disciplines, and industry. 
\end{itemize}

The work in the Topical Groups has culminated in the identification of a variety of specific R\&D challenges and technologies that need to be pushed in order to reach the performance parameters required by the envisioned future physics program. They are summarized in the following:

\medskip

\noindent{\bf Quantum Sensors}   Interest in quantum sensors has undergone explosive growth since Snowmass 2013 and they have been used in dark matter searches, searches for new forces (fifth force), electric dipole moments, variations in fundamental constants, and gravitational waves. The technologies include
atom interferometers and atomic clocks, magnetometers, quantum calorimeters, and superconducting sensors. Thanks to the National Quantum Initiative, there is significant support for the development of such devices, but mostly outside of HEP. It is highly desirable to develop mechanisms to support interactions with other disciplines to enable collaborations with fields with expertise in quantum sensors and to facilitate interactions to support theoretical work to address issues of materials and measurement methods.

\smallskip\noindent{\bf Photon Detectors}  In the last decade, major developments have occurred in the detection of photons, in particular the counting of single photons over a wide frequency range from the infrared to the ultraviolet. Many of the technologies being developed need additional work, which will be the focus of R\&D for the next decade, to enable large area arrays and to improve the energy resolution, timing, noise (dark counts), and wavelength coverage.  A significant achievement has been the solution for the detection of photons in liquid argon by use of Arapuca light traps followed by readout using SiPMs. Additional new ideas are being developed based on novel light collectors called dichroicons, based on Winston cones using dichroic mirrors, allowing Cherenkov and scintillation light to be separated and focused on separate photomultiplier tubes, individually optimized for their respective  wavelengths. Another significant achievement of the last decade is the development and commercialization of Large Area Picosecond Photodetectors, LAPPDs.  Sensors being developed include: 
\begin{itemize}
\item Superconducting sensors:  Microwave Kinetic Inductance Detectors (MKIDs); Transition-Edge Sensors (TESs); and Superconducting Nanowire Single Photon Detectors (SNSPDs);
\item Semiconducting sensors: Skipper-CCDs (including in CMOS); Photon-to-Digital Converters (PDCs); and extended wavelength coverage Germanium semiconductors  with enhanced UV sensitivity.
\end{itemize}

\smallskip\noindent{\bf Solid State Detectors and Tracking}
As collision rates increase and the number of interactions per crossing grows, tracking based on solid-state detectors, mainly silicon, provides the segmentation and radiation resistance necessary to meet the needs of the next generation of experiments. The focus is now on developing 4-dimensional trackers with timing resolutions of 10-30 ps and spatial resolution of a few $\mu$m. Improvements in the readout electronics, including radiation hardness and   management of the material and power budget, are important areas where R\&D is necessary.

\smallskip\noindent Candidates for novel sensors are silicon or diamond detectors with ``3D technology"; sensors with Four-Dimensional Tracking -- tracking detectors with 10-30 ps timing resolution, in addition to excellent position resolution. The overall detector system should contain a sensor with short drift time, high signal-to-noise, limited thickness in the path of a MIP to reduce the Landau fluctuations, and small TDC bin size; monolithic Active Pixel Sensors (MAPS), in which charge collection and readout circuitry are combined in the same pixel, which has been shown to be a promising technology for high-granularity, 
low material budget, and low-noise detector systems. Concerning integration, conventional bump bonding has been used for many years to connect high-density pixel sensors to readout ASICs. Much smaller pixel pitches and consequently finer connections are needed for future experiments. Advanced packaging techniques will be required. 3D integration can solve many of these problems and is being developed in industry. Detector mechanics will play a significant role in the performance of future detectors. Material savings using novel approaches have the potential of reduction on the order of 30-50\%.

\smallskip\noindent{\bf Trigger and DAQ} The challenge for future data-intensive physics
facilities lies in the reduction of the flow of data through a combination of sophisticated event selections in
the form of high-performance triggers and improved data representation through compression and calculation
of high-level quantities. These tasks must be performed with low latency (i.e. in real-time) and often in
extreme environments including high radiation, high magnetic fields, and cryogenic temperatures. Developing the trigger and data acquisition (TDAQ) systems needed by future experiments will rely on
innovations in key areas, including the application of Machine Learning (ML) to TDAQ systems, particularly in the
co-design of hardware and software to apply ML algorithms to real-time hardware and for other novel
uses to improve the operational efficiency and sensitivity to new physics of future experiments. One key recommendation is to create a dedicated (distributed) R\&D facility that can be used to emulate detectors and
TDAQ systems, offer opportunities for integration testing (including low- and high-level triggering, data readout, data aggregation and reduction, networking, and storage), and the development and maintenance of  an
accessible knowledge base that crosses experiment-project boundaries.

\smallskip\noindent{\bf Micropattern Gaseous Detectors (MPGDs)} Gaseous Detectors are the primary choice for cost-effective instrumentation of large areas and for continuous tracking of charged particles with minimal detector material. MPGDs are gas avalanche devices with order O(100 $\mu$m) feature size, enabled by the advent of modern photolithographic techniques. Current MPGD technologies include the Gas
Electron Multiplier, the Micro-Mesh Gaseous Structure, Thick GEMs,
also referred to as Large Electron Multipliers, the Resistive Plate WELL, 
the GEM-derived architecture, the Micro-Pixel Gas Chamber, and the integrated pixel readout. MPGDs are used in many currently operating experiments and are  planned for future experiments in particle physics and nuclear physics. The global HEP community would benefit from coordination of U.S. strategy with the ECFA Detector R\&D implementation process in Europe; and in  order to maintain and expand U.S. expertise on MPGDs, the NP and HEP communities would benefit strongly from a joint MPGD development and prototyping facility in the U.S..

\smallskip\noindent{\bf Calorimetry}   The precision energy measurement requirements of future physics programs are stimulating innovation in particle flow and dual readout systems. The increasing availability of precise timing is adding an important new dimension to system implementation, while the development of a range of fast, radiation-hard active materials is leading to
increased flexibility and exciting possibilities for calorimeter system designs.
Future calorimetry at fixed target and colliding beam experiments should be fundamentally multidimensional, providing shower position, time, energy, and a detailed look at shower constituents through
the exploitation of an application-specific combination of particle flow techniques, materials with intrinsically good time or energy resolution, and/or dual readout techniques. Sustained R\&D is needed to move multidimensional calorimeters from the prototype stage to realistic detectors.
Scaling to hundreds of thousands or tens of millions of channels while maintaining the required quality is a huge challenge.
Effective partnerships with chemists, materials scientists, industries large and small, and radiation facilities must be continued and strengthened to explore the landscape of materials that enable precision calorimetry and to lower the cost for future applications.
\smallskip\noindent{\bf Readout Electronics and ASICs} Increased channel count and granularity, driven by the requirements of particle flow calorimetry as much as by tracking, and the need to operate in very harsh environments, for example in high radiation or at cryogenic temperatures, has increased to need for custom readout electronics and ASICs. The development of ASICs requires access to very highly specialized electronics engineers and technicians and physicists with some training in this area. There are some overarching goals for advancing the field of readout and ASICs. Among those are to improve mechanisms for shared access to advanced technology providing broader access by
the community to foster real exchange of information and accelerate development; create frameworks and platforms for easy access to design tools;  develop specialized online resources for the HEP community (e.g. system simulations and design repositories);and    and  provide the basis
for true co-design R\&D efforts from simulation to verification and implementation.

\smallskip\noindent{\bf Noble element Detectors} Many neutrino and dark matter experiments employ noble elements in liquid, gaseous, or solid phases because they can provide large target volumes, with low backgrounds, and provide multiple signal paths for event detection and classification, including  ionization as well as scintillation and Cerenkov light (photons).  Key needs are to 
enhance and combine existing modalities (light and charge) to increase 
signal-to-noise and reconstruction fidelity; develop new modalities for signal detection in noble elements, including methods based on
ion drift, metastable fluids, solid-phase detectors, and dissolved targets; and address challenges in scaling technologies including material purification, background mitigation, large-area readout, and magnetization.

\smallskip\noindent{\bf Cross-cutting and Systems Integration} Presently U.S. funding for Advanced Detector R\&D is institute-based rather than collaboration-based. Yet collaborations are more essential than ever to leadership in Detector R\&D technology and funding constraints have limited the opportunity to establish significant collaboration.
Funding for the DOE’s KA25 Detector R\&D program should be increased significantly (2-5X) to enable the establishment of collaborative R\&D programs that can address Grand Challenges and pursue blue-sky concepts.
 We recommend that funding for key technical
personnel essential to these capabilities be considered on equal footing with the physicists.
It is essential that adequate resources be provided to support more speculative
blue-sky R\&D. 
The U.S. HEP community should establish a robust collaborative research program that includes both participation in international RD collaborations and the establishment of domestic Detector R\&D consortia. Facilities are required to advance all detector technologies. Test beams and irradiation facilities allow users to test the performance and lifetime of their detectors under realistic conditions. It is important that the energy, intensity, particle composition, and time structure of the beams are adequate for the detector needs of the next generation. Cryogenic Test Facilities are also required. The operation of sensors and systems at ultra-low temperatures is a major growth field, but one which necessarily has a high barrier for entry due to the relatively high cost of equipment, not just for the cooling platform itself but also for associated measurement equipment and electronics, and a knowledge gap in the operation and engineering of devices and systems at cryogenic temperatures. 
Facilities (mainly at U.S. National Labs) are a vital element of detector technology development and should be supported. Present gaps should be eliminated. Finally, multidisciplinary problems are prevalent in many areas of instrumentation development.  Multidisciplinary work has funding, recruiting, and career development challenges that new R\&D frameworks could alleviate.

\smallskip\noindent{\bf Radiodetection}  Cosmic rays, neutrinos, and gamma rays are all important probes of astroparticle physics and high energy physics.  They require  large volumes of natural material that can be monitored to study these very rare events. Standard targets are mountains, the Earth's crust, large volumes of ice, or the atmosphere, the latter two of which are being actively pursued. Ice is transparent to radio-frequency signals over distances of order 1km.    Approaches under study for radio detection of cosmic particles include Askaryan emission, geomagnetic detection, and radar. R\&D is required to optimize designs and reduce cost, optimize power consumption and simplify electronics design, especially for large extended arrays, and prepare the way for  future large experiments by undertaking pathfinder experiments for new mm-wave detectors with high channel density.

\medskip 
Detector instrumentation is at the heart of scientific discoveries. Cutting-edge technologies enable U.S. particle physics to play a leading role worldwide. The Instrumentation Frontier report summarizes the current status of instrumentation for High Energy Physics, the challenges and needs of future experiments, and indicates the high-priority research areas.

\subsection {Neutrino Frontier~\cite{Huber:2022lpm}}

Neutrinos, the neutral electroweak partners of the charged leptons, are predicted to be massless in the SM. The discovery that neutrinos of one flavor  oscillate into another flavor as they propagate is a clear indication that there is physics beyond the SM. The study of neutrinos and their properties is therefore an essential component of the U.S. HEP program in the coming decades. The physics questions which drive these investigations include:

\begin{itemize}

\item {\bf How are the neutrino masses ordered?} In particular, is the electron neutrino mostly made up of the heaviest mass eigenstate or not?

\item {\bf What are the values of the neutrino masses?} While oscillation measurements only constrain neutrino mass-squared differences, how well can laboratory-based or cosmological observations constrain the overall mass scale?

\item {\bf What is the origin of the neutrino masses?} Do neutrinos get mass via the same mechanism as the other quarks and leptons or does their mass-generation mechanism involve new physics?

\item {\bf Are neutrinos their own anti-particles?} Are neutrinos Dirac or Majorana particles?

\item {\bf Do neutrinos and antineutrinos oscillate differently?} Is there CP violation in the neutrino sector and, if there is, could it play a role in the generation of the matter-antimatter asymmetry in the early universe?

\item {\bf Is the three-flavor picture of neutrino oscillation complete?} Are there more neutrinos, potentially with different electroweak properties?

\item {\bf Do neutrinos interact in novel ways, or with new and so-far undiscovered particles?} Are there additional neutrino properties related to dark matter or new physics at low energy scales?

\item {\bf Using neutrinos as a window on the universe, what will we see?} How can we use neutrinos as probes into some of the most energetic and still-mysterious processes in the universe?

\end{itemize}

Since neutrinos interact so weakly, their properties can only be studied in sensitive experiments with a large target mass where the few neutrino interactions can be confidently measured and precisely characterized in the presence of backgrounds. Neutrino oscillation measurements in particular can be effectively carried out using GeV-scale neutrino beams generated using high-intensity protons.  Neutrino fluxes are measured at near detectors before propagating over long baselines to large underground detectors.

The DUNE collaboration was formed to realize the 2014 P5 vision of a best-in-class long-baseline experiment based at Fermilab.
The DUNE R\&D program, propelled by the development of large-scale liquid-argon detectors in the U.S. and Europe, in particular through the CERN Neutrino Platform, has demonstrated the power and feasibility of this technique. 

The DUNE program will unfold in parallel with the Hyper-Kamiokande (HK) long-baseline experiment in Japan. The HK experiment plans to utilize the upgraded J-PARC 1.3 MW proton beam, an upgraded near detector, a
proposed new intermediate detector to be installed 
$\sim 1$ km away from the neutrino production target, and a new 260 kt
water-Cherenkov far detector located 295 km away from the neutrino source. HK will have
a strong sensitivity to CP violation for much of the parameter space, assuming the neutrino mass ordering is known. While
HK’s design parameters will make mass-ordering determination difficult using its accelerator neutrino data sample alone, a
simultaneous fit to the accelerator and atmospheric neutrino data is expected to significantly enhance sensitivity to oscillation parameters. Civil construction for the HK tank is underway, with data-taking scheduled to start in 2027.

DUNE will be built in two phases. Phase I  includes completion of the Long-Baseline Neutrino (LBNF) and the Sanford Underground Research Facility (SURF) in South Dakota, 1300 km from the neutrino source at Fermilab and about 1475 m underground, including far site facilities to accommodate four far detector modules (FDs), each with a mass of at least 10-kt (fiducial), and installation of the first two FD modules, as well as near-site facilities to support the full (Phase II)  near detector (ND), installation of a minimal suite of near detectors, and a 1.2 MW source proton beam at FNAL that can be upgraded to 2.4 MW. LBNF will be completed before the beginning of DUNE Phase I data-taking in the late 2020s. The Phase I configuration is sufficient for early physics goals, including the determination of the neutrino mass ordering. 

{\bf DUNE Phase II is required to achieve the precision neutrino oscillation physics goals laid out by the last P5 and is the U.S. HEP neutrino community’s highest priority project for 2030–2040.} The Phase II project has three components: an upgrade or replacement of the Fermilab 8 GeV Booster to deliver 2.4 MW to the DUNE target and possibly to provide beams for other experiments; the construction of an additional 20 kt (fiducial) of far-detectors at SURF; and the full, highly capable, near-detector complex at FNAL to control the systematic uncertainties for the far detector measurements. The full DUNE program will perform definitive studies of neutrino oscillations, test the three-flavor paradigm, search for new neutrino interactions, resolve the mass ordering question, and have the ability to observe CP violation if it is present in neutrino oscillation.

The Neutrino Frontier  program includes a broad scope of activities spanning the full range of scales, including DUNE as
a major international program with more than a thousand collaborators; medium-scale experiments such as those
making up the short-baseline neutrino program; contributions to international experiments; smaller-scale experiments
at accelerators, reactors, and spallation neutron sources; down to tabletop experiments and blue-sky R\&D activities.
Many examples are described in the Topical Group reports and a comprehensive list of Neutrino Frontier experiments
is provided in the Frontier Report from information provided by the community
members.
All of these efforts have the potential to make paradigm-changing discoveries or innovations or provide necessary
inputs to experiments that will make these discoveries, and they will work in synergy to address the science Drivers
described above. 

{\bf A neutrino program incorporating the breadth and diversity of efforts at different scales does more than just
increase the chances for major discoveries.} Such a program is healthy for future scientific progress by making space for
creative thinking, and also provides training opportunities to ensure a capable workforce in the long term. The last
P5 report recommended: “Maintain a program of projects of all scales, from the largest international projects to mid
and small-scale projects”. The Neutrino Frontier endorses continued and enhanced support of this recommendation.

Cross-pollination between Frontiers and other fields of science offers further opportunities. Physics topics within
the Neutrino Frontier overlap strongly with each other, and also with other Frontiers. Examples include instrumentation for dark matter and neutrinoless double beta decay searches; the study of neutrino-nucleus interactions; the detection of high-energy neutrinos with far forward experiments at the LHC; and the Cosmic Frontier programs that provide insight into neutrino properties
from cosmological observables. 

The Neutrino Frontier includes the following Topical Groups:  NF01, Neutrino Oscillations; NF02, Understanding Experimental Neutrino Anomalies; NF03, Beyond the Standard Model (BSM);
NF04, Neutrinos from natural sources; NF05, Neutrino properties; NF06, Neutrino Interaction Cross Sections; NF07, Applications; NF08 $\rightarrow$ TF11, Theory of Neutrino Physics (a topical group shared with the Theory Frontier); NF09, Artificial Neutrino Sources; and NF10, Neutrino Detectors.

To summarize, the Neutrino Frontier recommends:

\begin{itemize}
    \item A future neutrino program with a healthy breadth and balance of physics topics, experiment sizes, and timescales, supported via a dedicated, deliberate, and ongoing funding process, is highly desirable.

\item Completion of existing experiments and execution of DUNE in its full scope is critical for addressing the NF science Drivers.

\item  Directed R\&D needs to be supported to exploit the new and broader opportunities presented by DUNE Phase II.

\item In order to reap the full scientific benefits from future neutrino experiments, strong and continued support for neutrino theory is needed.

\item The Neutrino Frontier, with its strong connections across HEP and beyond, has a special responsibility to contribute to leadership for a cohesive, HEP-wide strategic plan for DEI and community engagement.

\end{itemize}

\medskip

A description of the work of the Topical Groups that contributed to the final Neutrino Frontier vision is given below:

\smallskip\noindent {\bf Theory and Motivation} A robust neutrino theory and phenomenology effort is required in order to exploit the unique probes of fundamental
physics provided by neutrino experiments to interpret the data, build models to accommodate new phenomena, provide guidance for future experiments, and connect the new discoveries in neutrino physics to other areas of particle
and nuclear physics, astrophysics, and cosmology. For precision measurements of the three-flavor picture, for example, theory provides the necessary work to produce a precise description of relatively low-energy neutrino-nucleon and neutrino-nucleus scattering.  
Theory is also making links between neutrino properties, established or conjectured, and cosmology. An increase in support for the neutrino theory effort is well justified.

\smallskip\noindent {\bf Three-Flavor Neutrino Oscillation} The three primary goals of the coming years of neutrino oscillation measurements are: to determine the ``neutrino mass ordering," which is defined by the sign of $\Delta m^{2}_{31}$; 
to determine whether $\theta_{23}$ is more than or less than 45$^\circ$ and how close to maximal it is, known as the ``octant"
question; and  to measure $\delta_{CP}$ and determine whether $\sin \delta_{CP} \  = \  0$ can be excluded or not. The next-generation experiments, DUNE/LBNF at Fermilab/SURF and the Hyper-Kamiokande  experiment in Japan are designed to achieve the necessary statistical precision to answer the above questions over the full parameter space. JUNO, a large liquid scintillator experiment in China, and IceCube-Gen2 will also contribute to the knowledge of oscillation parameters.

\smallskip\noindent {\bf Physics Beyond the Standard Model and Understanding Experimental Neutrino Anomalies} The rapidly expanding field of searching for BSM  neutrino physics now includes: Heavy Neutral Lepton (HNLs) searches, expanding the continuing effort to understand the neutrino ``anomalies";
neutrino flavor and neutrino scattering; and dark matter and baryon number violation searches (BSM physics searches that can be accomplished using neutrino detectors). The last P5 supported  experiments  aimed at understanding the LSND, MiniBooNE, and reactor anomalies. 
Should the interpretations converge, the short-baseline experimental and theoretical landscape will shift, with subsequent priorities depending heavily on what is learned about the various hypothesized anomaly origins. 
It is therefore essential that resources are available  to enable adaptation to developments during the next P5 period.

\smallskip\noindent {\bf Neutrinos as Terrestrial and Astrophysical Messengers} Neutrinos, by virtue of their weak interactions, probe otherwise inaccessible sources. They serve as terrestrial and astrophysical messengers, providing unique information about the nature of the sources themselves. In conjunction with other types of emissions (photons, gravitational waves, charged particles) astrophysical neutrinos provide a deeper and richer scientific story. Cosmological studies also provide important contributions to our understanding of neutrino properties, with sensitivity to the number of neutrinos, the sum of their masses, and potential new neutrino interactions. Current-generation detectors have measured neutrinos from a few hundred keV up to the PeV regime. Expanding outside of this energy interval is unexplored territory, requiring novel technology.

\smallskip\noindent {\bf Neutrino Properties}
KATRIN and cosmological surveys will eventually be sensitive to the absolute mass scale of neutrinos down to   $\sim$0.01 eV. 
The coming ton-scale generation of neutrinoless double beta-decay experiments will probe effective Majorana neutrino masses as small as 10 meV. The neutrino community is pursuing a thriving R\&D program for beyond ton-scale searches;
there is a strong consensus for building a ton-scale double beta-decay experiment with U.S. leadership, multi-agency, and international support, as well as interest in continuing participation in other programs worldwide. 
Neutrinos should have small electromagnetic interactions induced by radiative corrections that can be probed 
in the next generation of elastic neutrino-electron scattering and CE$\nu$NS experiments at accelerators and reactors. Other exotic neutrino properties, including those that violate fundamental SM symmetries or have gravitational interactions that violate the equivalence principle, are all potentially observable.

\smallskip\noindent {\bf Neutrino Interactions} A thorough theoretical and phenomenological understanding of neutrino cross sections over a wide range of energies and targets is crucial for the successful execution of the entire neutrino physics program. 
This understanding must be captured in event generators; these event generators must be maintained and updated whenever new measurements or theories become available, but this work is often undervalued relative to other activities.  Low-energy ($<0.1$  GeV) neutrino cross sections are relevant for studying solar, reactor, and supernova neutrinos. Stopped-pion neutrinos can be used to study, through processes like CE$\nu$NS, fundamental properties of neutrino weak interactions.  Long-baseline experiments make use of neutrinos in the few-GeV range and the need for understanding the cross sections in this regime is critical for the interpretation of oscillation experiments. New information will come from experiments using  electrons
as well as direct neutrino cross-section
measurements by
short-baseline experiments, near detectors in long-baseline experiments,  and dedicated neutrino scattering experiments.
High-energy cross sections (from a few hundred GeV to greater than the TeV scale) are relevant for high-energy astrophysical neutrinos. No laboratory neutrino
cross-section data are yet available for energies above about 350 GeV (although some measurements with astrophysical neutrinos have been done). 
Detection of neutrino interactions in the far forward region at the LHC will extend the laboratory reach into the TeV scale.

\smallskip\noindent {\bf Computing} Many of the neutrino community's computing issues are shared by experiments on the other Frontiers, such as the change to heterogeneous computing hardware and the dependence on HPCs.
The need for data preservation is also crucial since cross-section measurements will be taken over time but the most detailed use of them will be after DUNE has acquired enough data to reach the 1\% systematic uncertainty level. 
Special considerations include the need for strong and responsive support for GEANT4 (or successors), the integration of new experimental results and theory into event generators, and in general the maintenance of software over the long duration of the experiment and beyond. Machine learning has been embedded in the neutrino community for a long time but a new element is a reliance on externally developed frameworks, which provides many advantages but requires work to develop and maintain interfaces. DUNE also has a different data profile than many other HEP experiments so its storage requirements and access patterns may present special challenges. 

\smallskip\noindent {\bf Artificial Neutrino Sources} The primary artificial neutrino sources of wide use in the Neutrino Frontier are accelerator-produced beams.  Beam powers of hundreds of kW have been achieved at FNAL and J-PARC and upgrades to MW beams are planned for DUNE and HK.  Nuclear reactors have also had a central role in experimental neutrino physics ever since the discovery of these elusive
particles in 1956; they are the most intense source of MeV-scale electron antineutrinos.
Spallation neutron sources such as the Spallation Neutron Source at Oak Ridge National Laboratory provide, as a by-product, the most intense accelerator-based sources of neutrinos in the
world, and are particularly important for a wide variety of neutrino physics measurements. 
The LHC produces neutrinos 
and, with instrumentation downstream of the collision region at Point 1, the FASER$\nu$ collaboration recently reported the first observations of neutrinos from a collider.
There are many other novel neutrino sources  under consideration, including neutrinos produced from muon beams and neutrinos produced by proton cyclotrons.

\smallskip\noindent {\bf Detectors}
Neutrino physics spans an enormous range of energies and scales from keV to TeV. This means that a broad spectrum of detection technologies is required. Several areas have attracted large community interest, including expanding  programs aimed  at improving liquid and gaseous noble element TPCs; pursuing  hybrid Cherenkov/scintillation detectors; optimizing low-threshold neutrino detectors to achieve  even lower energy thresholds and better  background rejection techniques as well as moving toward larger detector masses;
developing technologies for neutrino detection at the PeV scale and beyond; and developing in collaboration with the Cosmic Frontier new dark-matter detectors.  Both the development of new noble liquid and gas
detectors and low-threshold detectors of various types  can produce devices  that are useful for both the Neutrino Frontier and the Cosmic Frontier.

\smallskip\noindent {\bf Applications} Somewhat surprisingly, given
the inherent difficulty of neutrino detection, the direct application of neutrinos to advance other fields of research or solve societal problems is also feasible. Better knowledge of reactor neutrino emission will improve our understanding of the physics of reactors and enables applications to nuclear non-proliferation safeguards, nuclear security, and arms control. Neutrino emission can also be used to probe the Earth's interior.
Given that neutrino physics represents a major component of the U.S. program, there are opportunities for leadership
in addressing societal issues. These topics are discussed in detail in the Community Engagement Frontier report.

\medskip

To summarize, if the three-flavor paradigm is sufficient to fully describe the neutrino sector, it will take approximately the next two decades to fill in the main features of the picture. In parallel, studies of the anomalies and the search for BSM physics will continue. Astrophysical neutrinos will continue to bring new information.  Neutrino physicists will continue to develop new instrumentation, new sources, new analysis methods and tools, and new concepts to position themselves for physics well beyond DUNE.

The physics of neutrinos and the physics that can be done with neutrinos extends from sub-eV to EeV – across eighteen orders of magnitude – and touches nearly every other area of particle physics.
The U.S. neutrino program is poised for an exciting future. The path is clear for the next P5 period and the decade beyond it.

\subsection{ Rare  Processes and Precision Measurements Frontier~\cite{Artuso:2022ouk}}

The Rare Processes and Precision Measurements Frontier
(RPF) encompasses searches for extremely rare processes or tiny deviations from the SM that can
be studied with intense sources and high-precision detectors. In Snowmass 2013, these topics were combined with neutrino physics to form the ``Intensity Frontier". For Snowmass 2021, neutrino physics was designated as a frontier in its own right, with the remaining very broad and diverse collection of topics constituting RPF. 
Searches for rare 
flavor transition processes and precision measurements are indispensable probes of 
flavor and fundamental symmetries and provide insights into physics that manifests itself
at higher energy or through weaker interactions than those directly accessible at high-energy colliders. This program includes very small-scale experiments, as well as medium- and large-scale experiments.  

The Rare Processes and Precision Measurements Frontier is currently working on two mid-sized U.S. projects at Fermilab endorsed by P5 in 2014,  namely the Muon $g-2$ experiment, which is nearing completion, and the Mu2e experiment, which is under construction. The program also has important investments in flavor physics through the support of the Belle II experiment in Japan and LHCb at CERN. Priorities for the next few years are to complete the analysis of the Muon $g-2$ experiment, begin taking data with Mu2e, and continue taking and analyzing data at Belle II and LHCb.

Since Snowmass 2013, several important results have emerged from RPF experiments. The initial results from the Muon $g-2$ experiment are consistent with the result obtained at BNL in 1997, now with increased precision, and much more data has since been taken and will soon be analyzed. Taken at face value, the gap between experiment and (current) theory is now greater than $4\sigma$, a tantalizing hint for physics beyond the Standard Model. However, lattice QCD calculations of the hadronic corrections to muon $g-2$ are rapidly improving in precision. In light of the first results indicating puzzling tensions with the current SM prediction, it will be important to see if lattice QCD results, once they have been cross-checked against each other, agree with the data-driven evaluation on which the current SM prediction is based. Since 2013, several measurements from LHCb and Belle II in $B$-meson decays that involve leptons in the final state have also shown tensions with predictions of the SM. These include $b \rightarrow  sl^{+}l^{-}$ and $b \rightarrow c\tau\bar{\nu}$ transitions at BABAR, Belle, and LHCb, and in rare and forbidden decays such as $b \rightarrow s\nu\bar{\nu}$ or $B_{s}  \rightarrow l^{+}l^{-}$ in LHCb.
While no single measurement has reached 5$\sigma$, there seems to be an emerging pattern that could be explained by the existence of a new scalar particle or leptoquarks. This highlights the potential  of RPF for the discovery of BSM physics.

Looking ahead to the next decade and beyond, the Frontier developed a broad community consensus consisting of six central points (unordered and intertwined)
\begin{itemize}
\item  
The threefold  replication of the generations, combined with the unexplained pattern of masses of quarks and leptons, spanning  several orders of magnitude, is a central mystery of particle physics. 
Flavor physics has two main goals. The first objective aims
at uncovering the underlying reason for family replication and its different properties. The second encompasses the study of the rich of decays of different flavor species available for experimental observation and  allows us to probe deviations from SM expectations with a multiplicity of approaches. 
The search to understand the physics of flavors and generations is central to both physics programs articulated by the Neutrino and Rare Processes and Precision Measurements Frontiers, but this centrality is hidden in the structure of our current science Drivers.
As a result of the breadth of flavor physics, its potential for discovering BSM physics, and the hints in the current data, the Rare Processes and Precision Measurement Frontier proposes that the upcoming P5 adds a {\bf new  science Driver: flavor physics as a tool for discovery}. 

\item The U.S. should support the LHCb Phase-II upgrades and Belle II. These experiments are broad, powerful, and
irreplaceable probes of flavor physics along with topics relevant to all the Frontiers. These experiments pursue
complementary research programs that utilize different beams and detection techniques to explore a vast array
of new physics in beauty, charm, and rare tau decays. In addition, they investigate the patterns of bound states
as the manifestation of the richness of QCD in its non-perturbative regime and contribute significantly to the
exploration of the dark sector.
\item We should select a portfolio of accelerator-based dark sector experiments that are well-motivated, unique,
and affordable. Many possibilities have been identified and studied during the Snowmass process. P5 needs to
support this physics, with a consequent process through which the community, DOE, and NSF select an efficient
and effective subset of the opportunities using the Dark Matter New Initiatives (DMNI) studies as input.

\item Experiments investigating charged lepton flavor violation and lepton number violation in the muon sector can probe mass scales far beyond the direct reach of colliders, as well as explore the nature of the flavor physics issues central to HEP. The PIP-II 800-MeV  SRF proton linac under construction at FNAL, if upgraded to CW beam capability with a total average beam power of up to  O(1 MW), would enable a new muon program at an unprecedented intensity that could increase the discovery potential of such experiments by at least an order of magnitude. The U.S. should support vigorous R\&D in order to realize this program.

\item The theory efforts that guide and enable these investigations, while not ``projects" in themselves, should be vigorously supported by P5. RPF relies on the techniques of and guidance from Effective Field Theories (EFTs), as well as calculations
of quantities that provide precise and
unbiased results that can be systematically improved as needed, such as Lattice QCD. Precision measurements
of deviations from the SM, instead of the direct observation of new particles, require these methods.
In addition, theory-experiment collaboration in developing new experimental approaches has been at the foundation of much of the progress in dark sector searches and is vital to this growing field.
\item A portfolio of experiments of different cost and time scales is an integral part of the RPF physics program. Promising experiments include:
\begin{itemize}
\item experiments measuring electric dipole moments, in particular, the proton EDM measurement in a storage
ring, along with experiments exploiting synergies with AMO techniques to examine fundamental symmetries;
\item experiments probing rare light meson decays, such as REDTOP;
\item involvement of the U.S. experimental community in the vibrant international program at both CERN and J-PARC studying rare K decays;
\item  the PIONEER experiment at PSI which will study lepton universality in the muon and electron generations.
\end{itemize}
\item  We also stress that small and medium-sized multipurpose experiments can provide training for early career
researchers over many stages of an experiment, from proposal-writing to design and construction to publication, exemplified by the Fermilab g$-$2 or the quark flavor experiments, that allow junior scientists to combine
instrument or computationally intensive tasks with physics analysis contributions in their portfolio. In addition,
the relatively small sizes of these experiments allow for less hierarchical organization, with a broad array of
leadership opportunities. A commitment to mentor early career scientists in a supportive and inclusive
environment should help to guide our programmatic choices.
\end{itemize}

This Frontier  connects to several other Frontiers. It connects to Nuclear Physics and AMO, especially in regard to EDM searches, which can be done with nucleons, atoms, or molecules and are really low-energy experiments. Since some of the experiments use storage rings or special accelerator facilities, there is a strong link to the Accelerator Frontier. The diversity of the experiments leads also to the need for specialized instrumentation, providing a strong connection to the Instrumentation Frontier.

The material developed in the Topical Groups that is used as input to this consensus is very briefly summarized in the paragraphs below. Many more details are given in the RPF summary report and the individual Topical Subgroup reports and white papers. 

\medskip

\smallskip\noindent {\bf Weak decays of $b$ and $c$ quarks} Studies of heavy-flavored hadrons provide rich, diverse, and model-independent probes for new physics at energy scales far beyond what is directly accessible. Heavy-flavor physics is crucial to our search for new physics in the upcoming 10--20 years, important advances are expected through a highly synergistic program of experiments: LHCb and its planned upgrades at the LHC, Belle II at the SuperKEKB asymmetric $e^+e^-$ collider, and the $e^+e^-$ charm factories BESIII and STCF. In particular, Belle II and LHCb have the unique potential to unveil new physics by confirming intriguing hints of deviations from the SM that have been recently observed in $b \rightarrow s \ell^+ \ell^-$ and $b \rightarrow c \tau\bar{\nu}$  transitions at BABAR, Belle, and LHCb, or finding new unexpected outcomes in the study of rare and forbidden decays such as $b\rightarrow s \nu \bar{\nu}$ or $B_{(s)}\rightarrow \ell^+\ell^-$. In addition, the continuing refinement of the studies of quark mixing, flavor oscillations, and CP violation may unveil subtle deviations from SM expectations. Farther into the future, the experimental program can be extended at the $e^+e^-$ circular colliders  proposed for precision studies of the Higgs boson. The U.S.\  flavor community is well-positioned to lead key aspects of the physics, computing, and detector construction in all of these experimental programs. Theoretical efforts that inspire and elucidate the fundamental impact of experimental findings are crucial to success and should be supported.
The lepton universality and other stringent tests foreseen in these experiments and their upgrades are discussed in the $b$ and $c$ topical group reports.

\smallskip\noindent {\bf Weak decays of strange and light quarks}
Studies of the light quarks include precision measurements of both  flavor-conserving and flavor-violating decays of pions,
kaons, hyperons, and $\eta / \eta^{\prime}$ mesons. Tests of new physics through checks of the first column unitarity of the CKM
matrix, along with lepton flavor/number and lepton universality tests, have revealed experimental anomalies. These
anomalies require additional experimental and theoretical studies of the light quark systems to conclusively assess
their impact.

\smallskip\noindent {\bf Fundamental Physics in Small Experiments} The study of static properties of elementary particles (electric and magnetic moments) and the fundamental symmetries (C, P, T, and their combinations, along with basic probes of Lorentz symmetry), all probe energy scales approaching the Planck scale. The experiments addressing this physics are of small to intermediate size and sometimes involve methods not traditionally considered high-energy physics. Nonetheless, in subjects such as $n\bar{n}$ oscillations, EDMs, and neutrinoless double beta-decay, 
the Frontier worked to identify ways to develop a synergistic physics program that exploits the expertise of the high-energy physics community, the relevant AMO and nuclear physics communities, and the scientists performing such research.
Storage ring EDM experiments are an exciting opportunity, and proton storage ring experiments might reach 10$^{-29}$ e-cm.
Low-energy antimatter gravity tests using muonium could be performed at the AMF facility described in the RPF Frontier report\cite{Artuso:2022ouk}, and these developments are also synergistic with possible efforts for precision measurements of the muonium
spectrum and searches for muonium-antimuonium oscillations. 
PIONEER can study lepton universality in charged pion decay to 0.01\%
precision, probing heavy neutral leptons and dark sector particles up to PeV mass scales, 
and complement universality tests in b-decays.

\smallskip\noindent {\bf Baryon and Lepton Number Violating Processes}
Experimental and theoretical research in baryon- or lepton-number violation (such as proton decay, $n-\bar{n}$ oscillations, and
searches for $0\nu2\beta$ decays) is traditionally supported by nuclear physics (NP). These efforts share many of the same intellectual problems, methods, and people with HEP. Unfortunately, the NP/HEP separation produces barriers that make it difficult for researchers at the border.
Studies of neutrons are another avenue of opportunity for profound discoveries. One particular opportunity of note
is the study of $n\bar{n}$  oscillations. The proposed NNBar experiment at the ESS could reach a limit of $\tau_{n\bar{n}}\sim$ 10$^{9-10}$s. ``Mirror neutrons", $n\rightarrow n^{\prime}$ oscillations, have been ruled out as an explanation of the discrepancy between
neutron lifetimes measured with cold and ultracold neutrons, but improved sensitivity to this phenomenon is possible
at the HIBEAM program at the ESS.
It is worth noting that studies of baryon-number violating decays of heavy flavors are uniquely being carried out in dedicated flavor experiments, such as LHCb and Belle II.

\smallskip\noindent {\bf Charged Lepton Flavor Violation (electrons, muons, and taus)} In contrast to neutrinos and quarks, charged lepton flavor violating/violation (CLFV) interactions that do not conserve lepton family number, have never been observed. The archetypal CLFV decay is $\mu \rightarrow e\gamma$ (with no emitted neutrinos).
Muons play a unique role in CLFV searches because we can make intense beams of muons and achieve the highest statistical sensitivity. In muon-to-electron conversion, $\mu^{-}N \rightarrow e^{-}N$,the Mu2e experiment at Fermilab will reach its goal of $R_{\mu e} < 8\times 10^{-17}$ at 90\% CL by the end of the decade. A factor of ten upgrade from Mu2e, Mu2e-II, needs R\&D effort to both improve the limit and change the physics probed with a different target nucleus, most likely Ti. 
The proposed Advanced Muon Facility (AMF) exploits the PIP-II CW beam upgrade augmented by new rings. 
This facility would allow us to examine both  $\mu^{-}N \rightarrow e^{-}N$ and $\mu \rightarrow 3e$, as well as
$\mu^{-}N \rightarrow e^{+} N’$ and muonium oscillation studies,
at rates far beyond those achievable at any other planned facility; here Detector R\&D is essential to use the rates available at PIP-II. AMF would also enable a dark matter experiment.  AMF itself also requires significant R\&D.
 
\smallskip\noindent {\bf Dark Sector Studies at High Intensities}
Because dark sectors are generically weakly coupled to ordinary matter, and because
they can naturally have (light) masses ranging from roughly the electron mass to the proton mass and  respect the Standard Model's symmetries, they are only mildly constrained by high-energy collider data and precision atomic measurements. Intensity Frontier experiments provide opportunities to carry out experimental studies of the dark sector with possible access to specific dark
matter candidates.
Existing large multipurpose detectors, especially Belle II and LHCb, can be used to
study dark-sector states as well as  dedicated efforts at high-intensity
accelerators. The DarkMatter New Initiative (DMNI) report has given initial support to two experiments,
and a promising set of experiments has been identified since the last P5 report. A broader, coordinated program
is emerging from this initial work. Dark-sector theory will also provide critical guidance. 
A key consensus of the RPF is that the U.S. should
develop a  portfolio of DM experiments using intense beams.

\smallskip\noindent {\bf Hadron Spectroscopy}
Hadron spectroscopy examines the description of the spectrum of hadrons from non-perturbative QCD. With the availability of intense sources of hadrons, many new states composed of quarks and gluons have been
discovered by LHCb, Belle/Belle II, BESIII, and other experiments. Several of these do not fit among conventional hadronic
states composed of quark-anti-quark pairs (mesons) or three quarks (baryons). New exotic states, such as tetra- and pentaquarks, hybrids, etc., are added every year. The nature of these states is not clear. Configurations such as  compact multiquark
states, hadron molecules, or even linear combinations of such components, are possible and require careful experimental measurements
to disentangle. Detailed studies require very large samples of these hadrons and need high luminosity colliders or intense beams. 

\medskip

RPF uses the techniques of many accelerators and experiments to search for new physics and do precision SM tests with diverse and complementary approaches. The major themes of this Frontier that emerged during the Snowmass process have been summarized here; the range and power of the RPF program are unique and several experiments that we examined encompass the flexibility that enables them to elucidate unexpected observations regardless of where they are found.

\subsection{ Theory Frontier~\cite{Craig:2022cef}}

   Theoretical particle physics seeks to provide a predictive mathematical description of matter, energy, space, and time that synthesizes our knowledge of the universe, analyzes and interprets existing experimental results, and motivates future experimental investigation. Theory connects particle physics to other areas of physics and extends the boundaries of our understanding.  Together, fundamental, phenomenological, and computational theory form a vibrant interconnected ecosystem whose health is essential to all aspects of the U.S. high-energy physics program.
    
        {\bf The Snowmass Theory Frontier recommends vigorous support for a broad program of theoretical research as part of a balanced portfolio, augmented by targeted initiatives to connect theory to experiment; support and training for students and junior scientists; and strengthening the commitment to improve diversity, equity, inclusion, and accessibility. A general description of progress and promise in each of the three categories of theoretical high-energy physics research is given below.}

\smallskip

{\bf (1) Fundamental theory: }Fundamental theory (often also referred to as formal theory) seeks deep understanding of the theoretical and mathematical structures that underlie our modern description of Nature and includes directions that are not (or not yet) directly connected to experimentally testable consequences. The last century witnessed the triumph of quantum mechanics and classical relativistic gravity, both initially aspects of fundamental theory, whose impact on basic science and industry -- and on our daily lives -- have been enormous. Current research in fundamental theory explores the frameworks that arise at the intersection of relativity and quantum mechanics: quantum field theory and quantum gravity.

$\bullet$~{\bf Quantum field theory:} The historical approach to the union of quantum mechanics and special relativity led to the successful formulation of quantum field theory (QFT). Our current understanding of high-energy experiments and cosmological observations is largely based on a perturbative approach to these theories. In the past decade, however, extraordinary progress has been made in developing new non-perturbative approaches to quantum field theory (including the advent of diverse bootstrap methods); extending the reach of perturbative computations (particularly for scattering amplitudes); and more fully leveraging effective field theory descriptions of physical systems. This progress has been made in tandem with complementary advances in lattice field theory quantifying non-perturbative properties in theories of interest. A new understanding of symmetries (including higher form symmetries, higher group symmetries, sub-system symmetries, and non-invertible symmetries) has also played a role in this progress. Although such generalized symmetries have only been discovered quite recently, they appear to be abundantly realized in nature, with considerable relevance to both high energy and condensed matter physics. Finally, quantum information has provided an important new perspective on quantum field theory, in which entropy and entanglement play a prominent role. Using information content as the organizing principle has led to the recognition that the structure of entanglement sheds new light on the properties of QFTs.

$\bullet$~{\bf Quantum gravity:} The union of general relativity and quantum mechanics remains one of the outstanding challenges of theoretical physics. Yet the greatest challenges have a tendency to catalyze the greatest progress. Fueled by a deeper understanding of holography and insights from quantum information, thought experiments involving the quantum mechanics of black holes have had a transformative impact on our understanding of quantum gravity and the nature of spacetime. The emergent nature of spacetime implied by such considerations has been made concrete in string theory, where the AdS/CFT correspondence has shed light on the emergence of bulk gravitational spacetime from the field theory description on the boundary. Many open questions remain regarding the emergence of spacetime in the vicinity of black holes. Addressing these questions requires a deeper formulation of quantum gravity than we currently possess, comprising a clear goal for the coming decade.	

$\bullet$~{\bf Physics and mathematics:} From classical mechanics to general relativity, fundamental theories of physics have evolved hand-in-hand with mathematics. This is no less true for modern approaches to quantum field theory and quantum gravity. Recent implications for both geometric and algebraic aspects of mathematics have included direct connections between the classification of superconformal field theories and the theory of canonical singularities in algebraic geometry; wrapped brane states in string theory and the mathematics of enumerative invariants. Connections between QFT, string theory, and number theory are exemplified by the “moonshine program”. Further relations between number theory, algebraic geometry, and quantum field theory are emerging from the geometric Langlands program. Recent work on perturbative scattering amplitudes also revealed deep connections to active areas of work in mathematics.

{\bf (2) Phenomenology:}  Particle phenomenology provides the connection between fundamental theory and the physical description of the real world, testable by experiments. Phenomenology helps to formulate the physics goals of particle experiments and identify promising new avenues for experiments. Phenomenology provides the tools to perform the precision calculations necessary to compare experimental results to theoretical predictions.  Phenomenology also helps develop the methods used for successfully analyzing and interpreting the results of the experiments, and often identifies signals requiring novel analysis techniques.

$\bullet$~{\bf Model building:} There are plenty of indications pointing to the need for BSM physics: the origin of neutrino masses, the quark and lepton flavor structure, the absence of CP violation in the strong sector, the coexistence of the weak and gravitational scales, and the origin of dark matter and dark energy responsible for the acceleration of the universe. Model building attempts to synthesize these clues into the next set of principles that determine the laws of physics at the shortest distance scales. While all BSM models of physics approximately reproduce the SM at low energies, they introduce a plethora of testable ideas using a broad range of theoretical approaches and techniques. Recent developments such as cosmological selection, neutral naturalness, and models of light dark matter have led to new experimental search strategies at existing experiments, as well as to formulations of wholly new experimental programs, often co-led by theorists.

$\bullet$~{\bf Collider physics:} 
Collider phenomenology is an essential interface between the theoretical and experimental high-energy physics communities, serving various roles: from connecting formal investigations to experiments (such as providing guidance on the exploration of the physics possibilities), to supporting the experimental community with the essential tools to simulate and interpret data, to bringing information back to the theory community. An explosion of theoretical activity in collider phenomenology has led to many new collider observables including many forms of jet substructure and the emerging field of multi-point correlators, employing widespread innovations in computational theory to leverage machine learning and artificial intelligence. 

$\bullet$~{\bf Precision collider theory:} Theoretical techniques for precise experimental predictions are the backbone of a successful program in particle physics. They are necessary for the determination of the fundamental parameters of the SM to unprecedented precision and to probe beyond the SM  to very short distances. Much progress has been made in recent years in increasing the accuracy of phenomenological computations, including perturbative calculations at higher orders, resummation of perturbative expansions, improved parton distribution functions, and Monte Carlo simulation packages and event generators. This topic is a prime example where cross-fertilization between phenomenology and fundamental theory has accelerated progress in both areas. The discovery and characterization of the Higgs boson is a prominent example of the success associated with both collider phenomenology and precision collider physics. 

$\bullet$~{\bf Neutrino theory:} The discovery of nonzero neutrino masses requires new fundamental fields and new interactions. Theory played a central role in the development of the formalism of neutrino oscillations, including the solution to the twentieth-century solar and atmospheric neutrino puzzles. This role is expected to persist and evolve as the oscillation probes grow more sophisticated and diverse. The coming era of precision neutrino oscillation experiments promises precise measurements of fundamental neutrino mass and mixing parameters, providing important information on the nature of these particles. A theory-driven, coordinated program combining nuclear effective theory, lattice QCD, perturbative QCD, and event generation to quantify the multi-scale nuclear cross sections at the required level of precision has been launched, that will allow us to unlock the full potential of the present and future neutrino physics program. With their vastly disparate energy scales, observations of astrophysical and cosmological neutrinos provide a wealth of information that is complementary to terrestrial experiments, entwining astrophysics, cosmology, and particle physics phenomenology. Theory and phenomenology play a key role in using this information to deepen our understanding of neutrino properties and interactions. 

$\bullet$~{\bf Flavor physics:} The ongoing and planned quark and charged-lepton flavor experiments provide essential constraints and complementary information on BSM models. The richness of B physics and the large $b$-quark mass enable many complementary tests of the SM, and have been a driving force to develop new perturbative multi-loop and non-perturbative effective field theory techniques since the 1980s. Taking advantage of the large amount of experimental data and the growing number of precise lattice-QCD results for hadronic parameters, new techniques in flavor physics are now enabling advanced theoretical analyses to obtain constraints on promising BSM candidate theories, maximizing the discovery potential of the experiments.

$\bullet$~{\bf Cosmology and astrophysics:} Cosmology and astrophysics provide a wide range of opportunities to expand our knowledge of the fundamental laws of nature, both through direct searches for BSM physics and through tests of the SM in extreme conditions that are impossible to recreate in the laboratory. The impact of theoretical effort in cosmology and astrophysics over the past decade includes advancing our understanding of fundamental physics by forcing us to ponder extreme scenarios where, e.g., quantum effects and gravity must be considered simultaneously; developing new microscopic models that can potentially explain the outstanding problems facing our understanding of nature; and inventing new approaches to test our best-motivated theories, in addition to developing the theoretical tools needed to properly interpret the resulting data. Considerable progress has been made in understanding inflation and properties of the early universe from the cosmic microwave background and large-scale structure. Dark matter theory is undergoing a renaissance with the exploration of the full range of allowed dark matter masses, numerous portals to dark sectors, and novel interaction mechanisms. Advances in dark matter phenomenology have gone hand in hand with new proposals for dark matter experiments, often envisioned and implemented by theorists. The discovery of gravitational waves has catalyzed rapid progress in precision calculations via scattering amplitudes and inspired the use of gravitational waves to study particle physics inaccessible via planned colliders. 

$\bullet$~{\bf Quantum Sensing:} The development of quantum sensing technologies has opened the door to new opportunities to search for new particles or interactions that arise in well-motivated BSM theories, enabling novel experiments, detectors, and measurements. Here high energy theorists are playing a central role in proposing and guiding such experiments in the pursuit of searches for new light-matter particles (axions, axion-like particles, dark photons, milli-charged particles), dark matter, gravitational waves over a large range of frequencies, and tests of quantum mechanics and of gravity – the list is long. 

{\bf (3) Computational theory and quantum information:} Computational theory seeks to quantitatively test our theoretical descriptions of physical phenomena and gain new insights into fundamental aspects of the underlying theories, through developing and deploying computational methods. All areas of high energy theory benefit from or contribute to the development of computational methods to some extent. Computational theory is tied to enabling technologies, both driving and benefiting from ever more sophisticated and powerful platforms for classical computers, quantum computers, and quantum simulators, as well as algorithm development and code optimization. 

$\bullet$~{\bf Lattice QCD:} The development of computational methods for the study of lattice field theory was originally motivated by the desire to understand QCD in the non-perturbative regime. Since then, lattice QCD has been developed into a precision tool with applications to a broad range of observables. It plays a crucial role in interpreting experimental measurements in flavor physics yielding precise SM predictions that reveal surprising new tensions, while determinations of the strong coupling and the quark masses enable precision tests of the Higgs boson. The scope of lattice QCD is undergoing a rapid expansion, promising to provide quantitative access to important new observables in the coming decade, including, among others, PDFs, Transverse Momentum and Generalized Parton Distributions, multi-hadron systems, and inclusive scattering and decay rates, as well as observables involving nucleons and nuclei used in many low-energy searches for new physics. 

$\bullet$~{\bf Lattice Field Theory:} LFT provides access to non-perturbative properties of other QFTs with interesting strongly-coupled dynamics. Theoretical developments such as gradient flow RG and a novel lattice formulation of supersymmetric Yang-Mills are expected to yield valuable insights. Promising results have been obtained in studies of composite DM models and of emergent conformal symmetry in Yang-Mills theories with complex fermion content, among others. 

$\bullet$~{\bf Quantum Information:} Quantum computing and quantum simulation of QFTs hold great promise to overcome the limitations of Euclidean LFT for the study of systems such as real-time scattering dynamics and finite density systems. Recent dedicated efforts to develop the methods and theoretical foundations for quantum simulations of quantum field theories relevant to high energy theory are already yielding intriguing results on currently available hardware, offering great promise for computations of classically intractable problems in the decades to come.

$\bullet$~{\bf Event Generators:} 
 Event generators are vital for the success of experimental programs across all Frontiers, connecting experimental measurements to theoretical predictions by accounting for hadronization, initial and final-state evolution, and rescattering. State-of-the-art generators incorporate higher-order QCD and EW corrections, factorization theorems, parton evolution, and resummation of QCD and QED effects. Improvements to event generators (driven in part by modern machine learning methods) will lead to further reduction of systematic uncertainties, directly enabling future experimental success.

$\bullet$~{\bf Machine Learning and Artificial Intelligence:} Machine Learning and Artificial Intelligence describe a broad class of learning algorithms, including e.g. deep learning architectures based on complex neural networks with many layers. Machine learning is now playing an increasingly important role in all areas of high energy theory, in particular, those that rely on theoretical simulations or numerical analysis. In event generators, all modules can be improved through ML, including phase space sampling, scattering amplitudes, loop integrals, parton showers, parton densities, and fragmentation. End-to-end ML generators, which use generative networks to directly generate parton-level events, complement standard generators in important ways. In LFT, ML applications are being developed for all stages of the computations,  including the generation of gauge fields and correlation functions as well as the numerical analysis needed for the extraction of  the physical observables. The development of novel machine learning methods based on new symmetry-preserving ML architectures for efficient sampling of gauge fields in LFT or of phase space in event generators is notable for its cross-disciplinary impact. 

\smallskip

To summarize, as we enter an era with many promising experiments but few guarantees of discovery, theory is as important as ever. Theory unifies the Frontiers of particle physics. It is essential to the conception, execution, and interpretation of current experiments. Theory is an essential driver of the development and implementation of new, enabling technologies. It also extends well beyond by laying the foundations for future experiments and advancing our understanding of Nature in regimes that experiment has yet to reach. A robust theory program is vital to the success of current and future projects in particle physics.

\subsection{ Underground Facilities \& Infrastructure  Frontier~\cite{Baudis:2022qjb}}

Experiments that require low backgrounds from cosmic radiation often must be performed underground. Underground experiments address some of the most important topics in  particle physics, including the search for dark matter; neutrino physics, including experiments using solar and atmospheric neutrinos and neutrinos produced by accelerators; neutrinoless double beta decay; cosmic-ray physics; supernova detection; and searches for proton decay. The physics underlying the program  is varied and spans several Frontiers. For Snowmass 2021, each underground experiment is included in the frontier appropriate to the physics topic that it addresses and the availability of suitable underground space is considered by this Frontier.

 The Underground Facilities and Infrastructure Frontier  assessed the requirements of proposed and potential experiments for existing and planned underground facilities and compared them to the available underground lab space and characteristics  worldwide to see whether the needs of the proposed program can be met. General needs for underground  experiments include depth (overburden); space; availability of  environmental radiation monitoring (muon, neutron, radon, Kr/Ar, etc.); muon and neutron veto: water or liquid scintillator; availability of calibration sources, including gaseous sources, neutrons; material screening capability; access for equipment and people; space for and availability of needed infrastructure (power, cooling, ventilation, etc) and local support, and underground machine shops.

 Three important recommendations from Snowmass 2013/P5 2014 were to form a new international collaboration to carry out a world-class long-baseline neutrino experiment hosted by the U.S. using neutrinos produced at Fermilab and detected 1300 km ($\sim$ 820 miles) away  at a depth of 1475 m (4850 feet) at the Sandord Underground Research Facility in South Dakota; perform a program of second generation (G2) dark matter experiments, and support one or more third-generation  (G3) direct detection dark matter experiments.  The first recommendation is now being realized by the LBNF/DUNE project and the excavation of the underground caverns (150 m in length × 20 m in width × 28 m height) for it  is well along in South Dakota. The DOE Office of Science has participated in the construction of two G2 dark matter experiments, LZ, also at SURF in South Dakota, which has already reported world-class results, and SuperCDMS at SNOLAB in Canada, which will start taking data in 2023.  The third of these recommendations,  to embark on a G3 dark matter experiment, is only in its earliest stages of R\&D and proposals for experiments.  A G3 O(100-ton) detector will be required to fully explore the WIMP space. This and a suite of smaller (less expensive, and  very innovative technologies) provide the best opportunities in the coming decade for a major discovery and resolution to the DM problem and are likely to need additional underground space.

 \medskip
 
Key conclusions from the Underground Frontier are:
\begin{description}
\item[UF Conclusion 1:] Leverage the Long-Baseline Neutrino Facility excavation enterprise to increase underground space at SURF in a timely and cost-effective way to permit siting of next-generation underground high energy physics research experiments.
\begin{itemize}  
\item Excavate and outfit  one or more new underground caverns at SURF at the depth of 4850'  to house at least one large next-generation experiment and some mid-size and small experiments.
\end{itemize}

\item[UF Conclusion 2:] Designate the Sanford Underground Research Facility as a U.S. Department of Energy User Facility.

\item[UF Conclusion 3:] Provide full support for the underground facilities hosting the Long Baseline Neutrino Facility (LBNF) and the Deep Underground Neutrino Experiment (DUNE).

\item[UF Conclusion 4:] Following the 2014 P5 Recommendation 20, R\&D and decision  making for a third-generation direct detection dark matter program should commence immediately to enable a construction start in the late 2020s.

\item[UF Conclusion 5:] To ensure a robust collection of scientific programs in underground facilities, support the enabling capabilities, technique development, and expertise required for underground experiments.
\end{description}
Taken together, these form a strategy and action plan for underground physics for the next decade and beyond.

 The work of this Frontier  is  carried out by the following Topical Groups; UF01, Underground Facilities for Neutrinos; UF02, Underground Facilities for Cosmic Frontier; UF03, Underground Detectors (Absorbed into UF05); UF04, Supporting Capabilities; UF05, Synergistic Research; and UF06, An Integrated Strategy for Underground Facilities and Infrastructure. Key observations of the Topical Groups are:

\smallskip\noindent {\bf Underground Facilities for Neutrinos and Cosmic Frontier} 
The LBNF/DUNE program is the  flagship U.S. high energy physics research program using accelerator neutrinos from Fermilab to study neutrino oscillations.
Cavern space for both the Phase I and Phase II parts of the DUNE detector is being excavated now at SURF. Future neutrinoless double beta decay experiments supported by nuclear physics programs, possible next-generation dark matter experiments, and  proposals for future large-scale detectors targeting measurements of neutrinos from  natural sources (e.g., supernova, solar neutrinos, geoneutrinos, etc) are expected to require underground facility space and infrastructure in the coming decade and beyond. Multiple new underground dark matter experiments are being planned (at large, medium,  and small scales). It remains an open question whether deeper experimental locations, e.g. 7400' at SURF,  are required for future neutrinoless double beta decay experiments. New suitable spaces must be available by the late 2020s to meet the demand. This demand may be met in North America by a proposed new  (beyond what is being done for DUNE) underground space at SURF or SNOLAB.

\smallskip\noindent {\bf Supporting Capabilities and Synergistic Research}
Future, larger experiments will increasingly require underground assembly with stricter radioactivity requirements. There will need to be larger, cleaner clean rooms, often with better radon-reduction systems and increased monitoring capabilities for ambient contaminants. Methods to assay dust deposition and
radon-daughter plate-out will need to be improved. There will be an increased need for underground machine
shops. Additionally, many labs are considering adding underground copper electroforming to mitigate against
cosmogenic activation, as has been successfully demonstrated at SURF. A survey was sent to all experiments that submitted white papers that used underground facilities asking them what their requirements were.

 \smallskip\noindent A survey of synergistic research in underground facilities showed such facilities are used by a range of communities including nuclear astrophysics, experiments probing fundamental symmetries, gravitational wave detection, and geology and geophysics. Interestingly, it appears quantum information science (QIS)—in particular quantum computing—may become a user of underground facilities. In all the cases identified, the synergistic research was either complementary or had dedicated facilities. In this way, no obvious conflicts between underground facility use for high energy physics and other research communities are predicted.

\smallskip\noindent {\bf An Integrated Strategy for Underground Facilities and Infrastructure}  An important  result of the work of this group  is an updated list of existing and planned underground facilities, including their main characteristics of interest to experiments, their availability for mid-sized and small experiments, the space and infrastructure available for large experiments, and any plans for new space.  These summaries are given in the Frontier report. The existing facilities worldwide are found to be fully subscribed. Moreover, emergent applications have been identified beyond neutrino and dark matter searches, including quantum information science, quantum computing, and atom interferometry that can also benefit from underground operations and add to the demand for space. The Underground Frontier concluded that future experiments and programs  and their enabling R\&D require new space. They propose a possible addition of the underground space at a depth of 4850 feet at SURF in South Dakota and possible additional space there at a depth of 7400 feet. Possible locations, near the DUNE excavation, are shown in the report. These new caverns should be able to accommodate the next generation of underground experiments, which are estimated to need 25m x 25m x 25m.  These new underground enclosures  would open up space for new experiments and would provide the opportunity for SURF to host next-generation dark-matter or neutrinoless double beta decay experiments. These new underground spaces would position the U.S. to be a leader in underground physics.
 
 \medskip

From the properties and nature of the neutrino to the direct measurement of galactic halo dark matter, research performed at underground facilities tackles a collection of precision knowledge of the constituents of the Standard Model of particle physics to broadly open-ended search well beyond the Standard Model.  In the Underground Frontier, we have identified the needs and requirements for underground facilities to continue to support the breadth of particle physics research planned in the coming 10–15 years.

\subsection{Snowmass Early Career Report~\cite{Agarwal:2022ldf}}

The Snowmass 2021 strategic planning process provided an essential opportunity for the U.S. High Energy Physics and Astroparticle community to come together and discuss upcoming physics goals and experiments. As this forward-looking perspective on the field often reaches far enough into the future to surpass the timescale of a single career, consideration of the next generation of physicists is crucial. The 2021 Snowmass Early Career (SEC) organization aimed to unite this group, with the purpose of  educating the newest generation of physicists while informing the senior generation of their interests and opinions. The definition of early career used for this document is “people within 10 years of their most recent degree with time allowance for any long-term leave from the field”. In practice, an overwhelming majority of the participants in the SEC group and those identifying with the term Early Career (EC) are at a graduate or post-graduate (e.g. postdoc) equivalent career stage.

One major activity of the SEC was the Snowmass Community Survey, which was designed by the SEC Survey Core Initiative team between April 2020 and June 2021 with the aim of collecting demographic, career, physics outlook, and workplace culture data on a large segment of the Snowmass community (both junior and senior). The team reviewed questions from past Snowmass surveys, developed new topics, and came to a consensus on the survey questions. The survey was released to the community on June 28, 2021, and had nearly 1500 total interactions before it closed on August 26, 2021. The survey questions broadly fall into seven categories: demographics, physics outlook, careers, workplace culture, diversity and racism, caregiving responsibilities, and the impacts of COVID-19. A high-level summary of the key findings and recommendations from the survey report can be found in the SEC contribution to this volume.

SEC organized a number of sessions at the Community Summer Study in Seattle, with a workshop-wide plenary that contained talks on the SEC core initiatives, survey reports, and long-term EC organizations across the High Energy Physics and Astroparticle community . Early career physicists took the lead in many other areas, including several successful events focused on industry careers, networking, and perspectives. They also hosted community discussions on mental health and invisible disabilities and were represented in panel discussions on community topics such as COVID-19 and career development. The level of interest in the early career perspective for Snowmass and the future of SEC led to the scheduling of an additional feedback session on early career issues in the final days of the meeting. Various informal social gatherings also took place to connect early career scientists beyond professional capacities.

Additional recommendations from the SEC community given in the SEC contribution to this volume cover the areas of ``Increasing Early Career Representation in Decision-Making Bodies",  ``Addressing Accessibility and Economic Equity", ``Robust Equity, Diversity, and Inclusion", and ``Empowering SEC for the next Snowmass Process".

\section{Special Cross-cutting Topics}

\subsection{Prospects and Evaluation of Future Colliders}

Extensive cross-frontier discussions and deliberations have taken place in the e$^{+}$e$^{-}$ Collider (EF-IF-AF) and Muon Collider (EF-TF-AF) forums and in the collider Implementation Task Force  (ITF). The outcomes of these activities are summarized below.
	
\subsubsection{e$^{+}$e$^{-}$ Colliders Forum \cite{Llatas:2022goe}}

The Snowmass 2021  e$^{+}$e$^{-}$ Collider Forum discussions covered a broad range of future electron-positron colliders from e$^{+}$e$^{-}$   Higgs factories - linear and circular - capable of providing a rich scientific program with sub-percent Higgs boson coupling measurements, to potential discovery machines for the next New Physics scales at 
O(10 TeV) center-of-mass energy. A circular Higgs Factory will provide the best precision for most Higgs couplings, but direct probing of
Higgs self-coupling and ttH couplings is deferred to a future higher energy proton collider. By comparison, a  linear
Higgs Factory will provide access to the Higgs self-coupling and ttH coupling. 

Key findings and recommendations of the Forum report \cite{Llatas:2022goe}, related to accelerators, include: 
\begin{itemize}
\item[a)] Higgs factories: primary consideration for the delivery of physics results is the start time of the physics program. Given the maturity of the technology, the ILC and the large storage rings hold the advantage of a possible early start of the program.  The ILC and CEPC both discuss possible starts late in the 2030s.  The FCC-ee would follow the HL-LHC program and start in the mid to late-2040s.  C$^3$ discusses a start similar to that of ILC assuming the completion of a technology demonstrator.  There is no published timeline for Fermilab-based collider options at this time.

\item[b)]	FCC-ee and CEPC have a signiﬁcant luminosity advantage and are able to complete the required runs at various luminosities faster but their larger civil engineering work requires signiﬁcantly more time and cost. An early start of the civil engineering construction of a circular machine is therefore key to the timely realization of physics. 
\item[c)]	The ILC and C$^3$ have cost, energy-reach, and polarization advantages but with lower luminosity, needing signiﬁcantly longer running time to achieve the same level of precision for measurements compared to circular machines.  From a potential siting point of view, all but the C$^3$ and HELEN machines require greenﬁeld sites. 
\item[d)]	Given the strong motivation and existence of proven technology to build an e$^{+}$e$^{-}$  Higgs Factory in the next decade, the U.S. should participate in the construction of any facility that has a ﬁrm commitment to go forward. Awaiting such commitment, the U.S. should also pursue research and development of multiple
options in this decade. This ensures that the global community will be able to begin constructing at least
one such machine in the following decade. The potential siting of a facility in the U.S. should also be pursued.
\item[e)]	Development of an O(10)-TeV scale e$^{+}$e$^{-}$ machine based on wakeﬁeld acceleration with suﬃcient luminosity capability for O(10) ab$^{-1}$ and energy-recovery technologies for improved power-to-luminosity costs, requires continued R\&D investment. Corresponding accelerator R\&D should focus on the development of self-consistent machine design parameters and the feasibility of attainment of collider speciﬁcations for the energy frontier.
\item[f)]In order to address the Detector R\&D and preparation of a Technical Design an R\&D program that
goes beyond generic R\&D is needed to address the specific challenges posed by the detector required for
e$^{+}$e$^{-}$ colliders. Such a program needs to start now for the technology to build a full-scale e$^{+}$e$^{-}$ collider detector to be ready when the HL-LHC program is completed.
\end{itemize}

\subsubsection{Muon Collider Forum: main findings and recommendations \cite{Black:2022cth}}

There has been a recent explosion of interest in muon colliders, as evident from a ten-fold increase in the number of related publications submitted to arXiv in the last couple of years. The topic generated a lot of excitement in Snowmass at meetings of the Energy, Theory, and Accelerator Frontiers  and continues to attract a large number of supporters, including many from the early career community. The Muon Collider Forum invited many experts to give their perspectives and to educate the broader community about the physics potential and technical feasibility of muon colliders. It facilitated a strong bond and exchange of new ideas between the particle physics community and accelerator
experts. Synergies with the Neutrino and Rare Processes Frontiers, as well as overlaps with nuclear science and industrial applications, were also extensively discussed. Finally, the Forum served as an interface between the U.S. community and the International Muon Collider Collaboration (IMCC) hosted by CERN. 

Key findings and recommendations of the Muon Collider Forum report \cite{Black:2022cth} include: 
\begin{itemize}
\item[a)]	A multi-TeV muon collider offers a spectacular opportunity for the direct exploration of the energy frontier. Offering a combination of unprecedented energy collisions in a comparatively clean leptonic environment, a high-energy muon collider has the unique potential to provide both precision measurements and the highest energy reach in one machine that cannot be paralleled by any currently available technology. Not  only does a $\ge$10 TeV muon collider provide a compact energy frontier machine, but also because at high energies emission of $W$ and $Z$ bosons from the initial state muon is enhanced, vector boson fusion becomes dominant. This is the reason why a muon collider is often also referred to as  a “vector boson collider".
The vector boson fusion channel provides the dominant production mechanism not only for single but also for multi-Higgs final states, and a novel exploration of the electroweak sector.
\item[b)]	Given the LHC results, a 10+ TeV lepton collider going beyond the classic precision versus energy dichotomy is an ideal machine.  The accelerator challenges of a multi-TeV Muon Collider are now considered to be overcome based on the technological advances achieved over the past decade. Significant progress has been made in the development of high-power targets and of high-field HTS solenoids, in the demonstration of the operation of normal-conducting RF cavities in magnetic fields, and in the self-consistent lattice designs of the various subsystems. No fundamental show-stoppers have been identified. Nevertheless, engineering challenges exist in many aspects of the design, and targeted R\&D is necessary in order to make further engineering and design progress. 
\item[c)]	There is an established plan and funding for muon collider-related R\&D activities in Europe and it is imperative for Snowmass/P5 to reestablish R\&D efforts in the U.S. and to enable participation of U.S. physicists in the International Muon Collider Collaboration (IMCC). 
\item[d)]	The most fruitful path forward toward the development of a conceptual design of a Muon Collider would be the engagement of the U.S. community in the IMCC. The U.S. Muon Collider community is well positioned to provide crucial contributions to physics studies, further advance the accelerator technology and detector instrumentation, and explore options for the domestic siting of a muon collider. An Integrated National Collider R\&D Initiative discussed in Snowmass \cite{Bhat:2022ybk} can  provide a much-needed platform for R\&D funding for such accelerator and detector development.
\item[e)]	Fermilab could be considered a candidate site for a Muon Collider with a center-of-mass energy reach at the desirable 10-TeV scale. The synergy with the existing/planned accelerator complex and neutrino physics program at FNAL is an additional stimulus for such an investment of effort. A set of Muon Collider design options, with potential siting at FNAL, could be a  contribution to discussions at the IMCC and the international committees to eventually form a global consensus decision on siting and selection of the Muon Collider. Having a pre-CDR document summarizing the design for the FNAL-sited Muon Collider in time for the next Snowmass is a good goal. The preparation of such a document will require a substantial, yet affordable, investment. Such an investment will reinvigorate the U.S. high-energy collider community and enable much-needed global progress toward the next energy frontier.
\item[f)] Major advancements in collider detector technologies have the potential to  dramatically reduce the impact of the challenges posed by beam-induced background (BIB). These technologies originate from HL-LHC detector upgrades and rely on 
 precision timing and incorporation of particle flow into the tracking and calorimetry systems, among other techniques. Further improvements are certainly conceivable and should be explored.
\end{itemize}

\subsubsection{Implementation Task Force (ITF) analysis \cite{Roser:2022sht}}

The collider Implementation Task Force (ITF) was organized and charged with developing metrics and processes to facilitate a comparison between projects. The ITF comprises 15 world-renowned accelerator experts from Asia, Europe, and the U.S., members of the Snowmass Early Career, and the EF and TF liaisons. Corresponding metrics have been developed for uniform comparison of the proposals ranging from Higgs/EW factories to multi-TeV lepton, hadron, and ep collider facilities, based on traditional and advanced acceleration technologies. An additional group consisted of versions of the proposals that could be located at FNAL. More than three dozen collider concepts have been comparatively evaluated by the ITF in terms of physics reach (impact), beam parameters, size, complexity, power, environment concerns,  technical risk, technical readiness, validation and R\&D required, cost, and schedule. The ITF report  documents the metrics and processes and presents comparative evaluations of future colliders \cite{Roser:2022sht}. 

Table~\ref{table:itf_collider_parameters}
is an excerpt from the ITF report Tables $1-5$ and lists the main parameters along with four columns with a summary value for technical risk (years of pre-project R\&D needed), technically limited schedule (years until first physics), project costs (2021 B\$ without contingency and escalation), and environmental impact (the most important impact is the estimated operating electric power consumption). The significant uncertainty in these values was addressed by giving a range where appropriate. 
\begin{table}[t]
\begin{tabular}{|l|c|c|c|c|c|c|} \hline
Proposal Name & c.m. energy & Luminosity/IP & Yrs. pre& Yrs. to 1st & Constr. cost& Electr. power \\ 
       &  [TeV]  & 10$^{34}$ cm$^{-2}$ s$^{-1}$& project R\&D & physics & [2021 B\$] & [MW] \\ \hline
FCC-ee$^{(1,2)}$ & 0.24 & 7.7 (28.9) & 0-2 & 13-18 & 12-18& 290 \\
CEPC$^{(1,2)}$   & 0.24 & 8.3 (16.6) & 0-2 & 13-18 & 12-18& 340 \\
ILC$^{(3)}$-0.25 &  0.25 & 2.7 & 0-2& $<$12  & 7-12 & 140 \\
CLIC$^{(3)}$-0.38 & 0.38  & 2.3 & 0-2 & 13-18 & 7-12 & 110 \\
C$^3$$^{(3)}$ &  0.25 & 1.3 & 3-5 & 13-18 & 7-12 & 150 \\
HELEN$^{(3)}$ & 0.25 &  1.4 &  5-10 & 13-18 & 7-12 & 110 \\ \hline
CLIC-3 & 3  &  5.9  & 3-5 &  19-24  &  18-30 &  $\sim$ 550 \\ 
$\mu\mu$Collider$^{(1)}$-3 & 3  &  2.3(4.6)  & $>$10 &  19-24 & 7-12 & $\sim$230 \\
FNAL$\mu\mu$$^{(1)}$ & 6-10 & 20(40) & $>$10 & 19-24  & 12-18 & $\sim$300 \\
FCC-hh$^{(1)}$ & 100  & 30(60) & $>$10  &  $>$25 &  30-50 & $\sim$560 \\
SPPC          & 125  & 13(26) & $>$10  &  $>$25 &  30-80 & $\sim$400  \\ \hline
\end{tabular}
\caption{Main parameters of several collider proposals evaluated by the ITF: Higgs/EW factories, multi-TeV lepton collider proposals (3 TeV center of mass (COM) energy options), colliders with 10 TeV or higher parton COM energy (see the full list in \cite{Roser:2022sht}).  The parenthetical superscripts next to the name of the proposal in the first column indicate (1) total peak luminosity for multiple IPs is given in parenthesis; (2) energy calibration possible to 100 keV accuracy for $M_Z$ and 300 keV for $M_W$; (3) collisions with longitudinally polarized lepton beams have substantially higher effective cross sections for certain processes. For each proposal, the ITF estimates are given on the years of pre-project R\&D, years to first physics after the decision to proceed, construction cost (including explicit labor, no escalation, and no contingency), and facility electric power consumption. }
\label{table:itf_collider_parameters}
\end{table}
The years of required pre-project R\&D is just one aspect of the technical risk, but it provides a relevant and comparable measure\ of the maturity of a proposal and an estimate of how much R\&D time is required before a proposal could be considered for a project start (CD0 in the U.S. system). Pre-project R\&D includes both feasibility R\&D, R\&D to bring critical technologies to a technical readiness level (TRL) of $4-5$, as well as necessary R\&D to reduce cost and electric power consumption. The extent of the cost and power consumption reduction R\&D is not well defined and it was assumed that it can be accomplished in parallel with the other pre-project R\&D. Nevertheless this R\&D is likely most important for the realization of any of these proposals. (Note that by using the proponent-provided luminosity values, ITF chose not to evaluate the risk of failing to achieve this aspect of performance. However,  performance risk was included in the evaluation of technical readiness in the ITF report.) The time to first physics in a technically limited schedule is most useful to compare the scientific relevance of the proposals. It includes the pre-project R\&D, design, construction, and commissioning of the facility.

The total project cost follows the U.S. project accounting system but without escalation and contingency. Various parametric models were used by ITF to estimate this cost, including the cost estimated by the proponents. The cost estimate uses known costs of existing installations and reasonably expected costs for novel equipment. For future technologies, pre-project cost reduction R\&D may further reduce the cost estimates used by the ITF.

Finally, the electric power consumption is for a fully operational facility including power consumption of all necessary utilities. The ITF used the information from the proponents if they provided it, otherwise, it made  rough estimates based on expert judgment. Pre-project R\&D to improve energy efficiency and develop more 
energy-efficient accelerator concepts, such as energy recovery technologies, have the potential to reduce electric power consumption significantly from the values listed in the tables.

Any of the future collider projects will constitute one of the largest $-$ if not the largest $-$ science facilities in particle physics. The cost, the required resources, and, maybe most importantly, the environmental impact in the form of large energy consumption will approach or exceed the limit of affordability. The ITF suggests that the Snowmass 2021 final report recommends that R\&D to reduce the cost and the energy consumption of future collider projects be given high priority.

\subsection{Dark Matter Complementarity~\cite{Boveia:2022adi}}

The fundamental nature of dark matter is a central theme of the Snowmass 2021 process, extending across all Frontiers. In the last decade, advances in detector technology, analysis techniques, and theoretical modeling have enabled a new generation of searches while broadening the types of candidates we can pursue. Over the next decade, there is great potential for discoveries that would transform our understanding of dark matter. A strong portfolio of experiments that \textbf{delves deep, searches wide, and harnesses the complementarity between techniques} is key to tackling this complicated problem, requiring expertise, results, and planning from all Frontiers of the Snowmass 2021 process. In a cross-cutting contribution to the Snowmass process, participants across Frontiers collaborated to outline a road map for dark matter discovery. 

Complementarity drives discovery in multiple ways. The space of viable DM candidates and their properties is large, and a single experimental approach cannot test all the possibilities; a diverse range of techniques provides access to a much broader ensemble of DM scenarios and properties. Different approaches offer unique discovery sensitivity to distinct scenarios and regions of parameter space, and results from any one class of searches can continuously inform the interpretation of other measurements. Lastly, different DM experiments can be co-located and/or profit from the same or similar technological infrastructure. Each of these facets of complementarity is illustrated in the DM Complementarity Report through summaries of the efforts across all Frontiers and four case studies of models generating significant interest in the Snowmass process: minimal WIMP DM, BSM and vector portal DM, sterile neutrino DM, and wave-like DM including QCD axions.

The DM Complementarity Report proposes the following strategy for discovering the fundamental nature of DM:

\noindent \textbf{Experiments at all scales are needed} to cover the broad range of theoretically motivated parameter space. Existing and planned large-scale facilities across the HEP Frontiers have exceptional potential to discover fundamental properties of DM. We should commit to scaling up mature technologies that can promise significant sensitivity improvements, maturing potentially transformative new technologies, and supporting efforts to maximize and make accessible large projects' science output in the search for DM. We should support DM opportunities at multi-purpose experiments, including cosmology experiments, from the design stage through analysis. At smaller scales, execution of the existing Dark Matter New Initiatives (DMNI) program and similar future calls are necessary to build the most compelling portfolio of DM experiments, develop experience in project execution, and accelerate the pace of discovery.

\noindent \textbf{Understanding the fundamental nature of DM is a worldwide endeavor.} Coordination and cooperation across borders are critical for discovery. While building a strong U.S.-based program, we should pursue opportunities to leverage key U.S. expertise as a collaborative partner in international projects and play a leadership role in this major discovery area.  

\noindent \textbf{A strong theory program is essential} to make connections between experimental Frontiers and take full advantage of new developments in simulation and analysis techniques. Theorists' input has been and will be critical for developing innovative new approaches to better understand and detect DM, and for determining how to predict and relate signals across a range of experimental probes.

\noindent \textbf{Searches for DM benefit from cross-disciplinary expertise}, with examples ranging from nuclear physics to metrology, and astrophysics to condensed matter and atomic physics. Mechanisms to support such interdisciplinary collaborations should be established. 

\noindent \textbf{Research funding is critical to enable discovery} and build on new capabilities, across all Frontiers and project scales, in projects focused specifically on DM and to support DM analyses at multi-purpose experiments. The number of active efforts exploring DM has increased tremendously in the past decade, without a concomitant increase in research funding. Without such support, the community will not be able to execute the complementary dark matter program, decreasing the chances of solving the mystery of DM. 

DM presents a fundamental puzzle to particle physics. To make progress on this challenging problem, maximize the chances of a transformative discovery, and fully elucidate the properties of DM and related new physics in the event of such a discovery, we advocate for a cross-frontier effort incorporating multiple complementary approaches to the problem. A decade of coherent cross-frontier DM exploration is an opportunity that should not be missed.

\subsection{Synergies and Complementarities with Astrophysics
and Nuclear Physics 
}

Particle physics has very significant synergies with related fields of science, most notably nuclear physics and astrophysics. The practical and intellectual synergies between particle physics, astronomy, astrophysics, astroparticle physics, and cosmology have been growing exponentially since the early critical discoveries in cosmic rays: the existence of the positron, observed in cosmic ray emulsions in 1932 and quickly recognized to be the anti-particle of the electron required by the Dirac equation; the totally unexpected discovery of the muon in 1936 -- just the tip of the iceberg of flavor; in 1947, the completely surprising discovery of mesons ushered in hadron physics which, now in retrospect, provided the first hint of QCD's quark substructure.  
Experimental particle and nuclear physics emerged together, going back to Lawrence's invention of the cyclotron. 
Cosmology and General Relativity  became real experimental sciences with the discovery of Cosmic Microwave Background radiation (using instrumentation developed by physicists).

These seeds matured into new branches of physics: particle physics, astrophysics, and nuclear physics focused on related but distinct topics over the years.  Nonetheless, important areas of overlap remain.  Many scientists work on the interface between these fields and indeed participate in scientific endeavors led by several communities. There is also significant complementarity and common development of experimental and theoretical tools, instrumentation, and computational and analysis techniques. The issues of workforce development, equity, inclusion, diversity, and community engagement are also common among these fields and warrant common approaches.

Theoretical approaches in particle, nuclear, astrophysics, and cosmology are highly complementary. The fields start from the same fundamental Lagrangian and deploy common tools and techniques. Effective Field Theory -- developed into a powerful tool within particle theory -- is now a standard part of the toolkit for nuclear physics; recently, the application of the EFT technique has significantly increased the sensitivity of analyses of cosmological data. AdS/CFT, once the domain of formal particle theory, has produced a lower bound for the shear viscosity of the quark-gluon plasma that is in fact saturated in heavy-ion collisions. Advanced computational techniques such as AI and lattice gauge theory are other shared examples.
Contributions from several fields are often needed to properly interpret observations.  For instance, predictions of the neutron star mass-radius relation and other observables depend on both the equation of state at the cores of neutron stars, pushing the boundaries of our understanding of QCD, and on the properties of the nuclear neutron skin, which rely on nuclear theory and are verified by precision measurements of parity violation in electron scattering off heavy nuclei;  models of exploding supernova with verifiable predictions rely on measurements of nuclear reaction rates and properties of the neutrinos. The standard cosmological model is sensitive to the number of active neutrinos and their masses. 
At the same time, understanding nuclear reactions and nuclear properties are becoming increasingly important for precision measurements in HEP, including the reconstruction of neutrino interactions at DUNE. Quantum computing is another area where both nuclear and particle physicists have overlapping interests, and where algorithms and computational techniques developed by one field can benefit the other.

As within the HEP program, experiments and projects in nuclear physics, astrophysics, astroparticle physics and observational cosmology benefit from joint developments of scientific instrumentation for particle detection, high-performance computing, and development of novel computational paradigms including AI/ML and quantum computing. Low-background experiments in HEP, NP, and astroparticle physics use shared underground space, and all these programs could benefit from or even require additional underground infrastructure going forward. The HEP program has the advantage of having a dedicated funding source for ``blue-sky" Detector R\&D, which has seeded and supported many novel developments before they could be deployed by specific programs and projects. Such R\&D is crucial for the overall health and future prospects of all these fields.  
The coordinated development of novel technologies across fields and cooperation in their engineering and deployment would be broadly beneficial, and to other areas of science as well.

Two new initiatives that have direct support from the Office of Science  through both the HEP and NP programs perhaps warrant special mention. The development of AI/ML techniques and algorithms and their application to domain sciences such as HEP and NP are one of the new congressional initiatives. HEP and NP scientists have already been actively developing these algorithms for decades, but new directions such as the deployment of AI to improve the efficiency and reliability of accelerator and detector systems, for automatic fault detection, and other similar approaches to complex systems could find industrial applications and potentially benefit the broad society.  
Astrophysics and cosmology also make powerful use of AI/ML in simulating and interpreting astrophysical and cosmological measurements.  With limited funds  available for AI/ML, it is important that our approaches are complementary, coordinated, and cohesive -- instead of being competitive.   

A second new congressional initiative with joint HEP-NP-Astro interest is Quantum Information Science. Quantum computing could one day lead to an exponential increase in performance, and allow us to tackle calculations that would otherwise be impossible. The development of algorithms that demonstrate a quantum advantage for HEP and NP is an important part of the program. Similarly, the development and deployment of quantum sensor technologies directly benefit measurements of CMB, gravitational waves, searches for dark matter, and rare processes such as $0\nu\beta\beta$ decay and EDMs. One should also point out that HEP and NP experiments also develop technologies that benefit QIS directly. Very large cryogenic installations require engineering solutions that are pioneered by our fields. The development of cryogenic electronics, in particular using industry-standard CMOS processes, is another area with direct QIS applications. And since nuclear physicists have demonstrated that the coherence time of superconducting qubits is affected by the presence of ionizing background radiation, low-radioactivity techniques, and deployment of underground installations, developed jointly by HEP and NP, are also becoming increasingly important.

Additional examples of the synergies and complementarities among HEP, Astronomy, Astrophysics, and Nuclear Physics are  highlighted below. 

\subsubsection{High Energy Physics and Astrophysics }

High Energy Physics research areas straddling the Particle-Astrophysics-Cosmology boundaries include dark matter, dark energy, and inflation, and the use of multiple cosmic probes (high-energy neutrinos, gammas, cosmic rays, and gravitational waves) and cosmological structure observations to constrain fundamental particle physics.  The discovery of neutrino mixing and hence mass emerged from anomalies in the solar neutrino flux relative to predictions of the standard solar model; now, cosmological limits on the sum of neutrino masses are more constraining of neutrino masses than direct measurements in particle and nuclear experiments. Early Universe Cosmology is a central part of particle physicists' toolbox for revealing physics beyond the SM, for instance, the limits on the existence of not-yet-discovered light particles from the observed abundances of primordial nuclei and other cosmological probes are superb complements to accelerator studies.  An emerging very important input for understanding QCD and nuclear physics are the discoveries of neutron stars with masses distinctly above $2 M_\odot$ and breakthroughs in measuring the neutron star mass-radius relation by LIGO-Virgo with gravitational waves, and with NASA's X-ray telescope NICER; the results are challenging our understanding of QCD. The list goes on and on. 

Reciprocally, particle physics gives invaluable contributions to Astrophysics and Cosmology.  Two of the three themes  prioritized by the 2020 Astro Decadal Survey are strongly aligned with the goals articulated in the Cosmic Frontier: ``New Messengers and New Physics", with particular stress on multi-messenger astrophysics which will dramatically enhance the utility of cosmic particle and gravitational wave probes, as well as ``Cosmic Ecosystems", aimed at improving our understanding of the astrophysical aspects of the evolution of the universe.

The underlying scientific motivation to study common topics is often different in the HEP and astrophysics communities, with astronomers interested for instance in questions like understanding the properties of galaxies and how they have evolved, and the menagerie of objects they contain, or the nature of the sources of the highest energy cosmic rays, whereas particle physicists are more focused on how the nature of the dark matter particle can be inferred from features imprinted on astrophysical systems, or distinguishing a potential dark matter annihilation signal from astrophysical background, or on using ultrahigh-energy cosmic particles to probe BSM physics far beyond accelerator energies.   Since an essential part of discovering new phenomena through astrophysics and cosmology is understanding the baseline expectations of the standard model of cosmology and the many other aspects of astrophysics that produce foregrounds and backgrounds to some possible BSM signal, physicists and astrophysicists have a common purpose in these explorations.  

Technologically, there is also a strong synergy between particle physics and astrophysics.
A major triumph of the past decade in astrophysics was the discovery of gravitational waves by Advanced LIGO, largely funded by the NSF, which made essential use of developments spearheaded by HEP in vacuum systems, alignment challenges, distributed control, and data acquisition systems.  A proposed future much larger gravitational wave observatory would further benefit from HEP expertise in km-scale facility placement, design, and execution, as well as the operation of multi-billion-dollar facilities.  The scientific rewards of such an instrument for particle physics span from an independent, accurate measurement of the Hubble constant, to the determination of the QCD equation of state in neutron stars, to strong constraints on theories of modified gravity, to signatures of phase transitions in the Early Universe, and much more.  Another triumph has been the detection of ultra-high-energy astrophysical neutrinos with the IceCube detector at the South Pole.

Presently coming to fruition, the Rubin Observatory was the highest priority ground-based project of the 2010 Astro Decadal survey.  Among its goals are mapping the structure of the Universe and its expansion history in much greater detail and precision than at present, which is of key importance for particle physics because of what it can tell us about inflation and dark energy, and possible tensions within $\Lambda$CDM - as well as a multitude of applications in high energy and other parts of astrophysics.   Its remarkable camera, LSSTCam, has been built at SLAC with DOE funding and expertise.  LSSTCam presented a number of technical challenges that had not been encountered with other astronomical cameras: large modular arrays of 4-side buttable detectors with extremely tight positional tolerances; fast readout necessitating a high degree of parallelization with several thousand channels of readout electronics;  finally, the positioning of the camera within the beam exiting the secondary mirror implied a severely constrained envelope to work in, with tight thermal requirements.  These features are more characteristic of the inner detectors associated with collider experiments than they are of typical ground-based astronomical instrumentation.  The team assembled to design and build the camera (largely at SLAC, BNL, and the IN2P3 labs in France) mostly involved technical people from the HEP community who had experience with such issues.

A current example of astro-EPP technical synergy is in building CMB-S4, one of the key projects identified by the last P5, which will have unprecedented sensitivity to the physics of the Early Universe.  The CMB-S4 transition-edge-sensor bolometric detectors, the superconducting circuity to make them dual-polarization and dual-band sensitive, and the low-noise SQUID-based multiplexed readout were all advanced by DOE-funded research.  The laboratories involved include SLAC, LBNL, ANL, and FNAL.   Another crucial HEP contribution has been the capability of LBNL NERSC computing to support the large data sets and the computation of the large sets of simulations needed to properly conduct the analysis.

These examples hardly scratch the surface of the multitude of ways that prior and current investments in particle physics R\&D provide tools and expertise for astrophysics and other areas of science. 

\subsubsection{High Energy Physics and Nuclear Physics}

The Nuclear Physics community is currently developing a new Long Range Plan (LRP) for Nuclear Science, aimed to be released in 2023. The fact that it coincides with the Snowmass and P5 process is an opportunity to further develop synergies, define common strategies when necessary, and maximize the scientific and societal impact of both fields.

Heavy ion measurements at the LHC are performed with the ATLAS and CMS detectors, as well as a dedicated detector ALICE. This complementarity and collaboration between the scientists working on high-energy colliders will grow with the advent of the Electron-Ion Collider (EIC). 
EIC, the top priority new facility construction in the 2015 Long Range Plan for Nuclear Science, will be the premier facility for precision studies of nuclear and nucleon structure, including the polarization and angular momentum observables, and the transition between the perturbative and non-perturbative regimes of QCD. 
The fundamental interest in the understanding of quantum chromodynamics at all regimes is a common thread between HEP and NP. Precision studies of the QCD, measurements of the parton distribution functions with high precision, and computational techniques that would be tested by the EIC could inform future measurements at the Energy Frontier. Conversely, nuclear collider experiments, including at RHIC and at the EIC, benefit from the tools and techniques developed by HEP, in particular jet reconstruction, precision tracking, heavy flavor tagging and reconstruction, calorimetry, particle ID, and others.

The overlap of scientific interest between nuclear and high-energy physics is probably greatest in the Neutrino Frontier. Snowmass has discussed in detail the many overlapping interests of neutrino scientists, and the desire of the scientific community to find ways to enhance the collaboration in neutrino physics while respecting the specific interests of the various stakeholders. The most pressing questions -- the fundamental nature of the neutrino and the violation (or conservation) of lepton number, the number of neutrino species, the absolute scale of the neutrino mass and its fundamental electromagnetic properties, as well as precision studies of the ways neutrinos interact with matter -- are of great interest to both communities. In fact, studies of neutrino oscillations are performed by both communities and supported by both OHEP and ONP, depending on the neutrino sources; some experiments, including  SNO and KamLAND, have been supported by both OHEP and ONP. These are just a few examples of constructive and fruitful collaboration between the two fields. 

Neutrinoless double beta ($0\nu\beta\beta$) decay remains an important topic for both HEP and NP. The next-generation (``ton-scale") experiments will be stewarded by the Office of Nuclear Physics, as recommended by the 2015 LRP. It is also well recognized by the community and the funding agencies that the quest for, or perhaps precision measurements of, $0\nu\beta\beta$ decay must continue beyond these experiments, independently of whether the discovery is made at the ton scale. The Snowmass process has showcased a plethora of novel ideas and experimental techniques aimed at developing experiments with sensitivity beyond the parameter space defined by the inverted ordering of light neutrino masses. At this scale, it would be especially important to develop multi-purpose detectors and maximize the scientific return on investment. Synergies in technologies between $0\nu\beta\beta$ and dark matter experiments, $0\nu\beta\beta$ and CE$\nu$NS experiments, as well as $0\nu\beta\beta$ and the long-baseline neutrino program, were discussed at Snowmass. Deployment of such multi-purpose detectors may require cooperation between multiple communities and funding agencies. 

Measurements of other neutrino properties such as the absolute neutrino mass or the neutrino magnetic moment, and new technological developments that could be pursued by both fields to enhance sensitivity, have also been discussed at Snowmass. Searches for sterile neutrinos -- with different sources and detection techniques but in overlapping parameter space -- are done by both HEP and NP. 
Measurements of neutrino cross sections, and of other processes that could improve the precision of neutrino  reconstruction at DUNE, are examples where NP experiments could provide crucial input to a flagship HEP program. Examples of such  HEP/NP collaborative experiments include electron-nucleus scattering experiments at Jefferson Lab, as well as neutrino scattering experiments at Oak Ridge National Laboratory’s Spallation Neutron Source (a DOE Basic Energy Sciences facility), which produces very high-quality, high-intensity neutrinos.

Precision tests of the Standard Model is another area where nuclear and high energy physicists complement each other, and in some cases directly collaborate. The Muon $g-2$ experiment is a great example of a project funded primarily by HEP and operated at a HEP facility in which NP-funded scientists played a crucial role, brought specific expertise, and made substantial contributions. 
Precision measurements of electroweak couplings of electrons and quarks through parity-violating electron scattering at Jefferson Lab and at the EIC will provide an important window to search for new physics and will complement electroweak observables at the LHC and at the $Z$ pole. 
Measurements of the CKM parameters, in particular those connected to the first generation, are often done in nuclear processes and are of great significance for the tests of the unitarity of the CKM matrix. Searches for Lepton Number and Baryon Number violation, EDMs, tests of lepton universality, and searches for dark sector bosons are other areas of joint interest. Some of the experiments with HEP funding are performed at NP-funded facilities (e.g. dark sector searches at Jefferson Lab) and therefore require collaboration between the offices. 
Searches for EDMs, enhanced by the advent of new atomic techniques and the use of molecules and deformed, often unstable, nuclei could be performed at nuclear facilities, such as FRIB. 
Conversely, the Advanced Muon Facility proposed at Fermilab could enable measurements that would benefit nuclear physics experiments directly, e.g. through precision measurements of muon capture. 

The U.S. Nuclear Physics community (in particular the DOE Office of Nuclear Physics) operates multiple accelerator facilities. Among the larger ones are the Relativistic Heavy Ion Collider (RHIC) at BNL, the Continuous Electron Beam Accelerator Facility (CEBAF) at Jefferson Lab, the Facility for Rare Isotope Beams (FRIB) at Michigan State University, and a score of smaller accelerator facilities at national labs and universities. The highest priority for new facility construction in the 2015 LRP is the Electron-Ion Collider (EIC), to be built and operated at BNL by the beginning of the next decade. 
This facility is the only particle collider expected  to be operated in the U.S. in the 2030s.  
It will require significant innovations in accelerator physics and technology and will employ a substantial fraction of the accelerator physics workforce in the U.S. The developments stimulated by HEP -- high-field magnets, accelerating structures, power delivery systems, polarized sources, final focus systems, beam instrumentation, and precision control of the beam dynamics -- will be deployed, further developed, and thoroughly tested by the EIC project. EIC will play an essential role in recruiting, training, and supporting the future accelerator frontier workforce. This workforce will be essential for any U.S. involvement in future large-scale accelerator projects for HEP and other fields.

\section{Summary}

The aspirations of the U.S. HEP community as informed by the Snowmass 2021 process can be succinctly summarized as follows:
\smallskip
\begin{quote}
{\em Lead the exploration of the fundamental nature of matter, energy, space and time, by using ground-breaking theoretical, observational, and experimental methods; developing state-of-the-art technology for fundamental science and for the benefit of society; training and employing a diverse and world-class workforce of physicists, engineers, technicians, and computer scientists from universities and laboratories across the nation; collaborating closely with our global partners and with colleagues in adjacent areas of science; and probing the boundaries of the Standard Model of particle physics to illuminate the exciting terrain beyond, and to address the deepest mysteries in the Universe.}
\end{quote}
This volume  presents the main conclusions of  Snowmass 2021, whose goal was to examine the status and scientific goals of U.S. HEP and to propose essential scientific programs to pursue, and to thereby provide community input to the 2023 DOE/NSF ``Particle Physics Project Prioritization Panel" (P5) charged with updating the strategic plan for U.S. HEP for the coming decade and beyond.
The work included herein was the output of ten Snowmass Frontiers and their  Topical Groups, which comprised a broad array of ground-breaking scientific research topics and the underlying technology and infrastructure needed to execute them, as well as a forum to examine how the U.S. HEP community can become more representative of and responsive to all members of our community and can engage with society as a whole.

The consensus of the Snowmass 2021 community was that the five science Drivers formulated by the 2014 P5 panel continue to be appropriate for the next decade. Specifically, the questions that should guide the planning for the next decade and beyond include:
\begin{enumerate}
\item Use the Higgs Boson as a Tool for Discovery, 
\item Pursue the Physics Associated with Neutrino Mass, 
\item Identify the New Physics of Dark Matter, 
\item Understand Cosmic Acceleration: Dark Energy and Inflation, 
\item Explore the Unknown: New Particles, Interactions, and Physical Principals.
\end{enumerate}
As a result of the breadth of flavor physics, its potential for discovering BSM physics,
and hints of possible new physics in the current data, the Rare Processes and Precision Measurement Frontier proposes that the
upcoming P5 adds a new science Driver: {\it flavor physics as a tool for discovery}.

For the future U.S. HEP experimental program, an important general conclusion of Snowmass was that the completion, construction, and operation of currently approved projects, especially those prioritized by P5  in 2014, such as the HL-LHC, LBNF/DUNE, LSST/Rubin Observatory, and Mu2e programs, is essential. Furthermore, as the construction of approved large experimental projects come to a completion, U.S. leadership in a broad and complementary set of new large experimental programs covering the general scientific areas of the Energy, Neutrino, Cosmic, and Rare Process and Precision Measurements Frontiers is needed to provide opportunities for incisive new discoveries. The specific projects considered are listed in section \ref{section:major-projects} of this document (without regard to cost or priority, since that is to be done by P5), and described in greater detail in the Frontier Reports in this volume as well as in the Topical Group Reports and white paper references contained therein. In addition, the inclusion in the U.S. HEP experimental portfolio of a broad and complementary set of mid- and small-scale experiments is vital to maintaining the diversity of physics topics investigated and of timescales to completion that are essential for a vibrant program.  Medium- and smaller-scale projects are discussed more fully in section \ref{section:mid-scale-projects} and in the reports in this volume.

A healthy HEP program requires more than projects, however, and the Snowmass community emphasized the need for robust support for physics research programs at universities and national laboratories. This consideration includes the strong and continued support for all aspects of particle theory, comprising the interconnected themes of fundamental theory, phenomenology, and computational theory. 
Also needed is greater support for the infrastructure and enabling technologies, namely accelerators, instrumentation and detectors, and computation --- both targeted support for specific future projects and generic support to develop new enabling technologies. 

Input received during the Snowmass process highlighted that HEP projects and research benefit from collaboration with adjacent scientific disciplines, such as Nuclear Physics, Astronomy and Astrophysics, Gravitational Physics, Atomic and Molecular Physics, Quantum Information Science and Sensing, and Materials Science, and contributes to them in return. Opportunities to strengthen and expand such collaborations are mutually beneficial and should be pursued.

Of particular importance in considering the future of the U.S. HEP program is to understand and support the needs of its early career physicists. Snowmass participants recommended strengthening the connections between early career and senior researchers, fostering the professional success of early career physicists both within and outside of academia, and ensuring that the voices of younger scientists are well-represented in physics and community planning.

Mindful of the continuing lack of diversity in particle physics, a strong consensus exists in the HEP community demanding the development of  a cohesive, strategic approach to promoting diversity, equity, and inclusion in high-energy physics, in collaboration with funding agencies, universities, adjacent scientific disciplines, APS/DPF and others. Specifically, the community should institute a broad array of practices and programs to reach and retain the diverse talent pool needed for success in achieving our scientific vision and to address the persistent under-representation of women scientists, LGBT+ scientists, scientists who are Black, Indigenous, and people of color (BIPOC), and scientists with disabilities.

The U.S. HEP community is generously supported by and embedded in the nation as a whole. The HEP community must therefore engage in a coordinated way with five other interrelated communities: academia, the K-postdoc education community, private industry, government policymakers, and broader society. Snowmass participants identified the need for a structure for formulating a coordinated approach to achieve these goals which is provided with the resources needed for success.

The Snowmass 2021 process highlighted the breadth, depth, and effectiveness of the U.S. HEP program
in exploring the fundamental nature of matter, energy, space, and time. The remarkable progress that has been made in probing the boundaries of the Standard Model of particle physics since 2013 was analyzed, and different and complementary methods to move forward in uncovering the mysteries of nature were examined. Highlights and opportunities that were identified are described in detail in the reports in this volume, as well as in the numerous topical group reports and white papers summarizing the work done for Snowmass.  The members of the U.S. high-energy physics community left the Seattle Community Summer Study workshop with an  appreciation of the great opportunities present in each Frontier, the interconnections between  the Frontiers, and the connections to programs in the rest of the world.  Through the Snowmass process, the U.S. HEP community has created an integrated vision to make progress in the coming decades to advance our understanding of the Universe.

\section*{Acknowledgements}

Several thousand people participated in the Snowmass 2021 process and provided their scientific input. Their enthusiasm and dedication is what made Snowmass 2021 a success! We also acknowledge the crucial contributions of the host institutions of all the virtual and in-person preparatory meetings.
The Snowmass Steering Group, Advisory Group, and Frontier Conveners Group provided overall coordination of Snowmass 2021. The Community Summer Study and Workshop was held at the University of Washington in Seattle. We thank the University and the CSS Scientific Program Committee and  Local Organizing Committee for producing a well-organized and successful ten-day-long meeting. The  names of the members of all these groups are listed in the acknowledgments at the end of the full report. 
Finally, we thank the four units of the APS whose work is closely related to HEP, namely Astrophysics, Nuclear Physics, Gravitational Physics, and Physics of Beams, for their many contributions.

\vfill\eject

\end{document}